\pdfoutput=1
\documentclass[pra,floatfix,twocolumn,tightenlines,superscriptaddress,amsmath,nofootinbib]{revtex4-1}
\usepackage[T1]{fontenc}
\usepackage[bbgreekl]{mathbbol}
\usepackage[normalem]{ulem}
\usepackage{graphics,graphicx,amsthm,amsmath,amssymb}
\usepackage{color}
\usepackage{mathtools}
\usepackage{setspace}
\usepackage{mathrsfs} %Ralph's Smoth's Formal Script Font
\usepackage{float}
\usepackage[bottom]{footmisc}
\usepackage[caption = false]{subfig}
\usepackage{bm}
\usepackage{stackengine}
\usepackage{url}
\usepackage[english]{babel}
\usepackage{wasysym}
\usepackage{grffile}

\DeclareSymbolFontAlphabet{\mathbbm}{bbold}
\DeclareSymbolFontAlphabet{\mathbb}{AMSb}

\def\({\left(}
\def\){\right)}

\def\bra#1{\mathinner{\langle{#1}|}}
\def\ket#1{\mathinner{|{#1}\rangle}}
\def\braket#1#2{\mathinner{\langle{#1}|#2 \rangle}}
\def\avg#1{\mathinner{\langle{#1} \rangle}}

\def\de{\delta\epsilon}
\def\dmin{d_{\rm min}}

\def\dminmin{d_{{\rm abs \,min}}}
\def\cVt{V_{\rm typ}}
\def\stn{\{0,1\}^n}
\def\Mr{\Omega_d}
\def\dminr{d_{\rm min}^{\rm res}}

\def\PDF{\rm PDF}

\def\Sti{\Sigma_{*}^{\prime\prime}} 
\def\Str{\Sigma_{\rm typ}^{\prime}}
\def\Styp{\Sigma_{\rm typ}^{\prime\prime}} 
\def\Sr{ \Sigma^{\prime}}
\def\Si{ \Sigma^{\prime\prime}}
\def\Re{{\rm Re}} 
\def\Im{{\rm Im}}

\def\exp{{\rm exp}}
\def\cl{{\rm cl}}
\def\pt{t_{\rm PT }}
\def\B{{B_\perp}}
\def\Bs{{B_\perp^2}}
\def\e{{\epsilon}}

\def\cA{{\mathcal A}}

\def\cP{{\mathcal P}}

\def\cO{{\mathcal O}}
\def\cE{{\mathcal E}}
\def\cH{{\mathcal H}}
\def\cV{{\mathcal V}}

\def\scC{{\mathscr C}}
\def\scP{{\mathscr P}}

\def\scH{{\mathscr H}}

\def\scG{{\mathscr G}}
\def\scF{{\mathscr F}}

\def\scS{{\mathscr S}}

\def\pzw{{\mathpzc{w}}}
\def\pzp{{\mathpzc{p}}}

\def\fp{{\mathfrak p}}

\newcommand*{\medcap}{\mathbin{\scalebox{1.25}{\ensuremath{\cap}}}}%

\DeclareMathAlphabet{\mathpzc}{OT1}{pzc}{m}{it}

\DeclareMathOperator\arctanh{arctanh}
\DeclareMathOperator \arccoth{arccoth}

\def\disteq{\mathrel{\ensurestackMath{\stackon[1pt]{=}{\scriptstyle d}}}}

\makeatletter
\def\l@subsection#1#2{}
\def\l@subsubsection#1#2{}
\makeatother

\begin{document}

\title{Non-ergodic delocalized states for efficient population transfer within a narrow band of the energy landscape}

\author{Vadim N. Smelyanskiy}
\affiliation{Google, Venice, CA 90291, USA}
\author{Kostyantyn Kechedzhi}
\affiliation{Google, Venice, CA 90291, USA}
\affiliation{QuAIL, NASA Ames Research Center, Moffett Field, California 94035, USA}
\affiliation{University Space Research Association, 615 National Ave, Mountain View, CA 94043 }
\author{Sergio~Boixo}
\affiliation{Google, Venice, CA 90291, USA}
\author{Sergei V. Isakov}
\affiliation{Google, 8002 Zurich, Switzerland}
\author{Hartmut Neven}
\affiliation{Google, Venice, CA 90291, USA}
\author{Boris Altshuler}
\affiliation{Physics Department, Columbia University, 538 West 120th Street, New York, New York 10027, USA}

\date{\today}

\begin{abstract}

We address the long-standing problem of the structure of the low-energy eigenstates and long-time coherent dynamics in quantum spin glass models.  This problem remains challenging due to the complex nature of the distribution of the tunneling matrix elements between the local minima of the energy landscape. We study the transverse field induced quantum dynamics of the following spin model: zero energy of all spin configurations except for a small fraction of spin configuration ("marked states") that form a narrow band at large negative energy.  The low energy dynamics can be described by the effective down-folded Hamiltonian that acts in the Hilbert subspace involving only the marked states. We obtain in an explicit form the heavy-tailed probability distribution of the off-diagonal matrix elements of the down-folded Hamiltonian.  This Hamiltonian is dense and belongs to the class of preferred basis Levy matrices (PBLM). Analytically solving nonlinear cavity equations for the ensemble of down-folded Hamiltonians allowed us to describe the statistical properties of the eigenstates.  In a broad interval of transverse fields, they are non-ergodic, albeit extended. It means that the band of marked states splits into a set of narrow minibands. Accordingly, the quantum evolution that starts from a particular marked state leads to a linear combination of the states belonging to a particular miniband. Analytical description of this qualitatively new type of quantum dynamics is a key result of our paper.
Based on our analysis we propose the  population transfer (PT)   algorithm:  the quantum evolution  under constant transverse field $B_{\perp}$  starts at a low-energy spin configuration and ends up in a superposition of $\Omega$  spin configurations inside a narrow energy window.  
This algorithm crucially relies on non-ergodic nature of delocalized low energy eigenstates. In the considered model the runtime of the best  classical algorithm (exhaustive search) is   $t_{\cl}=2^n/\Omega$. For   $\sqrt{n} \gg B_\perp \gg 1$, the   typical runtime of the quantum PT algorithm  $ \sqrt{t_{\cl}} \,e^{n/(2B_{\perp}^{2})}$ scales with $n$ and $\Omega$ as that  of  the  Grover's quantum search, except for the small correction to the exponent.  
Unlike the Hamiltonians  proposed for analog quantum unstructured search  algorithms, the model we consider
is non-integrable and the transverse field delocalizes the marked states.
As a result, our PT protocol does not require fine-tuning of the transverse field  and may be initialized in  a
computational basis state.  We find that the runtimes of the PT algorithm  are distributed according to  the  alpha-stable Levy law  with tail index 1. We argue that our approach can be applied to study PT protocol in other transverse field spin glass models, with the potential quantum advantage over classical algorithms.
\end{abstract}
\maketitle

%\tableofcontents

\section{Introduction %and Summary of Results
}\label{sec:intro}

The idea to use quantum computers for the solution of search and discreet optimization problems has been actively pursued for decades, mostly notably in connection to Grover's algorithm \cite{grover1997quantum},  quantum annealing
\cite{kadowaki1998quantum,farhi2001quantum,brooke1999quantum,smelyanskiy2002dynamics,boixo_evidence_2014,knysh2016zero,boixo2016computational,denchev2016computational,albash2018adiabatic},
and more  recently, quantum approximate optimization
\cite{farhi2014quantum}.  Quantum tunneling of collective spin
excitations was  proposed and studied experimentally as  a mechanism
for moving between states in the energy landscape  that can lead to
shorter transition time scales  compared to classical Simulated
Annealing approaches under certain conditions \cite{brooke1999quantum}. Experimental evidence of the faster time scales was later corroborated numerically using an imaginary-time Quantum Monte Carlo (QMC) algorithm \cite{santoro_theory_2002,heim2015quantum}.
% However, in Ref.~\cite{heim2015quantum} it was  demonstrated that the quantum advantage found in \cite{santoro_theory_2002} was due to the discretization in the QMC implementation rather than being a genuine quantum effect. 
 Furthermore,  recent studies \cite{isakov2016understanding,jiang2017scaling} have shown that in QMC, the tunneling  corresponds to the Kramers escape through  the free-energy barrier in an extended spin system that includes spin replicas in an imaginary time direction. As a result, the incoherent quantum tunneling rate does not have a scaling advantage over such a QMC simulation. This happens because  incoherent tunneling dynamics  corresponded to sequential transitions 
connecting individual minima, where each transition is dominated by a single tunneling
path~\cite{isakov2016understanding}. 
In this paper we explore the qualitatively different tunneling dynamics where a large number of tunneling paths interfere constructively, giving rise to  "minibands" of the non-ergodic many-body  states delocalized in the computational basis (i.e. in the Fock  space). We demonstrate that  the  transport within the minibnads can be used  for efficient quantum search in spin glass problems.

%Previous studies focused on the case where the quantum dynamics corresponded to sequential incoherent quantum tunneling processes connecting individual minima, where each step is dominated by a single tunneling path. In this paper we  focus on understanding the computational role of coherent multiqubit tunneling in situations where large number of tunneling paths can interfere constructively, giving rise to the onset of many-body delocalization in  Fock space.
 
 %Additionally, quantum enhanced search and optimization algorithms implemented on analog devices \cite{Dwave} could potentially  provide a stop-gap solution for a limited class of applications prior to fault-tolerant universal quantum computation~\cite{NiclsonandChang}. 

%In a binary optimization problem, the goal
To describe the search task we start from the  binary optimization problem where the goal 
 is to find the minimum of a
classical energy function, $\cE(z)$, defined  over the set of $2^n$
configurations of $n$ bits (bit-strings) $z=(z^1,z^2,\ldots,z^n)$
where $z^k=\{0,1\}$. In quantum algorithms $\cE(z)$ is typically encoded in an
$n$-qubit Hamiltonian
\begin{align}
  H_\cl = \sum_z \cE(z) \ket z\! \bra z \label{eq:hcl}
\end{align}
diagonal in the basis of states $\ket z$  called the computational basis. Hard optimization problems have  their counterparts  in spin glass models of statistical physics \cite{mezard1987spin,fu1987application}.
The energy  function  of a  hard optimization problem is characterized by a large number of spurious local minima. Low-energy minima can be  separated by a large Hamming distance (number of bit flips  transforming  one to another). Such landscape gives rise to an interesting computational primitive: given an initial bit-string $z_j$ with atypically low energy, we wish to produce other bit-strings with energies in a narrow range $\Delta \cE_\cl$ around the initial one. In general, this can be a difficult search problem if the number of bit-strings of interest is exponentially small compared to $2^n$. 

Inspired by the Hamiltonian-based approaches to quantum search \cite{farhi1998analog} and optimization ~\cite{kadowaki1998quantum,farhi2001quantum,brooke1999quantum}
 we propose the following quantum population transfer (PT) protocol: first preparing the system in a computational state $\ket{z_j}$ with classical energy $\cE(z_j)$, we then evolve it under the Hamiltonian
\begin{align}
  H = H_\cl +H_D,\quad H_D =- \B \sum_{k=0}^n \sigma_x^k\;,\label{eq:H}
\end{align}
without fine-tuning the evolution time nor the strength of the
time-independent transverse field $\B$.  At the final moment we
projectively measure in the computational basis and check if the
outcome $z$ is a ``solution'', i.e.,  $z \ne z_j$ and the energy
$\cE(z)$ is inside the window $\Delta \cE_{\cl}$. The second term in
the Hamiltonian  (\ref{eq:H}) proportional to $\B$ is responsible for
the PT. It is usually referred to as a ``driver Hamiltonian'' in
the Quantum Annealing literature \cite{farhi2001quantum}.

We note that the  output of PT  $z$ can be used as an input  of a classical optimization heuristic such as simulated annealing  or parallel tempering in a ``hybrid'' optimization algorithm \cite{Neven2016} where quantum and classical steps  can be used sequentially to gain the complementary
advantages of both  \cite{chancellor2017modernizing}.

For random optimization problems  diagonal matrix elements $\cE(z)$ of the Hamiltonian (\ref{eq:hcl}) correspond to  a problem  instance  sampled from a particular statistical ensemble. 
Since off-diagonal matrix elements  
connect states separated by one bit-flip,  Eq.~(\ref{eq:H}) describes the Hamiltonian of the tight-binding model
with  diagonal disorder. The underlying lattice for this model is Boolean hypercube \cite{altshuler2010anderson} where individual sites correspond to bit-strings. The  model (\ref{eq:H}) can be viewed as  a generalization of the Anderson model  initially introduced in the context of transport in finite
 dimensional lattices ~\cite{anderson1958absence}.   In this model, as well as  in the  original Anderson model,  there exist  bands of localized and extended states separated in energy by a so-called 
``mobility edge''. Originally,  extensions of Anderson model appeared in a variety of many-body problems 
 in condensed matter physics \cite{basko2006metal,oganesyan2007localization}  giving rise to the concept of many-body localization (MBL). It  was demonstrated in  Ref.~\cite{altshuler2010anderson}   that MBL is responsible for the failure of  Quantum Annealing to find a solution of  the  constraint satisfaction problem (although,  the detailed analysis of this effect is still  needed  \cite{knysh2010relevance,knysh2016zero}).

 \begin{figure}[t]
   \includegraphics[width= 3.35in]{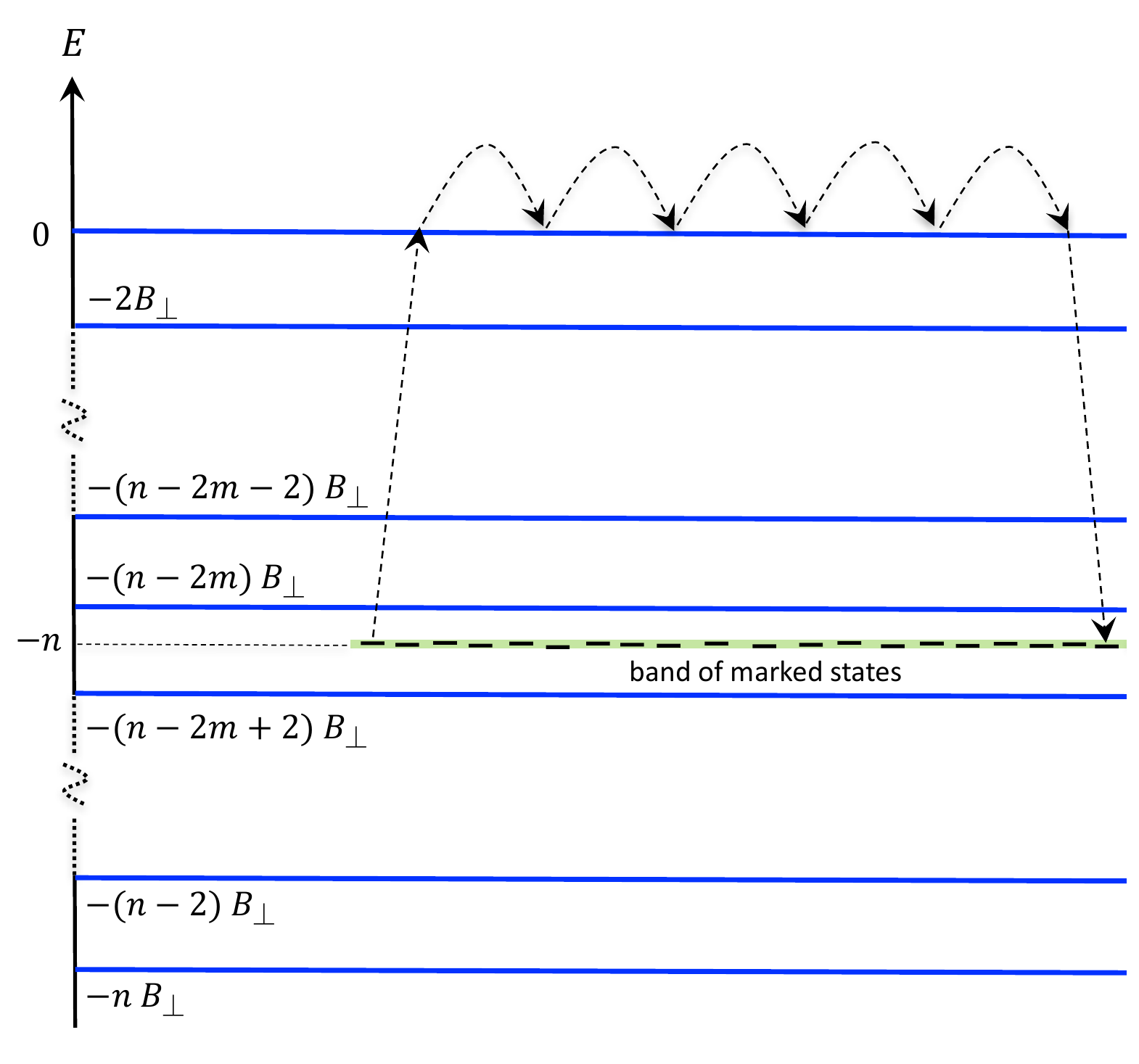}
  \caption{Cartoon of the level diagram. Horizontal blue lines depict the energy levels $-\B(n-2m)$ of the driver Hamiltonian $H_D$ in Eq.~(\ref{eq:H}) separated by $2\B$.  A narrow impurity band of width $W\ll \B$ is marked in light green. The sequence of short black lines depicts the energies of marked states $\cE(z_i)$. Dashed lines depict the elementary path to leading order perturbation theory in $\B$ for the tunneling matrix element $c_{ij}(E)$ given in (\ref{eq:c-HD-G}). In this paper we focus on the  case of relatively large transverse fields $\B>1$ so that the IB energies lie above the ground state of the total Hamiltonian (\ref{eq:H}) that corresponds  to  nearly all qubits polarized in $x$ direction.
    \label{fig:cartoon_levels}}
\end{figure}

   In models of quantum spin glasses the existence of the two types of eigenstates and the mobility edge were studied in Refs.~\cite{smelyanskiy2002dynamics,laumann2014many,mossi2017many}. We expect the Andreson models on Boolean hypercube have  an intermediate phase characterized by the onset of non-ergodic delocalized states forming narrow minibands.  Such a phase has been observed in tight binding models on Random Regular~\cite{altshuler2016nonergodic} and fully connected graphs~\cite{Kravtzov2015RzPr}. 

%For spin glass models (\ref{eq:H}) with $\B$ below the quantum spin glass transition, the PDF$(E_\beta)$ of the eigenvalues of $H$ is localized around the mean classical energy, with an exponentially decaying tail reaching towards the low energy states. %It remains quasi-continues function  at any finite $\B$ in the limit of $n\rightarrow \infty$.
%We choose the interval of energies $\Delta\cE_{cl}$ to be at the tail of the distribution, PDF$(E)\ll1$,  but sufficiently far from the ground state so that the typical  spacing of eigenvalues  is exponentially small  in $n$.
For spin glass models (\ref{eq:H}) with $\B$ below the quantum spin glass transition, the probability density function (PDF) of the eigenvalues $E_\beta$ of $H$ is localized around the mean classical energy, with an exponentially decaying tail reaching towards the low energy states. %It remains quasi-continues function  at any finite $\B$ in the limit of $n\rightarrow \infty$.
We choose the interval of energies $\Delta\cE_{cl}$ to be at the tail of the distribution, of $E_\beta$ but sufficiently far from the ground state so that the typical  spacing of eigenvalues  is exponentially small  in $n$. Under these conditions   classical  states inside the energy window $\Delta\cE_{cl}$  are located near   deep local minima of the classical energy landscape  $\cE(z)$. Hamming distances  between the minima scale with $n$ and the number of them is exponentially small compared to $2^n$ yet still exponentially large in $n$.

In this paper we apply the   PT protocol with Hamiltonian (\ref{eq:H}) to a simple yet nontrivial model  of $\cE(z)$  with the properties  mentioned above
\begin{align}
  H_\cl = \textstyle{\sum_{j=1}^M} \cE(z_j)\ket {z_j} \bra{z_j}\;.\label{eq:H_ib}
\end{align}
Here $M\gg 1$ {\it marked} states $\ket{z_j}$ ($n$-bit-strings $z_j$) are chosen uniformly at random from all bit-strings of length $n$, with energies $\cE(z_j)$ 
independently distributed around $-n$ within a narrow band of width $W \ll \B$. All other states $z$ have energies $\cE(z)=0$ and are separated by a large gap $\simeq n$ from the very narrow band of marked states (see Fig.~\ref{fig:cartoon_levels}). This model is inspired by the impurity band model in doped semiconductors \cite{shklovskii2013electronic}. It also corresponds to a classical unstructured search problem with multiple marked states.

We provide a detailed  description of the  PT dynamics in the above model by developing a microscopic analytical theory of the "minibands" of non-ergodic delocalized states \cite{PhysRevLett.113.046806}. We derived  an effective downfolded Hamiltonian in the energy strip associated with the PT. Its matrix elements correspond to the tunneling between the deep local minima and obey the heavy-tailed statistics. The ensemble of downfolded Hamiltonians for PT corresponds to the preferred basis Levi matrices (PBLM).  We use the  cavity method for Levi matrices~\cite{abou1973selfconsistent,cizeau1994theory,metz2010localization,tarquini2016level,facoetti2016non,monthus2016localization}
 to find analytically  the fractal dimension of the delocalized minibands, and  the probability distribution  of their spectral width. This allowed us to find the probability distribution and the scaling with $n$ of the PT times.  
 
  It is crucial that the dynamics within the IB of the model (3) in the transverse field can be  non-ergodic yet delocalized in computational basis. The model is by no means unique from this point of view. We  believe that the extended but non-ergodic quantum states exist for  quantum extensions of any problem Hamiltonian, which is characterized by a classical spin glass behavior: for Random Energy Model  \cite{derrida1981random}, Sherrington-Kirkpatrick model \cite{PhysRevLett.35.1792},  p-spin model \cite{kirkpatrick1987p}, K-Satisfiability \cite{mezard2002analytic},  etc.

 Indeed, the main difference between classical and quantum spin-glass models is the existence of the many-body localized (MBL) phase in the latter case. However we see no reason to expect a direct transition between the MBL and ergodic phases without intermediate non-ergodic phase similar to the case of ordinary Anderson localization in finite-dimensional space. This difference is due to the fact that the number of relevant bit-strings at a given Hamming  distance $d$ from a given one increases for spin-glass models exponentially with $d$, or even quicker, whereas for finite-dimensional models this increase is only polynomial.

A key challenge in developing  a theory of non-ergodic delocalized phase for  quantum spin glass models is the calculation of the statistics of the tunneling matrix elements between deep local minima separated by large Hamming distances $d$.  We derived  analytically its dependence on the transverse field $B_{\perp}$ and Hamming distance $d$ using WKB theory of collective spin tunneling in asymptotic limit of large $n$. We demonstrated that in the delocalized phase  it is qualitatively different from that given by the leading order perturbation theory in $B_{\perp}$, known as a forward scattering approximation (FSA) that has been previously used in these  problems~\cite{laumann2014many,baldwin_many-body_2015,BaldwinLaumannPSpin2017,BaldwinLaumannEnergyMatching2018}. As a consequence, our results for the scaling of the PT time with $n$ and the structure of the delocalized eigenstates in IB model are qualitatively different from the FSA predictions.

%We obtain an explicit analytical form for the statistical properties of the PT dynamics in the above model by deriving an effective down-folded Hamiltonian in the  energy strip associated with the PT and using the cavity method for random matrices
 
In model (\ref{eq:H_ib}), the most efficient classical algorithm is purely random search with running time   $\sim 2^n$. We find that  the  typical runtime of the PT  algorithm  $\pt$ displays the following scaling dependence  
 on  $n$ 
\begin{equation}
\pt\propto \(\frac{2^n}{ \Omega \log\Omega}\)^{1/2}e^{n/(2 B_{\perp}^{2})}\;.\label{eq:pt-intro}
\end{equation}
Here $\Omega\gg 1$ is the number of  computational basis states  within the target window of energies that contribute with comparable probabilities to the quantum state  at the end of PT.
The expression applies  in the range of transverse fields $n^{1/2}$ $\gg$ $\B-1$=$\cO(1)$ (for arbitrary $\B$ see Eq.~(\ref{eq:theta-x})). 

The dependence  of $\pt$ on $\Omega$ is the same as in the  multi-target  Grover quantum algorithm 
that searches for   $\Omega$  marked states  starting from the  fully-symmetric   state  $\ket{S}=2^{-n/2}\sum_{z}\ket{z}$. In the  Hamiltonian version of this algorithm~\cite{farhi1998analog}, one uses the projector to $\ket{S}$  as a driver, $H_D=w\ket{S}\bra{S}$. This algorithm is proven to be optimal for problems without structure. We emphasize that according to Eq.~(\ref{eq:pt-intro}) the exponential scaling of $\pt$ with $n$ differs from that in the  Grover  algorithm by a term $\sim B_{\perp}^{-2}$  that can be made arbitrary small at sufficiently large transverse fields.

 PT algorithm is qualitatively different from  the quantum annealing, adiabatic optimization  and Hamiltonian implementation of  Grover search because it  exploits   the structure of the excited energy spectrum. The  PT Hamiltonian $H$ (\ref{eq:H}) is non-integrable and its eigenstates are delocalized in the low-energy manifold.

In  analytically tractable example considered here   the PT algorithm has  new  and potentially advantageous features compared to  the Grover algorithm whose 
Hamiltonian is integrable and all of its eigenstates but one are localized.  Therefore the  quantum  evolution resulting from the Grover Hamiltonian cannot form a massive superposition of $\Omega\gg 1$ solutions if it starts from a computational basis state. The algorithm must always start from the state $\ket{S}$.  Moreover, Grover's algorithm performance is exponentially sensitive to fine-tuning of the weight of the driver $w$ on the scale $\delta w\sim 2^{-n/2}\sqrt{\Omega}$. In contrast, the scaling of the runtime of PT  (\ref{eq:pt-intro}) with $n$ is robust to the choice of $\B$ that can take on a broad range of values for $\B\gg1$.

The nearly optimal (Grover-like) performance of the PT protocol is the consequence of the asymptomatic orthogonality between the eigenstates in the marked state subspace to the rest of the Hilbert space. This  suppresses the population transport from the marked states to the $\mathcal{O}(2^n)$ of  states $\ket{z}$ with classical energies $E_z=0$  even at large $B_{\perp}$. Such "orthogonality catastrophe" cannot be obtained within the perturbative in $B_{\perp}$ approach such as FSA.

The paper is organized as follows. Sec.~\ref{sec:QualDisc} contains a qualitative discussion of the main results. In Sec.~\ref{sec:reduction} we develop a down-folding procedure to reduce the original problem to the nonlinear eigenproblem in the marked state subspace. In Sec.~\ref{sec:cd} we calculate the off-diagonal (tunneling) matrix elements of the down-folded Hamiltonian and studied their dependence on $n$ and Hamming distance   using Wentzel-Kramers-Brillouin (WKB) theory.
 In Sec.~\ref{sec:lin} we develop an expansion of the nonlinear eigenproblem near the center of the IB shifted by transverse field and obtain the effective Hamiltonian  $\scH$ of the PT problem.
% We obtain the uniform shift of the band due to the transverse field and the explicit form of the effective Hamiltonian $\scH$ for the PT problem. We also investigate the vicinities of resonant values of $\B$  when the levels of the driver ``cross'' the impurity band and establish the condition for when the resonant effects can be neglected.
In Sec.~\ref{sec:ensembleH} we study the statistical ensemble of Hamiltonians $\scH$. Sec.~\ref{sec:NumSim} discusses numerical results. In Sec.~\ref{sec:FGR} we study the PT within the Born approximation. In Sec.~\ref{sec:NumRes}
we estimate the number of states in the miniband.  In Sec.~\ref{sec:Cavity} we provide an overview of the cavity method for dense random matrices.  In Sec.~\ref{sec:sol} we solve the cavity equations and obtain the distributions of the real and imaginary parts of self-energy. In Sec.~\ref{sec:compl} we discuss the complexity of PT  algorithm.
In Sec.~\ref{sec:Grover} we provide a comparison between PT and Grover's algorithm with multiple target states and systematic errors in oracle phase and driver weight.  In Sec.~\ref{sec:conc} we provide a summary and concluding remarks.

\section{Qulatitative discussion of results}\label{sec:QualDisc}

Each marked state $\ket{z_j}$  is a deep local minimum of $\cE(z)$ separated from other minima by a typical Hamming distance  $n/2$  while the separation from the nearest market state is also extensive $\dmin$=${\cal O}(n)$ for $M$=$2^{\mu n}$ and $\mu<1$.

 The transverse field $\B$ gives rise to  multiqubit tunneling between  the states.   The tunneling amplitudes  from a given minimum to its neighbors located at a Hamming distance $d$ decrease exponentially with $d$ while the number of neighbors increases exponentially with $d$ for $d={\cal O}(n)$.  As a result, an eigenstate
$\ket{\psi_\beta}$ of $H$ associated  with the impurity band can become delocalized over a large subset  of marked states $\scS_\beta$  with size $1\ll |\scS_\beta|\propto M^\alpha$ and  $0<\alpha \leq 1$.  
For $\alpha=0$ the eigenstate $\ket{\psi_\beta}$ is localized, for $\alpha=1$ the eigenstate is delocalized in the  entier space of  marked states. For $0<\alpha<1$ the eigenstate can be considered "non-ergodic" and its support  set $\scS_\beta$ is sparse in the space of the  marked states.  
We express the transition probability  from $\ket {z_j}$ to $\ket z$,
\begin{equation}
  P(t,z|z_j) = \Big |\textstyle{\sum_\beta} \braket z {\psi_\beta} \braket{\psi_\beta} {z_j} e^{-i E_\beta t} \Big |^2\;,\label{eq:ptz}
\end{equation}
\noindent
in terms of the eigenstates  and corresponding eigenvalues  of  $H$, where $H\ket{\psi_\beta} = E_\beta \ket {\psi_\beta}$. In the delocalized phase, for a given state  $\ket{z_j}$ there exists a large set %$\cS(z_j)$ 
of eigenstates $\ket{\psi_\beta}$  that have peaks at  $\ket{z_j}$.    These eigenstates  possess  important properties~\cite{Kravtzov2015RzPr,altshuler2016multifractal,altshuler2016nonergodic}: they have largely overlapping supports $\medcap_{\beta}\scS_\beta$\;$\approx$\;$\scS(z_j)$, and they are  close in energy thus forming a narrow mini-band.  The mini-band width $\Gamma$ may be interpreted as the inverse scrambling time and determines the width of the  plateau in the  
Fourier-transform of the typical transition probability $\tilde P(\omega,z|z_j)$  \cite{Kravtzov2015RzPr}.\footnote{The same plateau width  characterizes the frequency dependence of the  eigenfunction  overlap correlation coefficient $K(\omega)=M\sum_{j=1}^{M}\sum_{\beta,\beta'}   |\braket{j}{\psi_\beta}|^2   |\braket{j}{\psi_{\beta'}}|^2 \delta(\omega-E_{\beta}+E_{\beta'})$~\cite{Kravtzov2015RzPr}.}
 %The marked states $\ket{z_j} \in \scS$ are precisely the resonances mentioned above.    %It follows from this  discussion  and Eq.~\eqref{eq:ptz},  that   
 In other words, the  significant PT of $P(t,z|z_j)$   from the initial marked state $\ket{z_j}\in \scS $ into the other states of the same miniband $\scS$ occurs over the time $\pt\sim 1/\Gamma$. The window $\Delta \cE_{\rm cl}$ is related to the miniband width $\Gamma$.

Understanding \textcolor{black}{ the} properties  of  non-ergodic delocalized  states  is crucial for describing the dynamics of quantum spin glasses driven by many-body coherent tunneling processes.
Developing its  microscopic theory is a challenging problem. This paper studies the  transport problem in an  "impurity band" (IB) model (\ref{eq:H_ib}) by making use of the down-folded   Hamiltonian   in  the marked state subspace derived in Secs.~\ref{sec:reduction}, \ref{sec:lin}.
While the original Hamiltonian (\ref{eq:H}) is sparse in the basis of states $\ket{z}$ (it couples only states  separated by  Hamming distance 1),
the   down-folded Hamiltonian $\scH$ (\ref{eq:IBH}) is a dense $M\times M$ matrix.

The transverse field leads to a uniform shift $\sim \Bs$ of the marked state energies as shown in Sec.~\ref{sec:lin},  (\ref{eq:E0})-(\ref{eq:Delta0a}).
Diagonal elements of $\scH_{ii}$ are given by the marked state energies counted off from the center of the shifted impurity band. 
Their PDF  is assumed to be exponentially bounded with some width $W$.

Each pair of  marked states is coupled via multi-qubit tunneling. 
The off-diagonal matrix elements  $\scH_{ij}=V(d_{ij})\cos\phi(d_{ij})$ are completely determined by the Hamming distance $d_{ij}$ between the marked states $z_i$ and 
$z_j$. The amplitude $V(d)$  decays steeply with $d$, inversely proportional to a square root of $\binom{n}{d}$ (see Eq.~(\ref{eq:V2})). The phase  $\phi$ shown in Fig.~\ref{fig:phases} monotonically increases  by $\cO(1)$ when $d$ is changed by 1. In the analysis of spectral properties of $\scH_{ij}$ the quantity $\cos\phi(d_{ij})$  can be replaced by a random sign. The explicit form of $V(d)$ and $phi(d)$ is obtained using WKB theory of collective spin tunneling. At $B_\perp > 1$ the tunneling paths correspond to long spin-flip sequences connecting the initial and final states. They  include many loops passing through the the states with $Ez=0$ that are neglected in FSA.

 The typical matrix element between the two marked states is $V_{\rm typ}\sim n^2 2^{-n/2}e^{-n/(4\Bs)}$. The typical matrix element between a given marked state and its nearest neighbor is also exponentially small in $n$ but it is exponentially larger than the value $V_{\rm typ}$. 
This fact corresponds to  a  strong hierarchy of the off-diagonal matrix elements of $\scH_{ij}$ 
which is  a signature   of  their  heavy-tailed probability density function  \cite{cizeau1994theory,monthus2016localization}.  Such matrices are called Levi  matrices.

The  PDF of the rescaled squared amplitudes $w_{ij}=V^2(d_{ij})/V_{\rm typ}^{2}$
  derived in Sec.~\ref{sec:tails} is
\begin{equation}
{\rm PDF}(w)=\frac{1}{w^2\sqrt{\pi \log w}},\quad w\in[1,\infty). \label{eq:poly}%P(scH_{ij}^{2})=\frac{1}{V^2(n/2)}g\(\frac{\scH_{ij}^{2}}{V^2(n/2)}\)
\end{equation}
The particular form of scaling is the direct consequence of the fact that our problem has no "structure":   the tunneling matrix elements depend only on Hamming distance and marked states are chosen at random.

The key difference  of the ensemble of matrices $\scH_{ij}$ from  Levy matrices studied in the literature \cite{tarquini2016level,monthus2016localization,metz2010localization,cizeau1994theory} is that the dispersion,  $W$, of the diagonal matrix elements is much larger than the typical magnitude of the off-diagonal elements  $V_{\rm typ}$. Therefore $\scH_{ij}$ can be called
preferred basis Levi matrices (PBLM). 

We note that the existence of heavy tails in the PDF of the off-diagonal matrix elements of the down-folded  Hamiltonian $\scH$ is due to  the infinite dimension of the Hilbert space of the original problem (\ref{eq:H}) for  $n\rightarrow \infty$. This happens because the exponential decay of the matrix elements with the Hamming distance $d$ is compensated by the exponential growth of the number of  states at the  distance $d$  from a given state. We believe that the  PBLM structure is a generic feature of the effective  Hamiltonians for PT at the tail of  the density of states in quantum spin glass problems.

Unlike the standard Levi ensemble, the eigenstates of PBLM allow for the existence of non-ergodic delocalized states when the width $W$ is much bigger than the largest off-diagonal matrix element in a typical row of $\scH_{ij}$ and much smaller than the  the largest off-diagonal element in a matrix 
\begin{equation}
V_{\rm typ} M^{1/2}\ll W\ll V_{\rm typ} M\;.\label{eq:VMV}
\end{equation}
For smaller dispersion $W\apprle V_{\rm typ} M^{1/2}$ the matrix eigenstates  are ergodic while for $W \apprge  V_{\rm typ} M$ the eigenstates are localized.
Such phase diagram resembles the one in the Rosenzweig-Porter (RP)  model   \cite{Kravtzov2015RzPr,facoetti2016non}. The difference of RP from PBLM is that  the statistics of the off-diagonal matrix elements in the RP ensemble are Gaussian \cite{rosenzweig1960repulsion} rather than polynomial (\ref{eq:poly}). In this paper we will focus on exploring PT transfer within the non-ergodic delocalized phase, which is more likely to generalize to other models. We note that the localized phase does not support population transfer.% , while the ergodic phase does not generally provide the desired selectivity of the transfer because the population is spread over a large fraction of all states.  
% \footnote{In our model we refer to ``ergodic'' eigenstates when they are spread almost over the entire set of $M$ marked states, - but not over the rest of the $2^n-M$ states of the computational basis. Full ergodicity requires very large transfer fields that scale with the number of spins $\B=\cO(n^{1/2})$.}

Because of the PBLM structure of the Hamiltonian $\scH$ one can expect that the runtime of the PT protocol $\pt$ will have  a heavy-tailed PDF  whose form is of practical interest. It is closely related to the PDF of the miniband widths  $\Gamma\sim 1/\pt$.
We obtained the PDF$(\Gamma)$  by making use
of the cavity method for random symmetric matrices \cite{abou1973selfconsistent,cizeau1994theory,burda2007free,tarquini2016level}.

In previous work the
cavity equations were solved only in their linearized form, i.e.,  near the localization transition. We were able to solve  fully nonlinear cavity equations in the delocalized non-ergodic phase. We obtained the boundaries of the phase in terms of the ratio of $W/V_{\rm typ}$ and also the form of  $\scP(\Gamma)$ inside the phase. It is given by the alpha-stable Levi distribution \cite{gnedenko1954limit,cizeau1994theory} with the tail index 1, most probable  value $\Gamma_{\rm typ}= V_{\rm typ}(\pi\Omega \log\Omega/4)^{1/2}$, and characteristic dispersion 
$\pi \Gamma_{\rm typ}/(4\log\Omega)$ where $\Omega$ is the typical number of states in the  miniband. This number $\Omega=(\pi M V_{\rm typ}/W)^2$ is a square function of the ratio of the typical tunneling matrix element $V_{\rm typ}$ to the level separation $W/M$.
In a non-ergodic delocalized phase $M\gg \Omega\gg1$ and  the typical PT time $\pt\sim1/\Gamma_{\rm typ}$ obeys the condition
\begin{equation}
(M\log M)^{-1/2}\ll \pt V_{\rm typ}\sim (\Omega\log\Omega)^{-1/2} \ll 1\;.%\label{eq:VMVOmega}
\end{equation}
%In the the non-ergodic delocalized phase, the inequalities become strong.

We build on the observations made in the IB model and provide qualitative arguments that PT will have a  quadratic speed up over QMC in some quantum search problems where tunneling is a computational bottleneck.

%We use the  parametrization similar to the RP case  \cite{Kravtzov2015RzPr} 
%\begin{equation}
%W=\lambda \cVt M^{\gamma/2}\;\label{eq:param}
%\end{equation}
%where $\lambda$   is a (redundant) number of order of ${\cal O}(M^{0})$.  

\section{\label{sec:reduction}  Downfolding into the subspace of the marked states and  nonlinear eigenproblem }

%\subsection{\label{sec:downfold}  Downfolding Hamiltonian}

The 
driver Hamiltonian $H_D$ in Eq~\eqref{eq:H}
%\begin{equation}
%(H_D+H_{\rm cl}) \ket{\psi}=E \ket{\psi}.\label{eq:Hst}
%\end{equation}
connects bit-strings that are separated by a Hamming distance $d$=1.  % (i.e., matrix of the total Hamiltonian $H=H_D+H_{\rm cl}$ is sparse in computation basis).  
We note that, on one hand,  marked states  are separated by large Hamming distances $d_{ij}$ with typical value $d=n/2$.
Therefore  a pair of marked states $\ket{i}$ and $\ket{j}$
is  coupled  by  elementary spin-flip processes corresponding to high
orders $(H_D)^k$  of the driver Hamiltonian with $k \geq d_{ij}$. 
On the other hand, the
resolvent of the driver Hamiltonian 
\begin{equation}
G(E)=\frac{1}{E-H_D}\;.\label{eq:GE}
\end{equation}
connects directly every pair of marked states. Furthermore,  because $H_D$ is invariant under permutations of bits,  the  matrix elements  $G_{ij}(E)=\bra{z_i}G(E)\ket{z_j}$ depend only on the Hamming distance $d_{ij}$ between the corresponding states. They are   exponentially small in $n$ for extensive $d_{ij}=\cO(n)$. Therefore, one might expect that under certain
 conditions the quantum evolution stays approximately  confined to the marked state subspace and can be naturally described by the  downfolded Hamiltonian whose $M\times M$ matrix representation is dense in  the basis of marked states.

We use the identity
\begin{equation}
G(E)H_\cl\ket{\psi}=\ket{\psi}\;,\label{eq:GHphi}
\end{equation}
where $E$ and $\ket{\psi}$ is an   eigenvalue and the corresponding
eigenvector  of $H$. We introduce a new vector
\begin{equation}
\ket{\cA}=\sqrt{H_\cl}\ket{\psi}
\end{equation}
that  has no support in the subspace orthogonal to that of marked states.  Then, multiplying both parts of equation (\ref{eq:GHphi}) 
by $\sqrt{H_\cl}$, we obtain after simple transformations
\begin{align}
\(H_\cl+\Lambda\) \ket{\cA} &=E\ket{\cA},\label{eq:Hc0}
\end{align}
where
\begin{align}
\Lambda =\sqrt{H_\cl}H_D &G(E)\sqrt{H_\cl}\;.\label{eq:Lambda}
\end{align}
The
operator $\Lambda$  plays the role of a ``driver Hamiltonian'' in the
downfolded picture, and it couples states in the marked subspace.

Equation~(\ref{eq:Hc0}) can be written in matrix form (see Appendix, Sec.~\ref{sec:dfH} for details)
\begin{equation}
\sum_{j=1}^{M}\cH_{ij}(E)\cA_j=E \cA_i,\label{eq:Hc1}
\end{equation}
where $\cA_i=\braket{\cA}{z_i}$ and $\cH_{ij}$ is a dense symmetric $M\times M$ matrix
\begin{equation}
\cH_{ij}(E)=\delta_{ij} \cE(z_i) +\sqrt{ \cE(z_i) \cE(z_j)}\,c_{ij}(E)\;.\label{eq:Hc2}
\end{equation}
Here   $\delta_{ij} $  is  the Kronecker delta and 
\begin{equation}
c_{ij}(E)=c(E,|z_i-z_j|)=\bra{z_i}H_D \frac{1}{H_D-E}\ket{z_j}\;,\label{eq:c-HD-G}
\end{equation}
is a coupling coefficient that depends only on a Hamming distance 
$|z_i-z_j|$  between the bit-strings $z_i$ and $z_j$.
%\begin{equation}
%c_{ij}(E)=c(E,|z_i-z_j|)\;.\label{eq:cc}
%\end{equation}

We note that (\ref{eq:Hc1}) has the form of a nonlinear
eigenproblem. A solution of (\ref{eq:Hc1}) for the $M$-dimensional vector $\ket{A}$  with nonzero norm requires
\begin{equation}
\det[\cH(E)-I E]=0,\label{eq:nep}
\end{equation}
where $I$ is the identity matrix.  Because the downfolded  Hamiltonian
$\cH(E)$ explicitly depends on the energy $E$,   different roots $E_\beta$ of the equation (\ref{eq:nep})  correspond to different Hamiltonian matrices $\cH_{ij}(E_\beta)$. %Therefore  the  eigenvectors %$\ket{A_\beta}$ 
%of the nonlinear eigenproblem  (\ref{eq:Hc1}) 
%\begin{equation}
%\ket{A_\beta}=\sum_{j=1}^{M}A_{\beta j}  \ket{z_j},\label{eq:Aket}
%\end{equation}
%are, in general,  not orthogonal.  
This can be understood from the fact that the original $2^n\times 2^n$
Hamiltonian (\ref{eq:H}) couples the $M$ dimensional marked state
subspace to the rest of the Hilbert space. Therefore, the projections
of the eigenvectors $\ket{\psi_\beta}$  of  $H$ onto  the  subspace
are, in general, neither normalized nor orthogonal. The same is true for the corresponding vectors $\ket{\cA_\beta}=\sqrt{H_\cl}\ket{\psi_\beta}$.
The normalization condition for the projections has the form (see Appendix~\ref{sec:dfH} for details)
\begin{equation}
\sum_{j,i=1}^{M}\frac{1}{Q_{ji}(E_\beta)} \psi_{\beta}( z_ j) \psi_{\beta}(z_ i) =1.\label{eq:normA}
\end{equation}
where
\begin{equation}
\frac{1}{Q_{ij}(E)}=\sqrt{\cE(z_i)\cE({z_j})}\ \frac{d}{d E}\left(\frac{\cH_{i j}(E)}{E}\right)\;.\label{eq:Qjk}
\end{equation}
This condition along with Eqs.~(\ref{eq:Hc1})-(\ref{eq:nep})  completely defines the eigenvector projections onto the marked state subspace and the corresponding eigenvalues.

We observe that there are exactly $M$  roots $E_\beta$ of (\ref{eq:nep})   that   originate from $M$ classical energies of the marked states $\cE(z_j)$ at $\B=0$. These eigenvalues and the corresponding eigenstates  will be the sole  focus  of our study. Here we just mention briefly that the rest  of the states  originate in the limit $H_{\rm cl}\rightarrow 0$ from the eigenstates  of the driver Hamiltonian whose energy levels
$-\B(n-2m)$ (shown in Fig.~\ref{fig:cartoon_levels}) correspond to the total spin-$x$ projections $n-2m
\in[-n,n]$. The levels $-\B(n-2m)$ have  degeneracy $\binom{n}{m}$,
which is partially  lifted  due to the coupling to the impurity band with $M$ states.
The splitting of the  driver energy  levels $-\B(n-2m)$   increases  as a function of  transverse field in the vicinity of  ``resonances''   with the levels
 of  the impurity band where $\B(n-2m)\approx -n$ for integer values of  $m$. At resonance, the eigenstates of the   driver with total spin-$x$ projection $n-2m$  are strongly hybridized with the marked states $\ket{z_j}$.  
As will be discussed  below, the width of the resonances remains exponentially small in $n$ for $\B={\cal O}(n^{0})$.
In Fig.~\ref{fig:spectrum} we plot the evolution of the energy spectrum of the Hamiltonian $H$ as a function of transverse field for the case of two impurity states $M=2$.

\section{\label{sec:cd} Coupling coefficients in the downfolded Hamiltonian}

\begin{figure}[ht]
 \includegraphics[width= 3.35in]{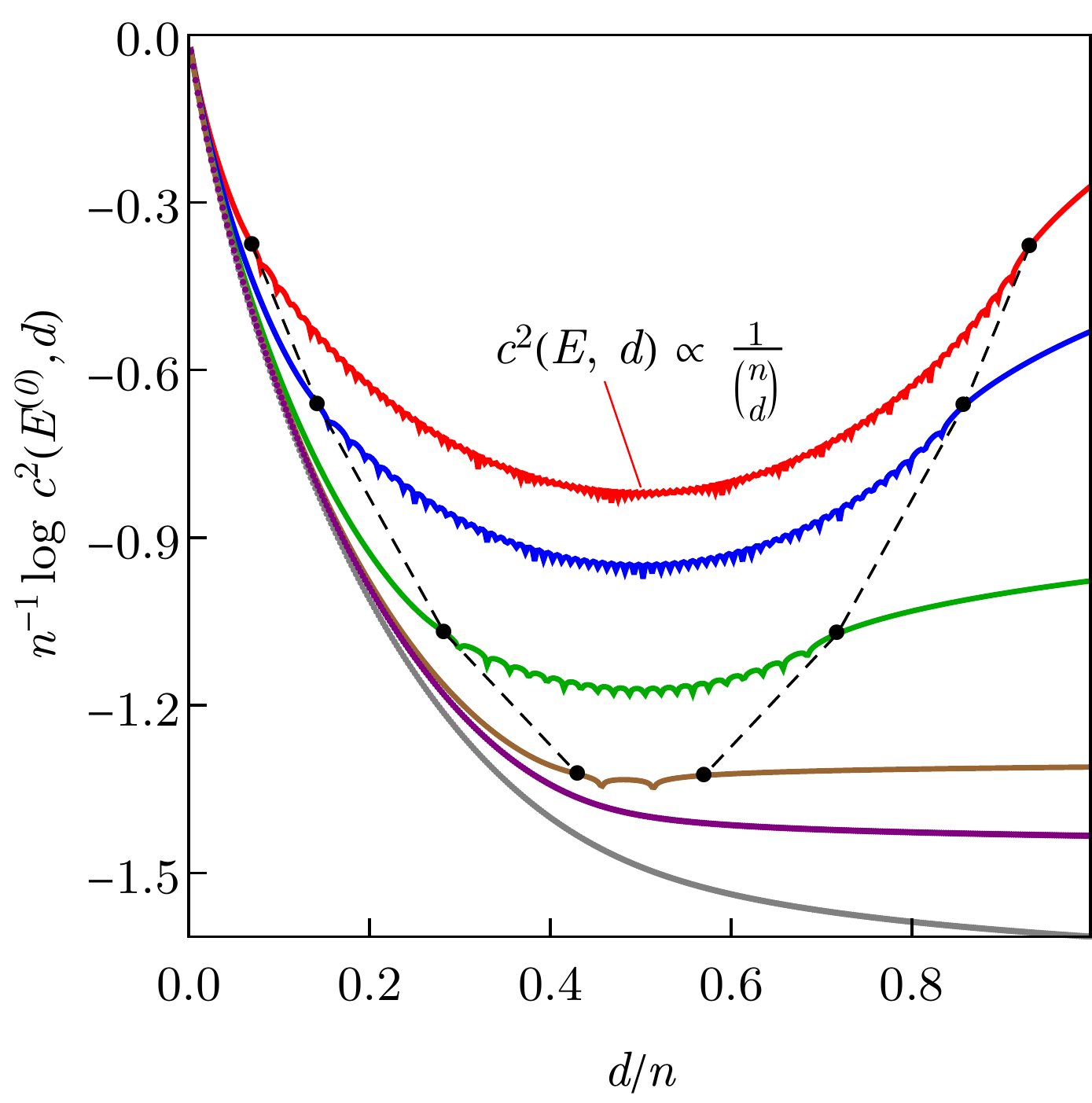}
  \caption{Colored lines show the dependence of the rescaled logarithm of the coupling coefficient $n^{-1}\log c^2(E,d)$, Eq.~ (\ref{eq:CEd}), on the rescaled Hamming distance $d/n$ for $n=400$. The energy $E$ is set to the value $E^{(0)}\simeq -n-\Bs$ that reflects the overall shift of the impurity band  due to the transverse field (cf.  (\ref{eq:E0}),(\ref{eq:Delta0a})). Different  colors correspond to different values of the transverse field  $\B$=1.93 (red), 1.43 (blue), 1.11 (green), 1.01 (brown), 0.99 (purple), 0.95 (gray). The scale along the $y$-axis suggests that $c(E^{(0)},d)$ scales exponentially with $n$ for $d/n={\cal O}(n^{0})$. The inset shows the leading order factor in the $d$-dependence of the coupling coefficient  for $\B>|E|/n$  (cf. (\ref{eq:c})). Black dots show the boundaries $d=n/2-m_0, n/2+m_0$ of the region of the oscillatory behavior of $c(E,d)$ with $d$ given by WKB theory (\ref{eq:m00}) (see Appendix~\ref{sec:det-wkb} for details).
    \label{fig:c10c}}
\end{figure}

The coupling coefficient  $c_{ij}(E)\equiv c(E,d_{ij})$ for $i\neq j$  determines the off-diagonal matrix element of the downfolded Hamiltonian (\ref{eq:Hc2}) corresponding to  the  tunneling transition that connects marked states $\ket{z_i}$ and $\ket{z_j}$. In the IB model, the tunneling matrix element   depends only on the Hamming distance $d_{ij}$ between the states. It can be calculated in explicit form from  Eq.~(\ref{eq:c-HD-G}). For this we use  the basis of eigenstates $\ket{x}$ of the driver Hamiltonian $H_D\ket{x}=H_D^x\ket{x}$ in Eq.~(\ref{eq:c-HD-G}). They correspond to bit-strings $x=(x^1,\ldots, x^n)$ of individual qubits polarized   in positive $x^a=0$ and negative $x^a=1$ direction of the $x$ axis. The eigenvalues of the driver  $H_D^x=-\B(n-2h_x)$ depend only on the Hamming weight of the bit-strings $x$. Therefore one can perform explicitly the partial summation over  basis vectors $\ket{x}$ in  
 (\ref{eq:c-HD-G}) under the conditions that  $\sum_{a}x^a =k$  for all bit positions   $a$ such that $z_j^a\neq z_i^a$, and $\sum_{a}x^a =l$ for all $a$ where  $z_j^a= z_i^a$. Finally the result  (\ref{eq:c-HD-G}) can be written as a double sum over $k\in(0,n-d_{ij})$ and $l\in(0,d_{ij})$ 
 \begin{equation}
 c_{ij}(E)=\sum_{k=0}^{n-d_{ij}}\sum_{l=0}^{d_{ij}}\binom{n}{k}\binom{n-d_{ij}}{l}\frac{(-1)^l \,2^{-n}}{1+\frac{E}{\B(n-2 k-2 l)}}.\label{eq:CEd}
\end{equation}
 Here $d_{ij}$ is the Hamming distance between bit-strings  $z_i$ and $z_j$. %Also  integers $k$ and  $l$ in the above sum are the number of $x$-bits $x^j=1$  at the positions where bit assignments  in the bit-strings $z_i$ and $z_j$ are the same or different, respectively.   
 Plots of coupling coefficients as a function of Hamming distance $d$ based on (\ref{eq:CEd}) are given in Fig.~\ref{fig:c10c}. They display qualitatively different behavior  depending on the value of the parameter $n\B/|E|$.

 For $n\B/|E|<1$ the coefficient $c(E,d)$ decays exponentially  with $d$ in the entire range of values $d\in[0,n]$.  For  $n\B/|E|>1$ the coefficient decays until $d\sim n/2$, corresponding to minimum overlap between the marked states, and then begins to grow. For large transverse field $\B\gg1$ the behavior  with $d$ is nearly symmetric  with respect to $d=n/2$ and to leading order it does not depend  on $\B$.  Unfortunately, the  expression  (\ref{eq:CEd}) is  quite involved and is not suitable for the  study of the  asymptotic properties of the  population transfer in the  limit of large $n$.

For a very weak transverse field $\B\ll n^{-1/2}$ using perturbation theory in $\B$ to the leading order  one can obtain a standard expression   \cite{laumann2014many} for the coupling coefficient, $|c(E,d)|\simeq d\,! (\B/n)^d$.   It is given by the sum of the transition amplitudes over the  $d!$ shortest paths between the states $\ket{z_i}$ and $\ket{z_j}$ separated by a Hamming distance $d$. Intermediate states $\ket{z}$ along each path correspond to $\cE(z)=0$ while energies of initial and final states are $-n$ (see Fig.~\ref{fig:cartoon_levels}).

For larger transverse field values  (but still $\B\ll1$) the perturbative expression in the small-$\B$ limit can be modified to include the range of $\B=\cO(n^0)$ but $\B\ll1$.
In that range %\[|c(E,d)|\simeq d\,!(\B/n)^d\,\exp [B_{\perp}^{2}(d^2(3n-2d)/n^2+n/2 )/6]\]
\begin{equation}
|c(E,d)|\simeq d\,!\(\frac{\B}{n}\)^d\,e^{B_{\perp}^{2}\(\frac{ d^2(3n-2d)}{6n^2}+\frac{n}{12}  \)}\;.\label{eq:smallB}
\end{equation}
One can see that  for small $\B$  matrix element falls down with $d$ extremely steeply despite the presence of the 
factorial factor $d\,!$ in (\ref{eq:smallB}). We note that this  perturbation (FSA) expression is qualitatively valid  in the range $B_\perp<\left|E/n\right|\ll 1$. It gives a correct leading order form of the mobility edge in quantum REM~\cite{laumann2014many,baldwin_many-body_2015,BaldwinLaumannPSpin2017,BaldwinLaumannEnergyMatching2018} at small $B_{\perp} \ll 1$. 

For transverse field, $\B>|E|/n$, the dependence  of $c_{ij}(E)$ on $d_{ij}$ changes qualitatively.  It becomes non-monotonic, reaching its  minimum at the point $n/2$ of minimum overlap between the bit-strings $z_i$ and  $z_j$.  In a certain region around the minimum  it has oscillatory behavior, as seen in Fig.~\ref{fig:c10c}. The boundary of this region is shown with black dots. The details of the behavior in the oscillatory region are shown in  Fig.~\ref{fig:c01}. 
The exponential dependence of  the envelope of $c(E,d)$ on $d$ is captured by  the factor $1/\binom{n}{d}$ and is  independent on the transverse field strength.  This region of $d$ and values of $\B>|E|/n$ are of the most relevance to the transport in non-ergodic minibands which is of central interest to in this paper.

\subsection{WKB calculation of coupling coefficients}

 In this paper we   develop an approach (described in the Appendix~\ref{sec:det-wkb})  based on the   WKB theory for large spin  \cite{garg1998application} to calculate the coefficient $c(E,d)$ for $n\gg1$ and arbitrary values of transverse  fields  $\B$ without relying on perturbation theory in $\B$.
The coefficient  $c_{i j}(E)$ can be expressed in terms of the operator of the total  spin-$x$ projection $S_x=1/2\sum_{j=1}^{n}\sigma_{x}^{j}$
\begin{equation}
c_{i j}(E)=\delta_{ij} - E \bra{z_i}( E+2\B S_x)^{-1}\ket{z_j}\;.\label{eq:c_ij-1}
\end{equation}

We will utilize   the basis of  eigenstates  $\ket {m}$ of the operator $S_z=\sum_{k=1}^{n}\sigma_{z}^{k}$ corresponding to  its eigenvalues   $m\in[-n/2,n/2]$ and  the maximum value of the total spin $S=n/2$
\begin{align}
S_z \ket {m} = m\ket{m}, \quad m=-n/2 ,\ldots,n/2\;.\label{eq:Sz}
%\sum_{s=x,y,z}S_{s}^{2}\ket{m_z}&=\frac{n}{2}\left(\frac{n}{2}+1\right)\ket{m_z}\nonumber
\end{align}
The  state  $| n/2-d\rangle$ 
 is a  normalized sum  of all computational basis states $\ket{z}$ with  $d$ spins pointing in the  negative $z$ direction and $n-d$ spins   pointing in the positive $z$ direction ($m=n/2-d$)
 \begin{equation}
| n/2 -d\rangle=\frac{1}{\sqrt{\binom{n}{d}}}\sum_{z\in\stn}\delta_{|z|,\,d} \ket{z}.\label{eq:sym}
\end{equation}
Here $|z|=\sum_{k=1}^{n}z^k$ and $\delta_{k,d}$ is a Kronecker delta.

Because the coefficients $c_{i j}(E)$ (\ref{eq:CEd}) depend only on the Hamming distance $|z_i-z_j|$ between the bit-strings $z_i$ and $z_j$, we can assume,  without loss of generality, that  in  Eq.~(\ref{eq:c_ij-1}) one of the bit-strings, e.g., $ \ket{z_j}$,  corresponds to all individual spins  pointing in the positive $z$ direction
\begin{equation}
\ket{z_j}=\ket{00 \ldots 0}\equiv \ket{n/2}\quad \left(m_z=n/2\right)
 \end{equation}
The main observation is that  we can  pick, instead of the state $\ket {z_i}$, any computational basis state $\ket z$ whose Hamming weight satisfies the condition $|z|=|z_i|$ without changing the value of the coefficient $c_{ij}(E)=c(E,|z_i|)$.  Therefore averaging both  sides  of the Eq.~(\ref{eq:c_ij-1}) over the states $\ket{z_i}$ that satisfy the condition $|z_i|=d$ for some integer $d\in[0,n]$  we obtain
%\begin{equation}
%c(E,d) =\delta_{d,0}- \sum_{z\in \stn} \frac{ \delta_{|z|,d}}{  \binom{n}{d}} \bra z \frac{E}{E+2\B S_x} \ket{n/2}. \nonumber %\label{eq:c_ij-2}
%\end{equation}
% By  comparing this equation with (\ref{eq:c_ij-1})  we can write the coefficients $c_{i j}(E)\equiv c(E,|z_i-z_j|)$  as  follows 
\begin{equation}
c(E,d)=\delta_{d,0}-\frac{E}{\sqrt{\binom{n}{d}}} G_{\frac{n}{2} -d,\frac{n}{2}}(E)    \;. \label{eq:c-g}
\end{equation}
Here $G_{m,\frac{n}{2}}(E)=\bra{m}(E+2\B S_x)^{-1}\ket{n/2}$  are the  matrix elements of the resolvent  (\ref{eq:GE}) of the transverse field Hamiltonian  $H_D$ between the states (\ref{eq:sym}) that belong to  a maximum total spin subspace $S=n/2$.

As will be shown below, for typical instances  of the ensemble of Hamiltonians $H$, the Hamming distance from  a randomly selected marked state to its closest neighbor is an extensive quantity $\cO(n)$. Therefore the  above off-diagonal  matrix elements of the resolvent can be analyzed in a semiclassical approximation  corresponding to  $S=n/2 \gg1$. This  approximation for the quantum propagator of a large spin and {\it diagonal} elements of the resolvent was considered in \cite{garg2004bohr,novaes2005semiclassical} using the spin coherent state  path-integral representation. The analysis in these papers was   quite involved because the path-integral formulation requires a careful treatment of the fluctuation determinant and a so-called   Solari-Kochetov correction in the action. Also, these results were  focused on  a general case of  large spin Hamiltonian
 and  only considered  {\it diagonal} elements of the resolvent. Because of this,  instead of trying to extend the results in \cite{garg2004bohr,novaes2005semiclassical}  to our case, we follow a different path.
 
\begin{figure}[ht]
  \includegraphics[width= 3.35in]{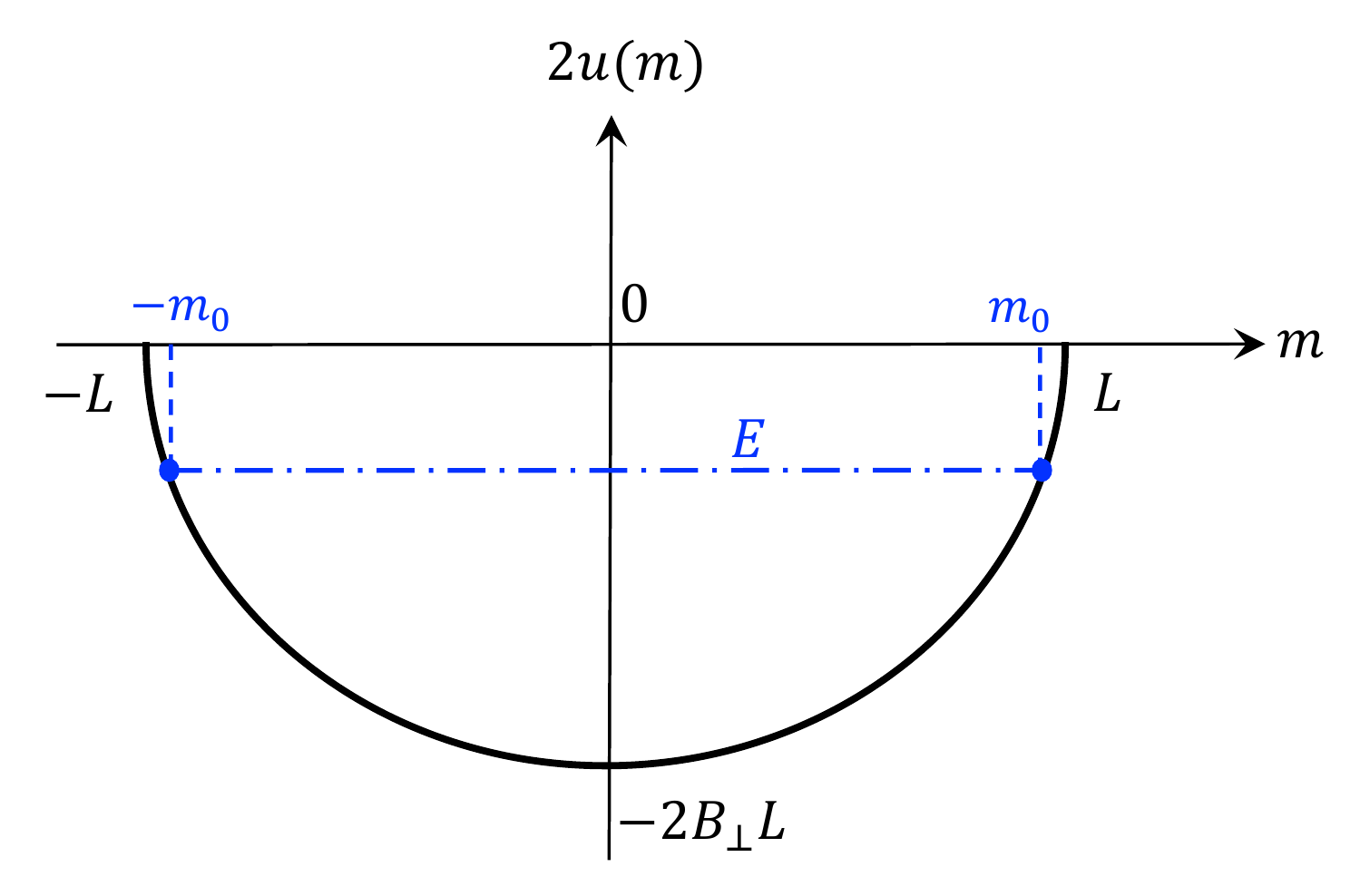}
  \caption{The black line shows the plot of $2u(m)$ (\ref{eq:u}) vs $m$ between the  interval boundaries $\pm m$=$L$=$(n+1)/2$.
 The horizontal dashed-dotted blue line depicts the region of oscillatory behavior of $G_{m,n/2}(E)$ with $m$ for a given $E$ described by the WKB solution (\ref{eq:c}) (see also Eq.~(\ref{eq:wkb-form}) in Appendix) and shown in Fig.~\ref{fig:c01}. The boundaries of this region are the turning points $m=\pm m_0(E)$ given by Eq.~(\ref{eq:m00}) and depicted with blue dots.  The regions of $m\in[m_0(E),L] \cup[-L, -m_0(E)]$ correspond to the exponential growth of $G_{m,n/2}(E)$ with $m$ (or decrease with $d=n/2-m$). The  WKB solution for the right region is given  in Eq.~(\ref{eq:f-wkb}).
   \label{fig:eff-pot}}
\end{figure}

\begin{figure}[ht]
  \includegraphics[width= 3.35in]{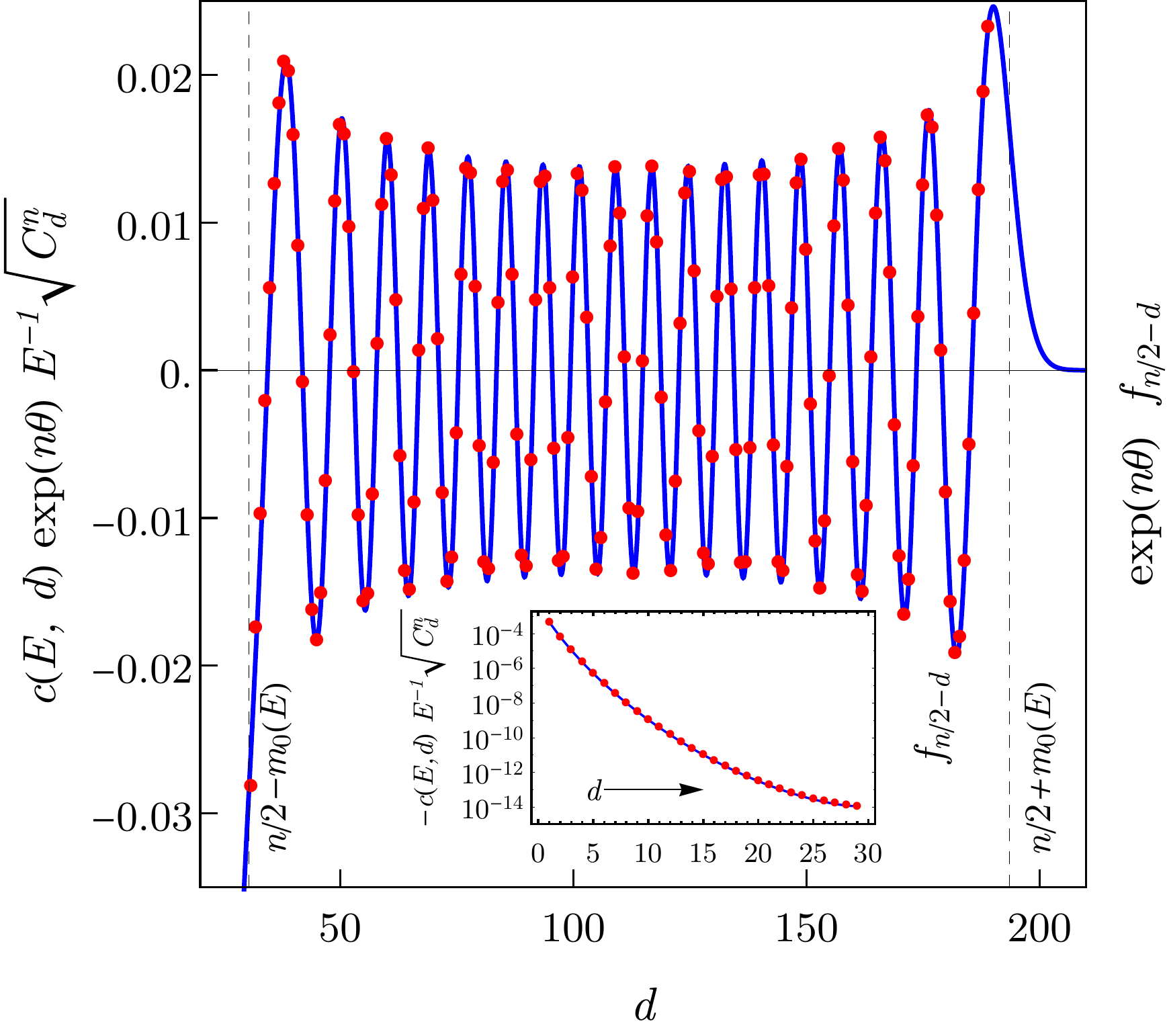}
  \caption{The blue curve shows the  $d$-dependence of the (rescaled) coupling  coefficients $c(E,d)$  computed from the exact expression (\ref{eq:CEd}) with $n=224$ and $E=-226.15$. We denote the binomial coefficient as $\binom{n}{d}\equiv C_{d}^{n}$.  The transverse  field is $\B=1.459$. For this  value of $\B$ the impurity band levels ${\cal E}(z_j)$ lie approximately in  the middle of the interval between the $p=34$th and $p=35$th excited energy levels  $-\B(n-2p)$ of the driver Hamiltonian. Red points depict the $d$-dependence of the same rescaled  coefficients  $c(E,d)$ given by  $G_{n/2-d,n/2}\, \exp(n \theta)$ and determined  by the asymptotic  WKB expressions given in Appendix (see (\ref{eq:f-wkb}),(\ref{eq:f-wkb-1})). Dashed lines indicate the boundaries of the oscillatory behavior of the WKB solution
   (\ref{eq:allowed}). The inset shows the plot for the  exponential $d$-dependence of the rescaled coupling coefficient   $-c(E,d)$  in the  region of its monotonic behavior
  $d\in[1,n/2-m_0(E)]$ (cf. Eqs.~(\ref{eq:f-wkb-1}),(\ref{eq:c-g})).  The solid blue line corresponds  to   the exact expression (\ref{eq:CEd}),  while the approximate WKB solution  is shown with red points.
\label{fig:c01}}
\end{figure}

The resolvent satisfies the equation \[I-2\B S_x G(E)=E G(E)\] where $I$ is the identity operator. We write this equation in the basis of states $\ket m$    (\ref{eq:Sz}). From (\ref{eq:GE})  we obtain
\begin{equation}
\delta_{m,\frac{n}{2}}+\sum_{s=\pm 1}u(m-s/2)  G_{{m+s},\frac{n}{2}}=E G_{m,\frac{n}{2}},\label{eq:3term}
\end{equation}
\vspace{-0.1in}
\begin{equation}
u(m)=-\B\sqrt{L^2-m^2},\quad L=\frac{n+1}{2}.\label{eq:u}
\end{equation}
In the limit of large $n\gg1$ we  solve this equation using the discrete Wentzel-Kramers-Brillouin (WKB) approximation  method \cite{braun1993discrete,garg1998application}. In the WKB analysis of Eq.~(\ref{eq:3term}) the function {\color{black} $2u(m)$ }plays the role of an effective potential for the classical system with coordinate $m$ and energy $E$. For {\color{black}$2u(m)>E$} the WKB solution for the resolvent $G_{m,n/2}(E)$  displays an oscillatory behavior with $m$ while for {\color{black}$2u(m)<E$} it exponentially increases with $m$. The boundaries of the oscillatory region $m\in[-m_0(E),m_0(E)]$ are ``turning points'' of the classical motion and  are given by the condition {\color{black}$2u(m_0)=E$} (see Fig.~\ref{fig:eff-pot}) where 
{\color{black}\begin{equation}
m_0=\sqrt{L^2-\(\frac{E}{4\B}\)^2}\;.\label{eq:m00}
\end{equation}}
In Fig.~\ref{fig:c01} we plot the comparison between the coefficient $c(E,d)$ computed based on the exact expression (\ref{eq:CEd}) and the WKB asymptotic  (details of the WKB analysis are given in Appendix~\ref{sec:det-wkb}).

In what follows we will be interested in the region $d\in[n/2-m_0,n/2+m_0]$ with  $m_0\simeq \sqrt{(n/2)^2-(E/\B)^2}$  defined by the condition $u(m_0)=E$.
This is the region of oscillatory behavior of $c(E,d)$ with $d$ where  the  leading order exponential dependence on $n$   is given by the expression
\begin{equation}
c(E,d) \propto  \frac{1}{\sqrt{\binom{n}{d}}}  e^{-n\theta(\B)}  \sin\phi(E,d)\;,\label{eq:c}
\end{equation}
with the prefactor  given in Appendix, Eqs.~(\ref{eq:c-app}),(\ref{eq:A-app}).

The function $\theta(\B)$ in (\ref{eq:c}) equals
\begin{equation}
\theta(\B) =\frac{2\arctanh\left(B_{\perp}^{-1}\right)+\B\ln\(1-B_{\perp}^{-2} \)}{4\B}\;.\label{eq:theta-x}
\end{equation}
It behaves  at large argument  as $\theta\simeq 1/(4\Bs)$.

\begin{figure}[ht]
  \includegraphics[width= 3.35in]{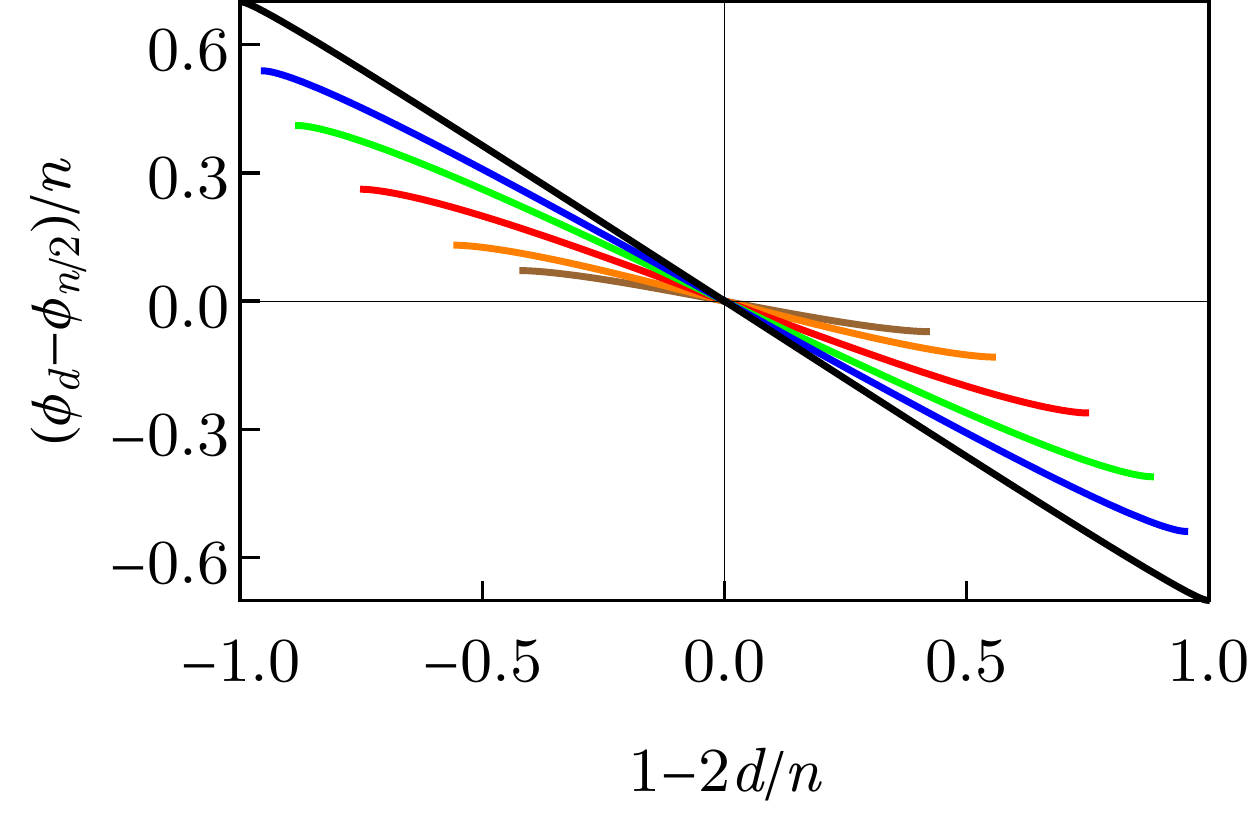}
  \caption{Plots of the  WKB phase $\phi_d\equiv \phi(E,d)$ of the oscillations of the coupling coefficient $c(E,d)$ with the  Hamming distance $d$ for a number of qubits $n=1000$.    Both axes are rescaled by  $n$.  The phase is plotted relative to its value at $d=n/2$. We set the energy $E=E^{0)}$ where
 $E^{(0)}\simeq -n-\Bs$ reflects   the overall shift of the impurity band  due to the transverse field (cf.  (\ref{eq:E0}),(\ref{eq:Delta0a})). Different color curves correspond to different values of $\B>|E|/n$ with $\B=$1.1 (brown), $\B=$1.2 (orange), $\B=$1.5 (red), $\B=$2.1(green), $\B$=3.2 (blue), $\B=$ 10 (black).   Each curve varies in its own range  $n/2-d \in[-m_0,m_0]$ where  $m_0$ is given in (\ref{eq:m00}) and  determines  the region of oscillatory behavior of the coupling coefficients (see {Appendix}~\ref{sec:det-wkb} for details). For $\B\simeq$1 the region of oscillatory behavior  shrinks to a point $d\simeq n/2$. In the limit of large values of $\B\gg 1$ this behavior occupies almost the entire range $d\in [0,n]$.
    \label{fig:phases}}
\end{figure}

An explicit form of  the WKB phase $\phi(E,d)$ in (\ref{eq:c}) is given in Appendix, Eq.~(\ref{eq:phi-m}). The dependence of the phase on $d$  for different values of $\B$ is shown in the Fig.~\ref{fig:phases}. 
This phase varies by $\cO(1)$  when $d$ is changed by 1 and it is  responsible for fast oscillation of the coupling coefficient with the Hamming distance between marked states $d$. Its dependence on $d$ simplifies in the limit of large transverse field $\B\gg1$:
\begin{equation}
\phi(E,d)\simeq \frac{\pi d}{2}-\frac{\pi n}{4}\frac{\chi(1/2-d/n)}{\B}\;,\label{eq:phi-largeB}
\end{equation}
where  $\chi(x) \simeq 1-2 \arcsin(x)/\pi+\cO(n^{-1})$. The  second term  in (\ref{eq:phi-largeB})  is much smaller the the first one, and  varies very little when $d$ is changed by 1. A predominately linear dependence of $\phi(E,d)$ on $d$ at large fields can be seen in Fig.~\ref{fig:phases}.
This property will be important in the analytical study of population transfer.

For large transverse fields  the  magnitude of  the squared coupling coefficient  (\ref{eq:c})  can be estimated to exponential accuracy as $c^2(E,d)\sim \exp[-n/(2B_{\perp}^{2})] /\binom{n}{d}$. %Here we estimated $\theta$ by expanding (\ref{eq:kappa}) in powers of $1/\B\ll1$ and assumed  that $E=-n+\cO(n^0)$. 
We note that the number of marked states $M_d$ accessible via all possible $d$-bit flips from a given state is $M_d=M 2^{-n}\binom{n}{d}$. Therefore  the leading order dependence of the coupling coefficient on $d$ is 
 proportional to  $1/\sqrt{M_d}$. As will be shown later,  in the limit of large transverse fields this leads to a nearly Grover complexity of the PT algorithm, up to a factor 
$\sim \exp[-n/(4B_{\perp}^{2})] $, which gives very small correction to Grover scaling for large $B_{\perp}$. However  when $d$ decreases below the boundary value
$d<n/2-m_0$,  the coupling coefficient grows exponentially faster than $1/\sqrt{M_d}$, as follows from the discussion in Appendix (cf. Eq.~(\ref{eq:f-wkb-1})).

\iffalse

\subsubsection{Coupling coefficients in the region  $ \left | \frac{n}{2}-d \right | \ll  n$}

In this region $ |m|\ll n$ and we can write
\begin{align}
\int_{m}^{m_0(E)} p(k,E)dk &-\frac{\pi}{4} \simeq \phi_\infty(E)\\
& -m\arcsin \left(\frac{2m_0(E)}{n}\right) \nonumber
\end{align}
Plugging this into (\ref{eq:f-wkb}), (\ref{eq:f}) we obtain in leading order in $n$
 \begin{equation}
 f_m=f_0 \frac{\sin(\phi_\infty(E)-m\arcsin(\sqrt{1-B_{\perp}^{-2}})}{\sin\phi_\infty(E)}
 \end{equation}
 We consider the limit
 \begin{equation}
 \B \gg 1\quad (\B={\cal O}(n^{0})
 \end{equation}
 In this limit 
 \begin{align}
 \sin\left(\phi_\infty(E)-m\arcsin \left(\frac{2m_0(E)}{n}\right)\right)\simeq \nonumber 
 \end{align}
 \vspace{-0.15in}
 \begin{align}
 \simeq \left [\begin{array}{cc}(-1)^{\frac{m}{2}}\sin(\phi_\infty(E)+m/\B),& m-{\rm even}\\ (-1)^{\frac{m+1}{2}}\cos(\phi_\infty(E)+m/\B&m-{\rm odd}\end{array}\right.\nonumber
 \end{align}
 With this the expression for the squared coupling coefficients $c^2(E,d)$ has the form
 \begin{align}
 c^2(E,d)= \frac{A(E) \left(1-(-1)^m \cos(2\phi_\infty(E)+2m/\B)\right)}{\binom{n}{d}}.\label{eq:c2} 
 \end{align}
 
\fi

 \section{\label{sec:lin} Downfolded  Hamiltonian near the  center of the Impurity Band }

The coupling coefficients  $c(E,d)$ (\ref{eq:CEd}) decay exponentially with Hamming distances for $d={\cal O}(n)$ (see details in Sec.~\ref{sec:cd}).
Marked states are selected at random and Hamming distances between them are order $n$ when the number of the states $M$ is exponentially smaller than $2^n$. 
Because the off-diagonal matrix elements of the downfolded  Hamiltonian ${\cal H}_{ij}(E)\propto c(E,d_{ij})$ they are exponentially small in $n$.
At the same time  the width  of the distribution of energies  of the marked states $\cE(z_j)=-n +\epsilon_j$  is also assumed to be very small, $W\ll \B$  (it is exponentially small in $n$ for the cases of interest).
Therefore we  can solve the nonlinear eigenvalue problem  (\ref{eq:Hc1})-(\ref{eq:nep})  by an iterative approach  
treating the off-diagonal part of ${\cal H}(E)$  and terms $\propto \epsilon_j $ as a perturbation. Details are given in {Appendix}~\ref{sec:lin-H-app}.

\begin{figure}[ht]
  \includegraphics[width= 3.35in]{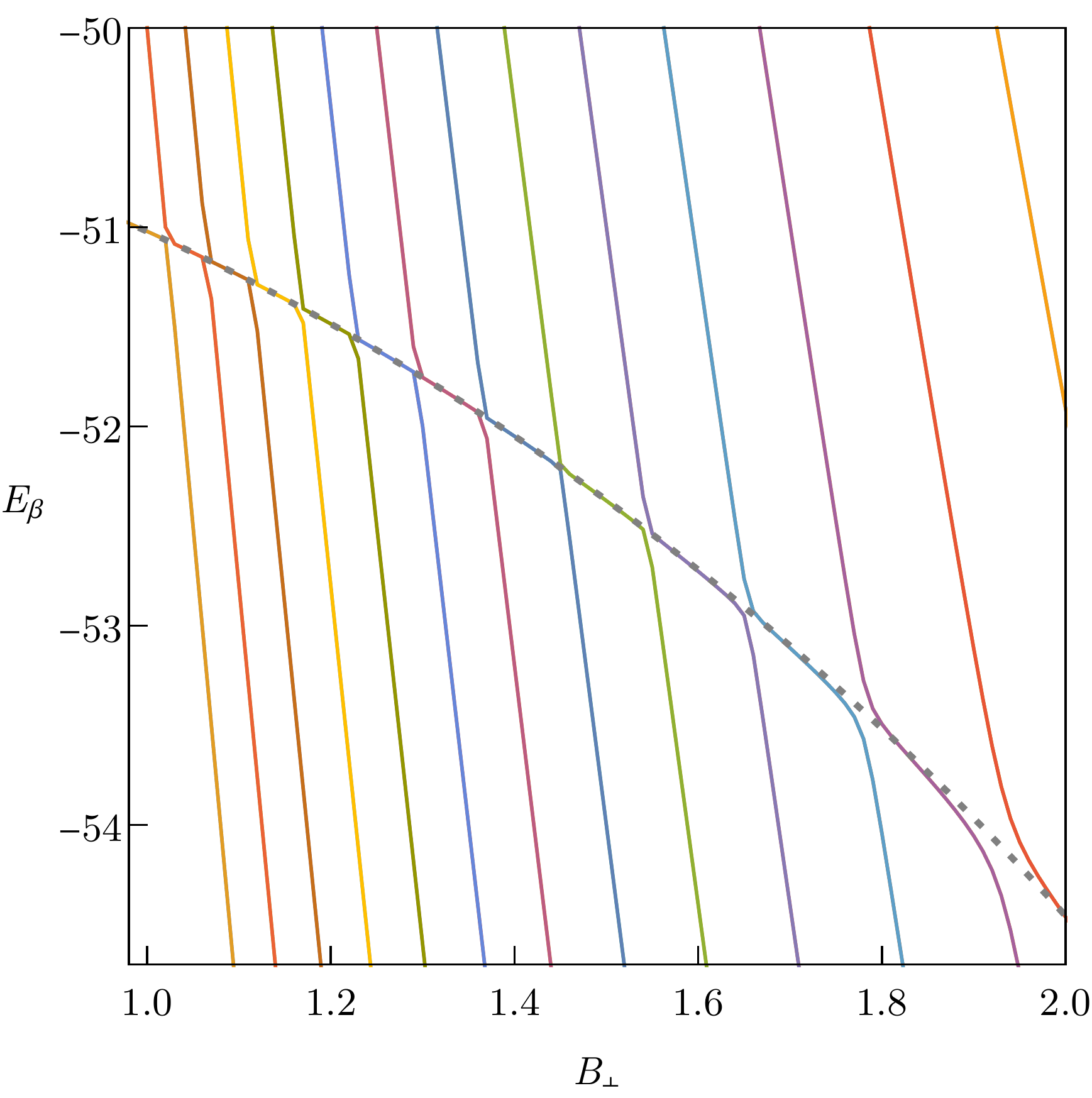}
  \caption{Solid lines show the  dependence on the transverse field  $\B$  of the eigenvalues $E_\beta$ of the non-linear eigenvalue problem with Hamiltonian $\cH(E)$  % (cf.Eqs.~(\ref{eq:Hc1}) ,(\ref{eq:Hc2}) ,(\ref{eq:nep})). 
  for the case of $n=50$ and $M=2$.  The plot shows the repeated avoided crossing between the two systems of eigenvalues.  One system (colored lines) corresponds to the eigenvalues of  the transverse field (driver) Hamiltonian $H_D=- \B \sum_{k=0}^n \sigma_x^k$ in the limit $H_{\rm cl}\rightarrow 0$.  The second system of eigenvalues corresponds to the  energies of the two marked states  in the limit $\B\rightarrow 0$. The splitting of the eigenvalues  is exponentially small in $n$ and not resolved in the plot. The asymptotic expression (\ref{eq:E0}),(\ref{eq:Delta0a}) for the two  eigenvalues $E_{1,2}^{(0)}=E^{(0)}$ neglecting the tunneling splitting and setting  $\cE({z_j})=-n$  for all $j\in[1,M]$ are shown with dashed  gray line.     \label{fig:spectrum}}
\end{figure}

At zeroth-order in the perturbation, the down-folded Hamiltonian  $\cH_{ij}^{(0)}(E)= \delta_{ij} n(c(E,0)-1)$ has one $M$-fold degenerate
energy level $E^{(0)}$ that is a root of the equation  $\cH_{ij}^{(0)}(E)=E$ that originates from the   marked  state energy, $E^{(0)}\rightarrow -n$,
in the limit of $\B\rightarrow 0$. %The  corresponding $M$ eigenstates $\psi_\beta(z_j)$ ($\beta\in[1..M]$) have support  over the  part of  computational basis   corresponding  to  
%marked  states:   $\psi_{z_j}^{\beta}\neq 0,\,\,j\in(1,M)$.  
Using $c(E,0)$ from Eqs.~(\ref{eq:c-HD-G}), (\ref{eq:CEd}) the explicit form of the equation  for $E^{(0)}$ is
\begin{equation}
E^{(0)} =-n-\Delta_0\;,\label{eq:E0}
\end{equation}
\vspace{-0.23in}
 \begin{equation}
\Delta_0=n 2^{-n}\sum_{d=0}^{n}\binom{n}{d}\frac{\B(n-2d)}{n+\Delta_0-\B(n-2d) }\;.\label{eq:Delta0a}
\end{equation} 
 Here    $\Delta_0$ is  the root of the  above  transcendental equation that satisfies the condition  $\lim_{\B\rightarrow 0}\Delta_0=0$. In general,  
  the sum (\ref{eq:Delta0a}) is dominated by the region of   values of $d$ such that $|d-n/2|={\cal O}(n^{1/2})$ where the factor  $2^{-n}\binom{n}{d}$  
reaches its maximum $\sim n^{-1/2}$. In that region we replace the binomial coefficient with a Gaussian function of $d$ and  the summation with the integral over $d$. Taking the integral 
we    obtain   $\Delta_0$  in a form of a series expansion in powers of $n^{-1}$ 
\begin{equation}
\Delta_0\simeq -B_{\perp}^{2}-\frac{B_{\perp}^{4}}{n}+{\cal O}(n^{-2}), \label{eq:Delta0as}
\end{equation}  
A comparison between the exact and asymptotic solutions for  $\Delta_0$ is shown in Fig.~\ref{fig:spectrum}. For $\B\ll n^{1/2}$ the overall shift of the energies of the marked states is negative and quadratic in $\B$.

According to Eq.~(\ref{eq:normA})   all  $M$  degenerate eigenstates  $\ket{\psi}_\beta$ have the same weight $Q(E^{(0)})=\sum_{j=1}^{M}|\psi_{\beta}^{(0)}(z_j)|^2$ on the marked state subspace.   In the  large $n$ limit we have
\begin{equation}
Q(E^{(0)})\simeq 1-\frac{B_{\perp}^{2}}{n}+{\cal O}(B_{\perp}^{4}/n^2). \label{eq:Qas}
\end{equation}
Under the condition   $\propto B_{\perp}^{2}/n\ll1$, the  eigenstates  are dominated by their projections on the marked state subspace. In the limit $n\rightarrow\infty$ they are 
 asymptotically orthogonal to the computational  basis states outside the IB.  Such "orthogonality catastrophe" cannot be obtained within the perturbative in $B_{\perp}$ approach such as FSA. 
 
\begin{figure}[ht]
  \includegraphics[width= 3.35in]{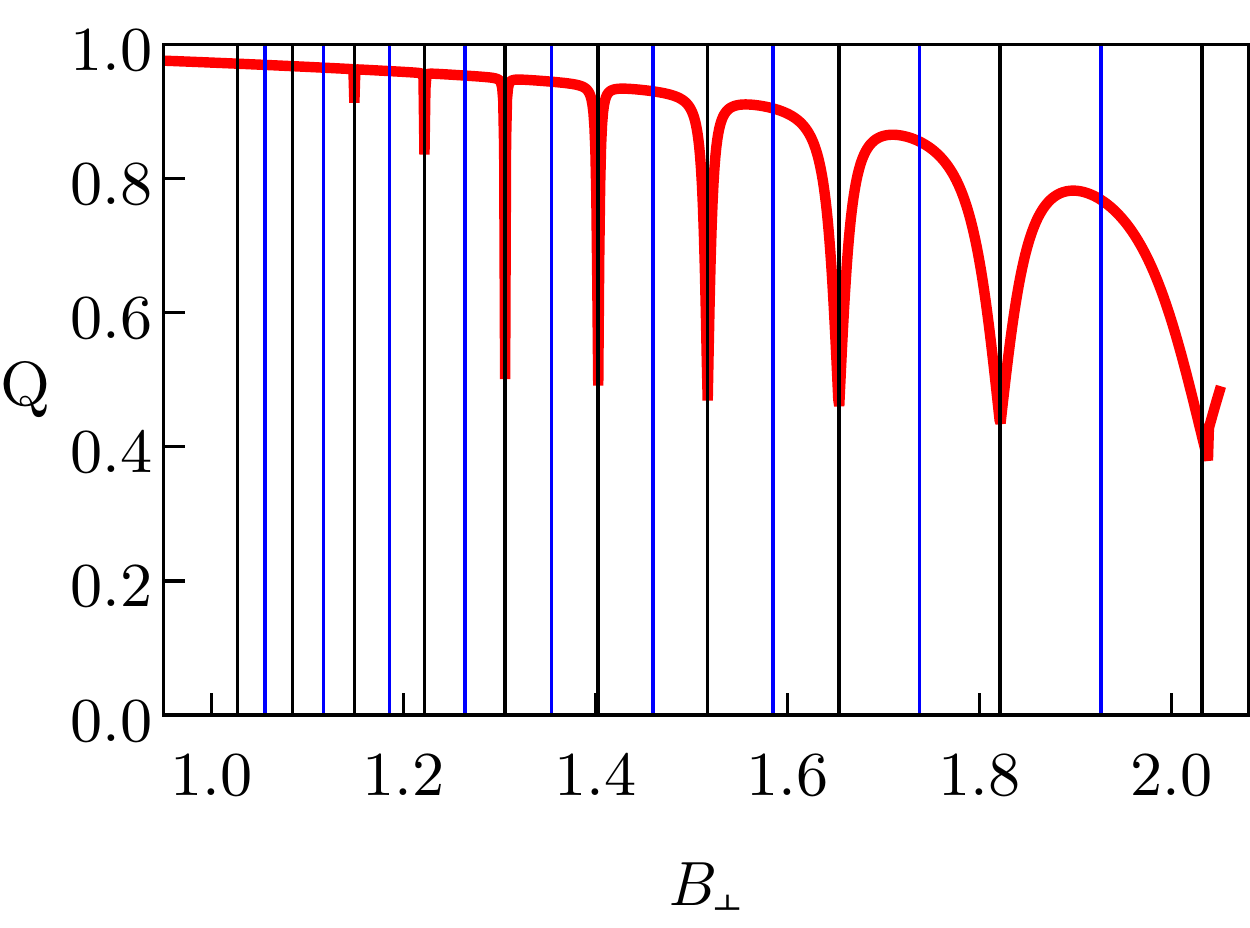}
  \caption{The solid red line shows the dependence of the total weight $Q$ {\it vs} transverse field $\B$ for $n=40$. Vertical black and blue lines, respectively, depict  the locations of $p$-even and $p$-odd resonances $\B=B_{\perp\, p}$ defined in the text. The total weight $Q$ undergoes sharp decreases in the vicinity of even resonances. For $p<5$ the resonance regions are so narrow that dips in $Q$ are not seen. The width of the regions grow steeply with $p$. 
   \label{fig:Q-res}}
\end{figure}

 The exact dependence of the weight $Q$ on transverse field $\B$ is given in Fig.~\ref{fig:Q-res}. The  expression (\ref{eq:Qas})
  is valid for $\B$ away from their "resonant" values $B_{\perp,p}\simeq n/(n-2p)$ where the $M$-fold degenerate energy level   "crosses''  the eigenvalues of the driver  Hamiltonian, $E^{(0)}=-\B(n-2p)$, for integer values of $p$, as shown in Fig.~\ref{fig:spectrum}. The width of such resonance regions $\Delta B_{\perp,p}\propto 2^{-n/2}\binom{n}{p}$  %grows quickly with $p$  but 
  remains exponentially small in $n$  for   $n/2-p\gg n^{1/2}$. 
  
\begin{figure}
 \includegraphics[width= 3.35in]{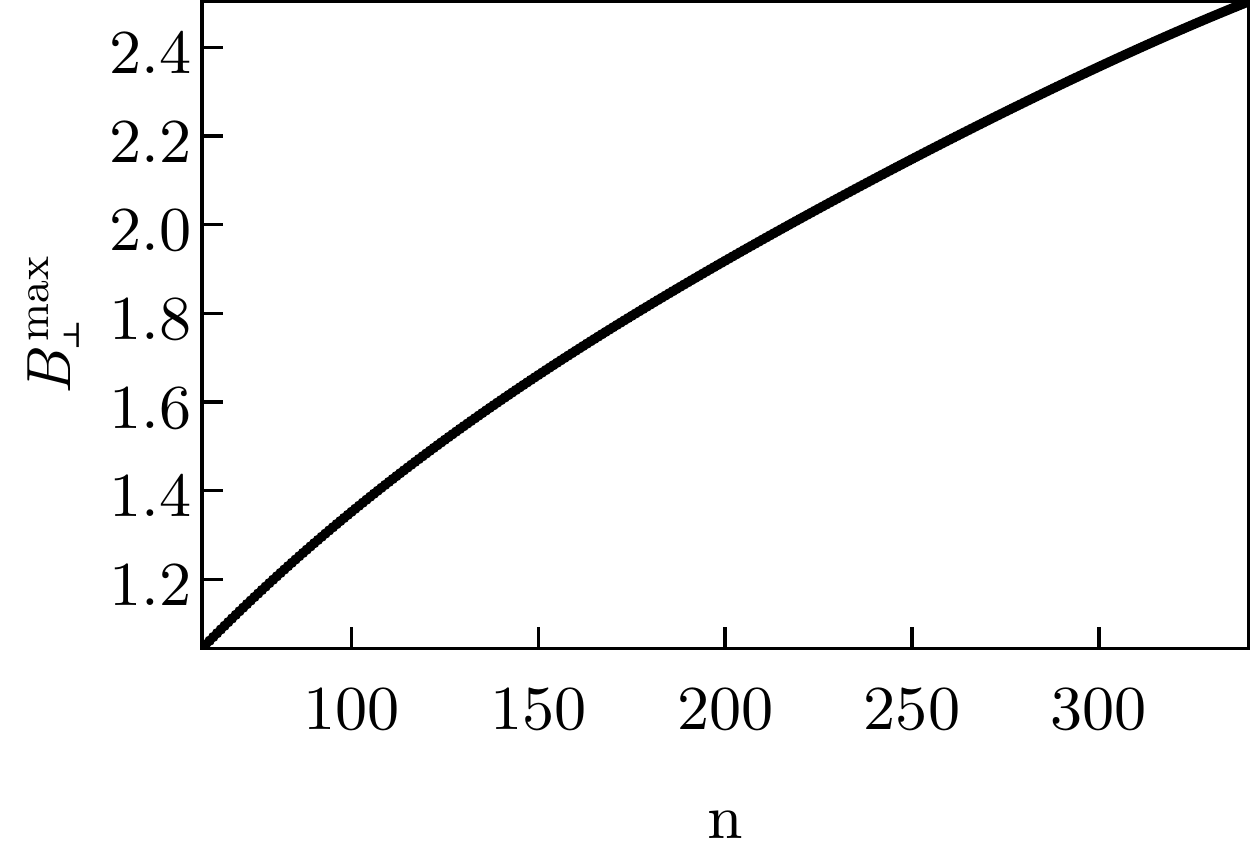}
  \caption{Plot of the  maximum value of the transverse field at mid-resonance point $B_{\perp}^{\rm max}$ as a function of $n$. We defined $B_{\perp}^{\rm max}=(B_{\perp\,p} +B_{\perp\,p+1})/2$ where  $B_{\perp,p}\simeq n/(n-2p)$ satisfies  the equation $E^{(0)}=-B_{\perp,k}(n-2p)$  and    the integer $p$ is equal to its maximum possible value $ p =p_{\rm max}$ for which   the weight factor $Q=Q(B_{\perp\,p} +B_{\perp\,p+1})/2)\geq 0.98$.
    \label{fig:Gamma-max}}
\end{figure}

 In this study we will  focus on the off-resonance case depicted in Fig.~\ref{fig:cartoon_levels}. One can  see from  Fig.~\ref{fig:spectrum} that $B_{\perp,p}$ increases with $p$ and so is the width of resonance region.  For $\B$  parametrically large  compared to unity one needs to make sure that $n$ is also large enough so that the width of the resonance regions  is small (cf.  Fig.~\ref{fig:Gamma-max}). Away from resonance, all $M$ impurity band eigenstates are  well localized in the marked states subspace (cf.  (\ref{eq:Qas})).

In the spirit of the degenerate perturbation theory, there exists an effective Hamiltonian $\scH$ that  determines the correct zeroth order eigenstates 
and removes the degeneracy of the energy levels
\begin{equation}
\scH \ket{\psi_{\beta}^{(0)}}=E_{\beta}^{(1)}\ket{\psi^{(0)}_{\beta}}\;,\label{eq:eigenIBH}
\end{equation}
Its matrix in the basis of the marked states has the form $\scH_{ij}=\delta_{ij} \e_i+n c(E^{(0)},d_{ij})$ where we neglected small non-important corrections (see Appendix~\ref{sec:lin-H-app}). {\color{black}Using the  expression for  the coupling coefficient 
(\ref{eq:c})  given in Appendix~\ref{sec:det-wkb}, ((\ref{eq:c-app}),(\ref{eq:A-app})) we have}
\begin{equation}
\scH_{ij}=\delta_{ij}\epsilon_j+(1-\delta_{ij})\cV_{ij} \sqrt{2} \sin \phi(d_{ij})\;.\label{eq:IBH}
\end{equation} 
%\vspace{-0.1in}
%\begin{equation}
%V_{ij}=V(E^{(0)},d_{ij})\equiv V(d_{ij})\;.\nonumber
%\end{equation}
Here   $\phi(d)\equiv \phi(E^{(0)},d)$ is a WKB phase shown in Fig.~\ref{fig:phases} that describes the  oscillation of the matrix elements with the Hamming distance.
{\color{black}Its explicit form is given in Appendix~\ref{sec:det-wkb}, Eq.~(\ref{eq:phi-m}) and also above in Eq.~(\ref{eq:phi-largeB}) for the case of large transverse fields.} The amplitude $\cV_{ij}$
equals 
\begin{equation}
\cV_{ij} \equiv V(d_{ij}),\quad V(d)=\sqrt{A(d/n)}\,\,\frac{ n^{5/4} \,  e^{-n\theta(\B)} }{\sqrt{\binom{n}{d}}}\;,\label{eq:V2}
\end{equation}
 where $i$ $\neq$$ j$ and the coefficient $A(\rho)$ equals (cf. 
  (\ref{eq:A-app}))
  \begin{equation}
A(\rho)=\sqrt{\frac{\pi}{32 }} \frac{e^{-\B\arccoth\B}}{(B_{\perp}^{2}-1)\upsilon(\rho)\sin^4(\phi(n/2))}\;,\label{eq:A}
\end{equation}
\begin{equation}
\upsilon(\rho)=\left(1-\frac{(1-2\rho)^2}{1-B_{\perp}^{-2}}\right)^{1/2}\;.\label{eq:upsilon}
\end{equation}
It is independent on $n$ apart from the phase
  $\phi(n/2)$ whose  explicit form is
\begin{equation}
\phi(n/2)=\frac{\pi}{4}\left(n(1-B_{\perp}^{-1})-\B\right)\;.\label{eq:phi0}
\end{equation}
 The function $\theta(\B)$ is given in (\ref{eq:theta-x}).
Expanding  (\ref{eq:theta-x}) in the limit  $\B\gg1$, 
\begin{equation}
\theta \simeq \frac{1}{4 B_{\perp}^{2} }+\frac{1}{24 B_{\perp}^{4}}+\frac{1}{60B_{\perp}^{6}}+\ldots.\; \label{eq:g-exp}
\end{equation}
In that limit $\theta\ll 1$. We note that even for modest values of transverse field, e.g., $\B\simeq 1.46$ (corresponding to that in the Fig.~\ref{fig:c01}) the first term provides a good estimate to the value of $\theta\simeq 0.13$ (error $9\%$). We shall refer to $\scH$ in (\ref{eq:IBH}) as the Impurity Band (IB) Hamiltonian.

{\color{black}The form of the IB Hamiltonian (\ref{eq:IBH}) only \textcolor{black}{applies} to  the region of oscillatory behavior  $d_{ij}\in[n/2-m_0,n/2+m_0]$  of the coupling coefficients $c_{ij}(E)$ with Hamming distance $d_{ij}$ where  $m_0$  is given in (\ref{eq:m00}). This above condition for $d_{ij}$ is always satisfied   in a typical row of the matrix $d_{ij}$ for the  values of $M$  considered in the paper (see the discussion in Appendix~\ref{sec:distV2} and  Eq.~(\ref{eq:cond-ossc})).}

\section{\label{sec:ensembleH} Statistical ensemble of the  Impurity Band Hamiltonians} 

Properties of the eigenstates and eigenvalues (\ref{eq:eigenIBH}) of the IB Hamiltonian $\scH$ (\ref{eq:IBH})
  determine the population transfer within the Impurity Band and are thus 
  of the central interest for us in this study.  They depend on the statistical ensemble of IB Hamiltonians. In the model  considered in this paper diagonal elements  $\epsilon_j$ of $\scH$   are  selected at random, independently  from each other and from  the choice of the corresponding marked states  $\ket{z_j}$.    In the present discussion we assume
  that  the   PDF 
    $p(\e)$  of  $\epsilon_j$   is exponential bounded with the width  $W$.   The results do not depend on the particular form  of $p(\e)$. For the sake of specificity in calculations we  will use  the window function form 
    \begin{equation}
    p(\e)=\theta\left(W/2-|\e|\right)\;,\label{eq:pA}
    \end{equation}
where $\theta(x)$ is a Heaviside theta function.
For the physical effects discussed in the paper to take place the width $W$ needs to scale down exponentially with $n$  
\begin{equation}
\lim_{n\rightarrow \infty} \log (W^{1/n})={\cal O}(n^{0}).\label{eq:eWn}
\end{equation}

\begin{figure}[ht]
  \includegraphics[width= 3.35in]{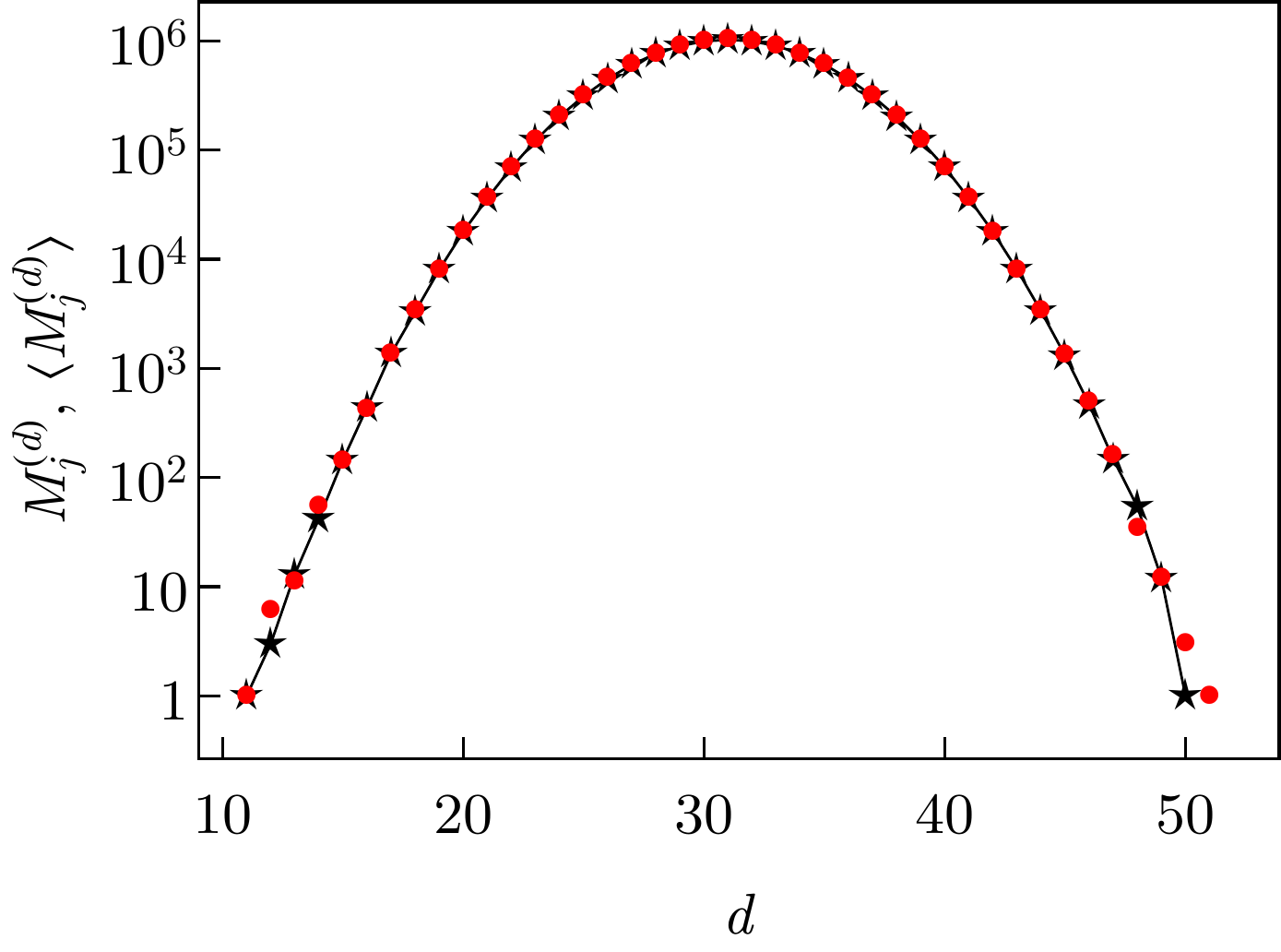}
  \caption{Red points show the empirical probability distribution $M_{j}^{(d)}$ {\it vs} $d$  with  $M_{j}^{(d)}=\sum_{j=1}^{M}\delta(d_{ij}-d)$. Here   $d_{ij}$ is a 
   matrix of Hamming distances $d_{ij}$ between the set of $M$ randomly chosen $n$-bit-strings (marked states) and $i$ is a randomly chosen marked state. The distribution corresponds to 
    $M=10^7$and $n=60$.  Black stars connected by a black line show the samples ${m_d}$ from multinomial distribution with mean values $\langle M_{j}^{(d)} \rangle = M p_d$ where $p_d$ is binomial distribution (\ref{eq:p_d}). 
  \label{fig:pd}}
\end{figure}

\subsection{Off-diagonal matrix elements}

%\subsubsection{\label{sec:stat}Statistical independence of matrix elements}

For fixed energies $\epsilon_j$ the  matrix of the IB Hamiltonian $\scH_{ij}$ is entirely determined by the symmetric matrix of Hamming distances $d_{ij}$ between the bit-strings corresponding to the marked  states. The set of $M$ bit-strings is randomly sampled from the full set of all possible  $2^n$ bit-strings $\stn$ without replacement, see Appendix~\ref{sec:stat-app}. Elements of the matrix $d_{ij}$ above or below the main diagonal will be considered independent from each other and taken from  the binomial distribution $p_d$,
\begin{equation}
p_d =\frac{1}{Z} \,2^{-n}\binom{n}{d},\quad Z=\sum_{d=1}^{n}2^{-n}\binom{n}{d},\; \label{eq:p_d}
\end{equation} 
under condition $1\ll M \ll 2^{n/2}$. Then, for a given row of the matrix $M\times M$ of Hamming distances $d_{ij}$,   the numbers of elements $M_{j}^{(d)}$ with $d_{ij}=d$ are samples from the multinomial distribution with mean values $\langle M_{j}^{(d)}\rangle=M p_d$ (see Fig.~\ref{fig:pd}).   According to  (\ref{eq:IBH}),(\ref{eq:c}) the statistical ensemble of  IB Hamiltonians (\ref{eq:IBH}) corresponds to that of symmetric random matrices whose associated graphs are fully connected and  matrix elements are  statistically independent.  

%  % \subsubsection{\label{sec:averaging}Oscillations of  $\left(\scH_{ij}\right)^{2}$ with $d_{ij}$}
%  Spectral properties associated with the local resolvent of  the random matrices of this type   can be  studied with  the cavity method    \cite{cizeau1994theory,metz2010localization,tarquini2016level,facoetti2016non,monthus2016localization} which is asymptotically exact in this case.
%In an analysis with the cavity method, the matrix elements  always enter in a {\it squared} form, $(\scH_{ij})^{2}$.   
As will be seen below  the spectral properties of  $\scH$ that are relevant for  our study  are determined by $\cV_{ij}^{2}$ and not by the oscillatory factor in (\ref{eq:IBH}). Therefore we will be interested in the PDF of  $\cV_{ij}^{2}$ 
\begin{equation}
P(\cV_{ij}^{2})= \sum_{d=1}^{n} p_d  \, \delta(V^2(d)-\cV_{ij}^{2})\;,\label{eq:V2dist}
\end{equation}
where $i\neq j$.% and  $\delta(x)$ denotes the Kronnecker  delta. 

\subsubsection{ \label{sec:typical}Typical and extreme values of the off-diagonal matrix elements $\cV_{ij}$}
  For a  randomly chosen row of the  matrix of Hamming distances $d_{ij}$ the most probable value (mean)  of its elements equals to $n/2$.   According to  (\ref{eq:V2}),  the    off-diagonal matrix elements   $\cV_{ij}$
  decrease rapidly with the Hamming distance $d_{ij}$, reaching the  minimum  value at $d_{ij}\simeq n/2$. Therefore a typical minimum value of the matrix elements $\cV_{ij}$  corresponds to  a typical value overall. We estimate it using Eq.~(\ref{eq:V2})  and Stirling's approximation 
  \begin{equation}
 V_{\rm typ}=V(n/2)\simeq \left(\frac{\pi A^2  }{2} \right)^{1/4}  n^{2} 2^{-n/2} e^{-n \theta}\;.\label{eq:typ}
 \end{equation}
 where coefficient $A=A(E^{(0)},1/2)$   (\ref{eq:A}) is essentially $n$-independent between the resonances  and $\theta$ is given in (\ref{eq:theta-x}). The matrix elements  $\cV_{ij}$ that  scale with $n$ as the typical value in (\ref{eq:typ})  correspond to  $| n/2-d_{ij}| ={\cal O}(\sqrt{n})$.  %Their number is  exponentially close to  $ M$  (i.e., almost all elements in a matrix). 
 
% \subsubsection{\label{sec:typical}  Typical largest  off-diagonal matrix elements $\cV_{ij}$}
 We note that in the Fig.~\ref{fig:pd} the plot points do not reach the boundaries of the interval $d=0,n$. In the matrix of Hamming distances   $d_{ij}$ the  typical  smallest off-diagonal  element in a  randomly chosen row can be estimated  as follows $M p_{\dmin}=1$ where $p_d$ is binomial distribution (\ref{eq:p_d})
\begin{equation}
\min_{j\neq i,\,1\leq j\leq M} d_{ij} \sim \dmin, \quad M 2^{-n} \binom{n}{\dmin} =1.\label{eq:dmin}
\end{equation}
  Using Stirling's approximation for  factorials in the limit $n\gg 1$ it is easy to show that minimum Hamming distance in a row is extensive for $ M=2^{\mu n}, \mu <1$. 
 
 The typical largest magnitude off-diagonal matrix element in a randomly chosen row of $\cV_{ij}$ is equal to $V(\dmin)$.  Using Stirling's approximation in (\ref{eq:V2}) we get,
 \begin{equation}
 \max_{j\neq i,\,1\leq j\leq M}|\cV_{ij}| \sim  M^{1/2} \, V_{\rm typ}\;. \label{eq:maxH}
 \end{equation}
 Using (\ref{eq:typ}) one can see   that the maximum off-diagonal matrix element in a randomly chosen row is still exponentially small in $n$.
 
 Similarly, one can estimate  the typical value of the absolute minimum $\dminmin$ of  a Hamming distance $d_{ij}$  between a pair of marked states. This distance remains extensive for $\mu<1, M=2^{\mu n}$. This distance corresponds to the overall largest in magnitude element  of the matrix $\cV_{ij}$ 
 \begin{align}
  \max_{1\leq i< j\leq M }|\cV_{ij}|  \sim  M  V_{\rm typ}\;.\label{eq:maxoverall}
 \end{align}
 Using (\ref{eq:typ}) the largest element is exponentially small  in $n$ provided that   $\mu< 1/2$ which corresponds to the condition of statistical independence of the elements of $\cV_{ij}$. A tight bound for the maximum eigenvalues of H can be obtained using Gerschgorin circle theorem~\cite{weisstein2003gershgorin}, see Appendix~\ref{sec:GerschgorinTh}.

 \subsection{\label{sec:tails}Heavy tails}
 
 It can be shown that the variance of $\scH_{ij}$  is not a good
statistical characteristic of its PDF and is  dominated by the extremely rare atypical instances of the
ensemble (see details in the  Appendix~\ref{sec:MeanValueSigmaHeff}). We observe that the relationship between the typical matrix element (\ref{eq:typ}), maximum matrix element in a randomly chosen row of $\cV_{ij}$ (\ref{eq:maxH}), and the  largest element of $\cV_{ij}$ overall (\ref{eq:maxoverall}) form a strong hierarchy that is a characteristic of the ensemble of    dense   matrices with broad non-exponential distribution of matrix elements (Levy matrices) \cite{cizeau1994theory}. The form of  the hierarchy \cite{monthus2016localization}  suggests  (up to a logarithmic factors)  the following asymptotic behavior at the tail  of the PDF of the matrix elements: 
 \begin{equation}
 \PDF(\cV_{ij}^{2})\propto |\cV_{ij}|^{-2},\nonumber
 \end{equation}
  for $|\cV_{ij}|\gg V_{\rm typ}$.

 We will build on the above observation and  obtain the explicit form of the  PDF of the matrix elements $P(\cV_{ij}^{2})$ (\ref{eq:V2dist}),  including its tails. 
 In the asymptotic limit of large $n\gg 1$ we consider $n$ to be a continuous variable (the validity of this approximation will be justified below). We replace the summation over $d$ in (\ref{eq:V2dist}) by an integral and Kronecker delta 
 $\delta(x)$ by Dirac delta
 \begin{equation}
 P(\cV_{ij}^{2})\simeq \int_{0}^{n} p_x  \, \delta(V^2(x)-\cV_{ij}^{2})dx\;.\label{eq:PVm}
 \end{equation}
 This expression is obtained  using the analytical continuation  of the binomial distribution  $p_d$ (\ref{eq:p_d})  from the integer domain $d\in (0,n)$  onto  the interval of a real axis $x\in (0,n)$ in terms of the Beta function and the resulting identity  $\int_{0}^{n}dx \,p_x=1$ (see {Appendix}~\ref{sec:distV2} for details).

In what follows we will study the   rescaled quantities 
\begin{equation}
w_{ij}=\frac{V_{ij}^{2}}{V_{\rm typ}^{2}}\equiv \frac{V^2(d_{ij})}{V_{\rm typ}^{2}}\;,\label{eq:w-m}
\end{equation}
\noindent
where   $i\neq j$ and  $ V_{\rm typ}$  is  given in (\ref{eq:typ}). We apply Stirling's approximation for  the binomial coefficient in Eq.~(\ref{eq:V2}) and (\ref{eq:p_d}) and obtain asymptotic expressions for   $V^2(d)$ and $p_d$, respectively. Plugging them into the  (\ref{eq:PVm}) and taking the integral there we  can obtain   the PDF
\begin{equation}
g(w_{ij})=V_{\rm typ}^{2}\,P(V_{\rm typ}^{2}\,w_{ij})\;.\label{eq:ginfP}
\end{equation}
whose form is given in Appendix,  Eqs.~(\ref{eq:g-ell}),(\ref{eq:ell}).

The  following assumption will be applied throughout the paper
\begin{equation}
M=2^{\mu n}, \mu \ll 1. \label{eq:log2M}
\end{equation}
According to Eqs.~(\ref{eq:V2}), (\ref{eq:dmin}),(\ref{eq:w})
a typical largest  element in a randomly chosen row of the matrix $w_{ij}$ is $\sim M$. Therefore based on (\ref{eq:log2M}) the following condition is satisfied in a randomly chosen row of $w_{ij}$
\begin{equation}
\frac{1}{n}\log_2 w_{ij}\ll 1\quad (1\leq w_{ij} \apprle M)\;.\label{eq:range-r}
\end{equation}
Under this condition,  the PDF of $w_{ij}$  takes  a particularly simple  form, $g(w)\simeq g_\infty(w)$  
 \begin{equation}
g_\infty(w)=\frac{1}{w^2\sqrt{\pi \log w}},\quad w\in(1,\infty)\;,\label{eq:g_0(w)}
\end{equation}
with normalization condition $\int_{1}^{\infty}g_\infty(w)dw=1$. Details of the derivation are given in Appendix~\ref{sec:distV2}. 

The  above analysis assumes the   scaling behavior (\ref{eq:V2}) of  $\cV_{ij}$  with $d_{ij}$ that requires  $|n/2-d_{ij}| <m_0$  with $m_0$ given in (\ref{eq:m00}). As shown in  Appendix~\ref{sec:distV2}  this condition is always satisfied for a typical row of $d_{ij}$ provided the constraint (\ref{eq:log2M}) on the values of $M$.

\subsection{Preferred basis Levy matrices (PBLMs)}

The problem of population transfer is reduced to the analysis of the described above ensemble of real symmetric $M\times M$ matrices  $\scH_{ij}$ of the down-folded IB Hamiltonian (\ref{eq:IBH}). %The ensemble is  described above and is defined over an $M$-dimensional subset of the computational basis associated with marked states.
%The non-oscillatory "amplitude" factor $\cV_{ij}$ in the expression for the  off-diagonal matrix elements   of the Hamiltonian  (\ref{eq:IBH})  obeys the Pareto-like distribution   with polynomial tail and typical value $\cVt$ (\ref{eq:typ}).  An explicit analytical form of the  PDF $P(\cV_{ij}^{2})$   is given in   (\ref{eq:w-m}),(\ref{eq:g_0(w)}). The PDF of diagonal matrix elements $p(\epsilon_i)$ is assumed to be exponentially bounded with some width $W$. 
 The matrices $\scH_{ij}$ form an ensemble of preferred basis Levy matrices (PBLMs), a generalization of Levy matrices actively studied in the literature~(cf., e.g., \cite{tarquini2016level,monthus2016localization,metz2010localization,cizeau1994theory}). Unlike Levy matrices PBLMs have a new control parameter: the ratio of typical diagonal to off-diagonal matrix elements $W/\cVt$ that controls the preferential basis (computation basis). This distinction is analogous to that between Gaussian Orthogonal ensemble (GOE) and the Gaussian ensemble with broken SU(N) symmetry, the Rosenzweig-Porter (RP) model~\cite{rosenzweig1960repulsion}.

Recent studies of RP ensemble~\cite{Kravtzov2015RzPr}  demonstrated two localization transitions that occur with varying parameter that controls the relative weight of the diagonal and off-diagonal matrix elements. One of them is the Anderson transition from localized to the extended states that are non-ergodic and posses distinct multifractal features. These states and the corresponding eigenvalues are organized in  "minibands'' so that the states within the same miniband mostly share the same support over basis states. The spectral width of the minibands is polynomially  small (in $M$) compared to $W$.  The second transition is from the extended non-ergodic states to the extended ergodic states similar to the eigenstates of the Gaussian Orthogonal Ensemble. We demonstrate analogous behavior in the IB model and analyze the population transfer in the non-ergodic regime.

\section{\label{sec:NumSim} Numerical simulations:  minibands of non-ergodic delocalized states}

In this Section we report exact diagonalization analysis of both the eigenvector statistics and the dynamical eigenstate correlator. Instead of the sparse $2^n\times2^n$ Hamiltonian Eq.~(\ref{eq:H}), it is efficient to diagonalize the dense $M\times  M$ matrices obtained by down-folding the Hamiltonian into the marked states subspace. This allows access to systems of $n=200$ qubits, reducing the finite size effects. The down-folded matrix Hamiltonian ensemble, is constructed as in Sec.~\ref{sec:ensembleH}, 
\begin{gather}
\scH_{ii}=\epsilon_i, \quad  \scH_{ij}=n c(E^{(0)},d_{ij}),  
\end{gather}
where the diagonal elements $\epsilon_m$ are distributed uniformly in the energy window $\left[-n-W/2, -n+W/2 \right]$, and the off-diagonal elements are constructed by sampling Hamming distances between uniformly random bitstrings of length $n$ and using Eq.~(\ref{eq:CEd}) with $E=E^{(0)}$ determined from Eqs.~(\ref{eq:E0}),(\ref{eq:Delta0a}). 

We introduce the scaling of the width of the distribution of $\epsilon_m$ with the matrix size $M$,
\begin{gather}
W= \lambda M^{\gamma/2} V_{\textrm{typ}}\,, \label{eq:NumericalBandWidth}
\end{gather}
where $\gamma$ is a real non-negative parameter that controls the scaling of the typical diagonal to off-diagonal matrix element $V_{\textrm{typ}}$ given in Eq.~(\ref{eq:typ}), and $\lambda$ is an auxiliary constant of order one.

\subsection{Eigenvector statistics}

\begin{figure}[ht]
\includegraphics[width=\columnwidth]{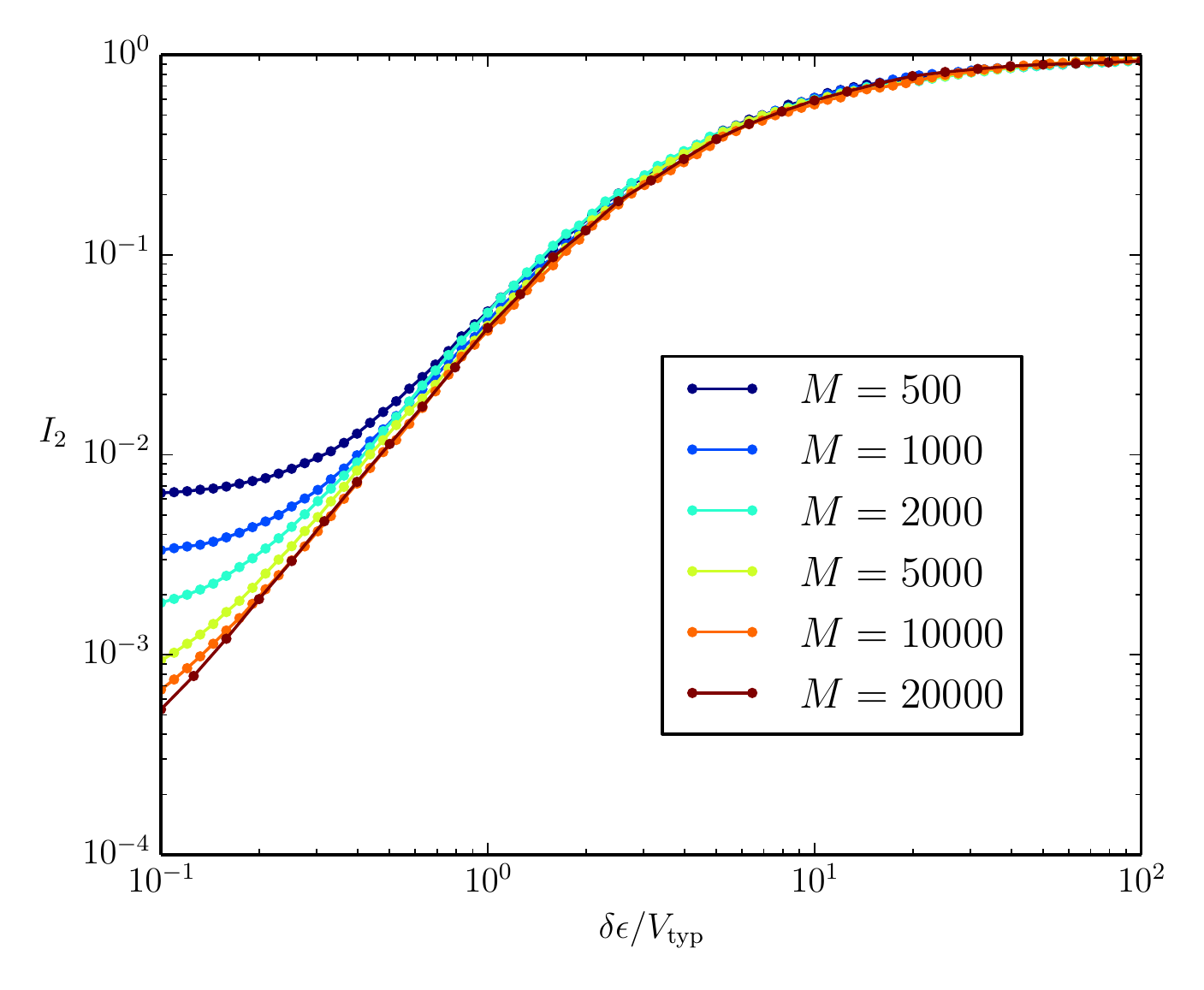}
\caption{The inverse participation ratio $I_2 = \sum_i |\braket{i}{ \psi_\beta}|^4$ as a function of the average classical (at vanishing transverse field) energy level spacing $\de$ in units of the typical coupling $V_{\rm typ}$ for different numbers, $M$, of states in the impurity band. We see that for $\de / V_{\rm typ} \ge 1$ the eigenstates become localized and $I_2 \to 1$ independent of $M$, indicative of eigenstates localized on single bitstring each.
\label{fig:I2_vs_delta}}
\end{figure}

 \begin{figure}[ht]
\includegraphics[width=\columnwidth]{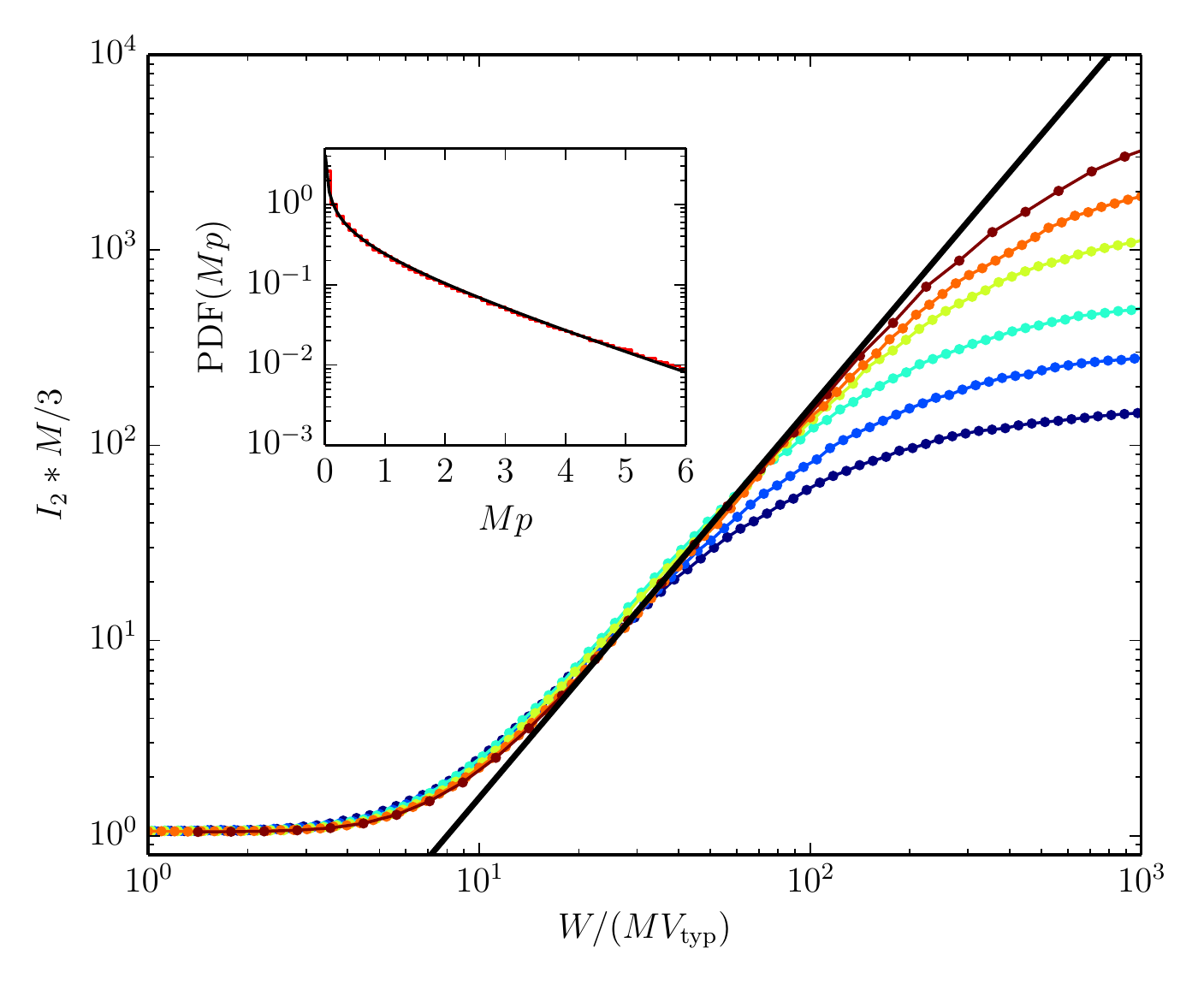}
\caption{ The re-scaled inverse participation ratio $I_2  M/3$ as a function of the re-scaled impurity band width $W/(M V_{\rm typ})$ for different numbers, $M$, of states in the impurity band. We see that in the ergodic regime, $W/(M V_{\rm typ}) \le 1$, we have $I_2 M/3 = 1$, corresponding to the orthogonal Porter-Thomas distribution of states in the impurity band. The inset shows the numerical probability distribution of normalized probabilities $M p$ for an eigenstate over computational states $z$ in the ergodic regime in black, and the analytical orthogonal Porter-Thomas distribution in red. Qualitative arguments in Section~\ref{sec:FGR} suggest that in the non-ergodic delocalized regime  $I_2  M/3 \propto (W/(M V_{\rm typ}))^2$. The black line is proportional to $(W/(M V_{\rm typ}))^2$ and we see that $I_2  M/3$ aligns with this quantity as long as we do not enter the localized regime $\de / V_{\rm typ} \ge 1$, see Fig.~\ref{fig:I2_vs_delta}.
\label{fig:rI2_vs_rdelta}}
\end{figure}

We define the inverse participation ratios (IPRs) $I_q$ and the entropy $H^z$ as,
\begin{align}
  I_q &= \sum_i |\langle \psi_\beta | i \rangle |^{2q},\\
  H^z &= -\sum_i |\langle \psi_\beta | i \rangle |^2 \ln |\langle \psi_\beta | i \rangle |^2,
\end{align}
where $\psi_\beta$ denotes an eigenstate with eigenvalue $E_\beta$. IPR $I_2$ is the second moment of the wave function probability distribution $|\langle \psi_\beta | i \rangle |^2$ in the computational basis (bitstrings) $\ket{i}$. The entropy $H^z$ characterizes the support set of an eigenstate in the computational basis~\cite{SupportSet}, i.e. the subset of bitstrings where the probabilities $|\langle \psi_\beta | i \rangle |^2$ are concentrated.

\begin{figure}[ht]
 
\includegraphics[width=\columnwidth]{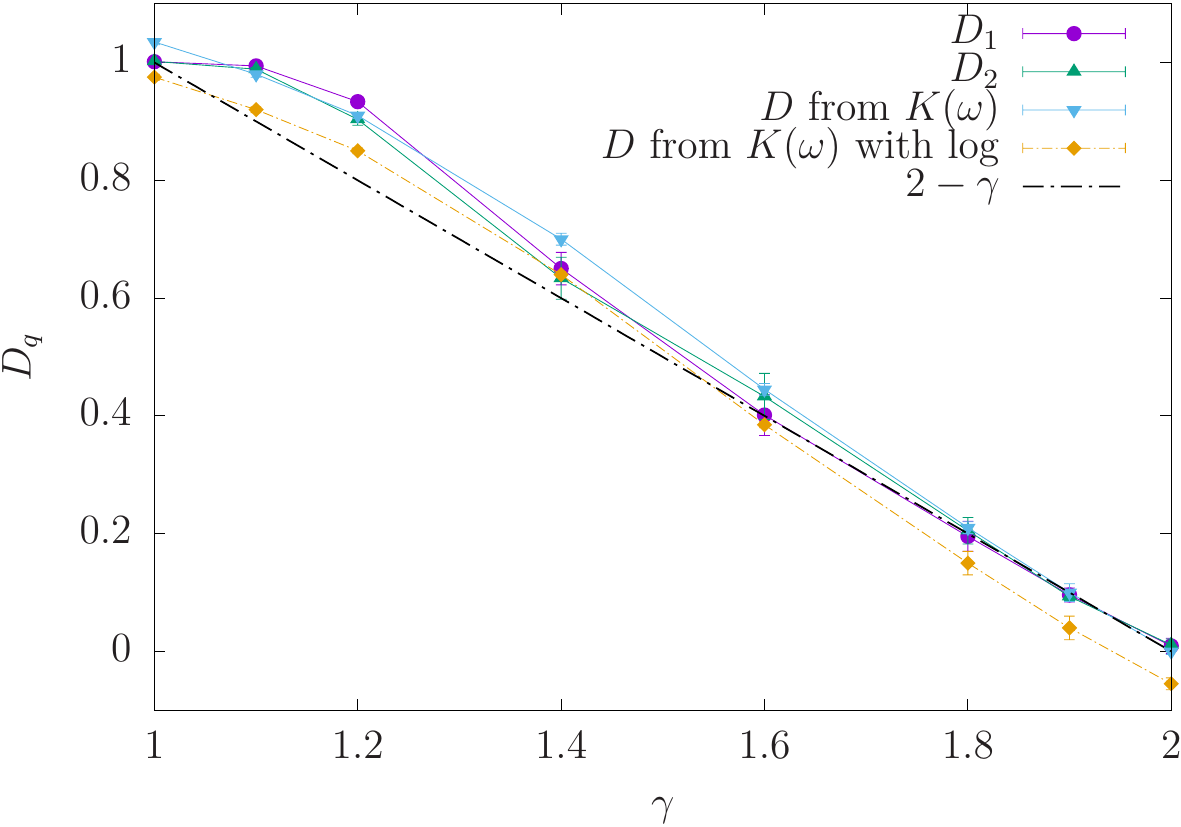}
  \caption{The multifractal dimensions $D_1$ (defined in
    Eq.~\eqref{eq:D1}) and $D_2$, (defined in
    Eq.~\eqref{eq:iq}) as functions of $\gamma$ for the ensemble of IB
    Hamiltonians with the dispersion of classical energies $W =
    \lambda V_{\rm typ} M^{\gamma/2}$, with $\lambda=3.3$. All the
    multifractal dimensions $D_q$ approach 1 in the ergodic regime
    ($\gamma =1$) and 0 in the localized regime ($\gamma = 2$). 
The difference between $D_1$ and $D_{2}$ is
also likely due to finite size effects. We also extract  a scaling exponent from the dynamical correlator (see Eqs.~(\ref{eq:p0}),(\ref{eq:Komega})).
Dot-dashed line corresponds to the analytical value in the Rosenzweig-Porter limit given by  Eq.~(\ref{eq:D12-RP}).
    \label{fig:ds_vs_a}}
\end{figure}

 Fig.~\ref{fig:I2_vs_delta} shows the participation ratio $I_2$ as a function of the ratio of mean level spacing $\delta \epsilon$ to the typical matrix element $V_{\rm typ}$, a measure of the number of states in resonance with a typical classical level $\epsilon_{i}$. The regime $\delta \epsilon \gg V_{\rm typ}$ corresponds to the localized phase, where the eigenstates have significant weight on a small number of bitstrings that are close to each other in Hamming distance. In this regime  $I_2 \sim 1$ and is system size independent. In our model marked states are separated by Hamming distance $d\approx  n/2+\mathcal{O}(\sqrt{n})$ with high probability and therefore most localized states have sharp peaks at exactly one bitstring, hence $I_2\approx 1$. As the ratio $\delta \epsilon/V_{\rm typ}$ decreases $I_2$ becomes system size dependent. Fig.~\ref{fig:rI2_vs_rdelta} indicates  that the combination $I_2 M/3\sim1$ becomes system size independent as level spacing becomes smaller than the typical matrix element, characteristic of the delocalized regime, where the wave function amplitude spreads over $\mathcal{O}(M)$ bitstrings, $|\langle \psi_\beta | i \rangle |^2\sim1/M$. Saturation value of $I_2 M \sim 3$ is consistent with approach to Porter-Thomas distribution of the wave function amplitudes. Both Figs.~\ref{fig:I2_vs_delta} and~\ref{fig:rI2_vs_rdelta} show a wide intermediate region between the localized and ergodic phases where non-ergodic dynamics is expected. This intermediate regime becomes apparent in Fig.~\ref{fig:ds_vs_a} where we introduce the multi-fractal dimensions  $D_q$  and $D_1$ which determine the scaling of $I_q$ and $H^z$ with $M$, respectively,
\begin{align}
	\ln I_q(M) & = -D_q(q-1) \ln M + c_q, \label{eq:iq} \\
  H^z(M) &= D_1 \ln M + c_1, \label {eq:D1}
\end{align}
where $c_q$ is a $q$-dependent fitting parameter. The extracted dimensions shown in Fig.~\ref{fig:ds_vs_a} as a function of the parameter $\gamma$ vary continuously between $D_q=1$ in the ergodic phase $\gamma\leq1$ and $D_q=0$ in the localized phase $\gamma\geq2$, with $1 <\gamma  < 2$ corresponding to non-ergodic regime for $q=1,2$.

\subsection{Eigenstate overlap correlator for non-ergodic minibands}

\begin{figure}
\includegraphics[width=0.99\columnwidth]{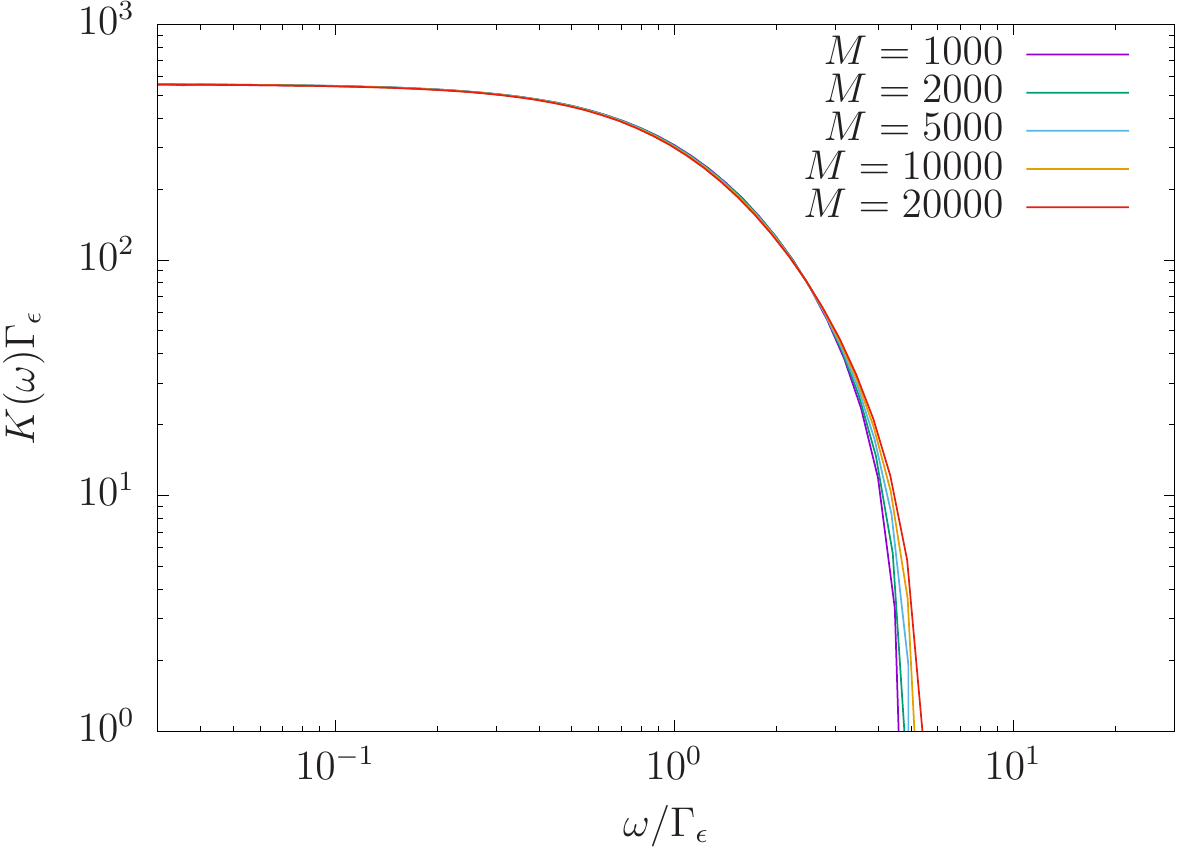}
\caption{
    We plot the rescaled overlap correlation function $K(\omega) \Gamma_\varepsilon$ {\it vs.}
  $\omega/\Gamma_\varepsilon$, where $\Gamma_\varepsilon = \Gamma_{\rm typ}
  M^\varepsilon$ and $\Gamma_{\rm typ}=2\Sigma''_{\rm typ}$ is the typical mini-band width and $\Sigma''_{\rm typ}\propto \cVt M^{1-\gamma/2} (\log M)^{1/2}$, Eq.~\eqref{eq:shift}. Different curves correspond to different values of
  $M$, and collapse well with $\varepsilon = 0.05$. We used the ensemble of IB
    Hamiltonians with a dispersion of classical energies $W =
    \lambda V_{\rm typ} M^{\gamma/2}$, with $\gamma
    = 1.2$ and $\lambda = 3.3$. }
\label{fig:k4}
\end{figure}

Population transfer dynamics in the non-ergodic regime can be characterized by the survival probability, see Section~\ref{sec:QualDisc}. The Fourier transform of the survival probability for a given initial marked state $i$ is given by,
\begin{gather}
p_i\left( \omega\right)=\textrm{Re} \int_0^\infty dt e^{i\omega t} \left| \langle i | \psi (t) \rangle \right|^2 \nonumber\\
=\pi\sum_{\beta,\beta'} \left| \langle i | \psi_\beta \rangle \right|^2\left| \langle  \psi_{\beta'}  | i \rangle \right|^2\delta (E_\beta-E_{\beta'}-\omega).
\end{gather}
Note that the limit $\omega\rightarrow 0$ gives the inverse participation ratio of a given bitstring in the basis of eigenstates,
\begin{gather}
p_i\left(0\right) =\pi\sum_{\beta} \left| \langle i | \psi_\beta \rangle \right|^4.\label{eq:p0}
\end{gather}  
The average of $p_i(\omega)$ over the initial state is related to the overlap correlation function $K(\omega)$ defined by~\cite{Kravtzov2015RzPr},
\begin{align}
  K(\omega) &\equiv \frac{1}{M}\sum_{i,\beta,\beta'} |\langle \psi_\beta|i\rangle|^2 |\langle \psi_{\beta'}|i\rangle|^2\delta (E_\beta-E_{\beta'}-\omega) \nonumber\\
                               &= \frac{1}{\pi M} \sum_i p_i\left( \omega\right)\;.\label{eq:Komega}
\end{align}
The fractal dimension extracted from the scaling of $K(0)$ with $M$ is shown in Fig.~\ref{fig:ds_vs_a}, it follows closely those extracted from the IPR in the computational basis. The collapse of the plots in Fig.~\ref{fig:k4} is achieved when the frequency is rescaled by the characteristic energy,
\begin{gather}
\Gamma_\varepsilon =\Gamma_{\textrm{typ}} M^\varepsilon, \;\; \Gamma \propto V_{\textrm{typ}} M^{1-\gamma/2} (\log M)^{1/2},  \label{eq:Gep}
\end{gather}
with a fitting parameter $\varepsilon \ll1$. The correlator $K(\omega)$ is constant for a range of energy differences $\omega  <\Gamma_\varepsilon  $ and decays quickly $\propto\omega^{-2}$ as $\omega > \Gamma_\varepsilon $. This can be interpreted in terms of the formation of non-ergodic mini-bands of eigenstates that share support in computation basis: for an average bitstring there is a range of eigenenergies $E_\beta$ within a width $\Gamma_\varepsilon$ around a bitstring dependent value $\epsilon_j$ where the eigenfunction overlaps with $z_j$ are relatively large, whereas for larger energy difference the correlation decays quickly below the value corresponding to uncorrelated case $K(\omega)< 1/M$ %M/\Omega^2$ (where $\Omega$ is the number of states in the mini-band),
 i.e. the amplitudes repel each other. 
The relation between the survival probability and eigenfunction overlap correlator, Eq.~(\ref{eq:Komega}) suggests that the characteristic population transfer is given by the inverse of the characteristic energy scale of the miniband width $\Gamma_\varepsilon$, the range of energy eigenstates with significant amplitude at the given bitstring. The auxiliary fitting parameter takes a small value $\varepsilon=0.05$ indicating only a small deviation from $\Gamma_{\rm typ}$ most likely due to finite size effects. In Appendix~\ref{sec:Num-Sim-Appendix} we show the results of direct simulation of dynamics of the model in the course of the PT protocol and confirm the scaling of the PT time.

\subsection{\label{sec:Numerical-Analytical-Comparison} Discussion of numerical results}

 The size of the matrix of marked states used in exact diagonalization $M\leq 20000$ is a small fraction of the size of the total Hilbert space Hamiltonian $2^n\times2^n$ with $n=200$. For such a small sample the distribution of Hamming distances $d_{ij}$ between marked states is dominated by $\left|d_{ij}-n/2\right|\sim\mathcal{O}(\sqrt{n})$. In this regime the square of the off-diagonal matrix element, see Sec.~\ref{sec:cd}, has approximately Gaussian dependence on $d_{ij}$ (cf. Eqs.~(\ref{eq:V2}),(\ref{eq:typ}))
\begin{gather}
\scH_{ij}^2 \approx V_{\text{typ}}^2 \,\exp\left(\frac{2}{n} \left(d_{ij}-\frac{n}{2}\right)^2\right),\label{eq:RPvar}
\end{gather}
and the probability to find a pair of bitstrings at a smaller distance $d_{ij}$ is strongly suppressed. The sign of $\scH_{ij}$ rapidly fluctuates as a function of $d_{ij}$ resulting in a negligible average $\langle \scH_{ij} \left(d\right) \rangle\sim \mathcal{O}(2^{-n})$. The distribution of off-diagonal matrix elements in Eq.~(\ref{eq:RPvar}) is non-Gaussian and instead has a heavy tail that cannot be fully characterized by the variance alone, see Section~\ref{sec:tails} and Appendix~\ref{sec:MeanValueSigmaHeff} where we introduced the class of Preferred Basis Levy Matrices and derived the asymptotic form of the distribution of matrix elements. For numerically accessible matrix sizes $M$ we expect the deviation from the Gaussian distribution in the observables to be very small.

The eigenstate statistics and the respective fractal dimensions for the model  Eq.~(\ref{eq:RPvar}) can be calculated using strong disorder perturbation theory. The calculation proceeds similarly to that in Ref.~\onlinecite{Kravtzov2015RzPr} resulting in,
\begin{gather}
D_1=D_2=2-\gamma.\label{eq:D12-RP}
\end{gather}
Comparison of the approximate Eq.~(\ref{eq:D12-RP}) with numerical results is shown in Fig.~\ref{fig:ds_vs_a} as the dot-dashed line. It appears that the $D_1$ and $D_2$ do not quite coincide with each other nor with Eq.~(\ref{eq:D12-RP}), which may be due to finite size effects.

%This approach cannot take into account the effect of delocalized states in the mini-band and therefore the result in Eq.~(\ref{eq:D12-RP}) is only an approximation. The solution of the non-linear cavity equations for the Green's function obtained in Sec.~\ref{sec:sol} together with expressions relating the IPRs and the Green's function~\cite{fyodorov1991localization,metz2010localization} allow a rigorous calculation of the fractal dimensions for our model. We leave this for future work. 

It is instructive to draw an analogy between characteristics of the PBLMs and that of the Rosenzweig-Porter (RP) model from random matrix theory, see Ref.~\onlinecite{rosenzweig1960repulsion,Kravtzov2015RzPr} and references therein, where the matrix elements are given by Gaussian random variable with zero mean and variance for diagonal and all off-diagonal matrix elements set $\langle \mathcal{H}_{ii}^2\rangle=1$ and $\langle \mathcal{H}_{ij}^2\rangle\propto M^{\gamma}$. Transition points between localized, delocalized and non-ergodic delocalized regimes as well as perturbative expressions for fractal dimensions Eq.~(\ref{eq:D12-RP}) are consistent in the two models. The dynamical correlator also shows similar behavior indicative of the formation of minibnads of non-ergodic eigenstates with the leading exponent $1-\gamma/2$ in the scaling of the population transfer time with $M$ coinciding in the two models. The prefactor $\left(\log M\right)^{1/2}$ however is affected by the heavy tail of the distribution of the matrix elements and needs to be calculated analytically. It is  difficult to extract it accurately from the numerical simulations due to the finite size effects.

\section{\label{sec:FGR} Born approximation for the transition rates}

In this section we develop a simple picture relying on Fermi Golden Rule (FGR) to study the  rates of population transfer away from a given marked state to a set of other marked states inside the same miniband. 
Assume that the system is initially prepared at  a randomly chosen marked state $\ket{z_j}$.  The probability amplitude  to remain in the initial state $\ket{z_j}$ equals
\begin{align}
\psi(z_j,t)= \sum_\beta \psi_\beta^2(z_j)  e^{-i E_\beta t}\;,\label{eq:ptz1}
\end{align} 
where $\ket{\psi(t)}$ evolves with the IB Hamiltonian $\scH$ (\ref{eq:IBH}) and $\scH \ket{\psi_\beta}=E_\beta \ket{\psi_\beta}$. If  the eigenstates  dominantly coupled to the  marked state  $\ket{z_j}$ are extended  then the amplitude $\psi(z_j,t)$ will  undergo decay in time. 

Here we calculate $\psi(z_j,t)$ using a simple effective Fano-Anderson model for the decay of a discrete state into a continuum \cite{mahan2013many}. This model captures the Born approximation for the ensemble of Hamiltonians introduced in Sec.~\ref{sec:ensembleH}. The model Hamiltonian $\tilde \scH$  is obtained from the IB Hamiltonian $\scH$ (\ref{eq:IBH}) by  zeroing  out all off-diagonal matrix elements except those in the $j$th column and the $j$th row  connecting state $\ket{z_j}$ to the rest of the marked states. The Hamiltonian $\tilde \scH$ has the form
\begin{align}
\tilde \scH = \e_j\ket{z_j}\bra{z_j}+ \sum_{m\neq j }(\e_m-i\eta)\ket{z_m}\bra{z_m}\;\label{eq:tVen}
\end{align}
\vspace{-0.23in}
\begin{align}
+ \sum_{m \neq j}\scH_{j m}(\ket{z_j}\bra{z_m}+\ket{z_m}\bra{z_j})\;,\nonumber
\end{align}
where the summation is over $m \in[1.. M], \, m\neq j$.  We consider the dynamics on a time scale when the population of the state $\ket{z_j}$ decays  into the other states and introduce a  small imaginary part  $-i\eta$ to  their energies. It is  assumed to be much bigger than the typical energy spacing,   $\eta\gg \de=W/M$ but smaller than the time scale on which the decay takes place.
 We introduce the parameterization similar to that in Sec.~\ref{sec:NumSim} for the distribution of energies $\e_j$, %the RP case  \cite{Kravtzov2015RzPr} 
\begin{equation}
W=\lambda \cVt M^{\gamma/2}\;,\label{eq:param}
\end{equation}
where $\lambda$ is a (redundant) number of order of ${\cal O}(M^{0})$.  
 %\cite{anderson1958absence}. 

\begin{figure}[ht]
\includegraphics[width= 3.35in]{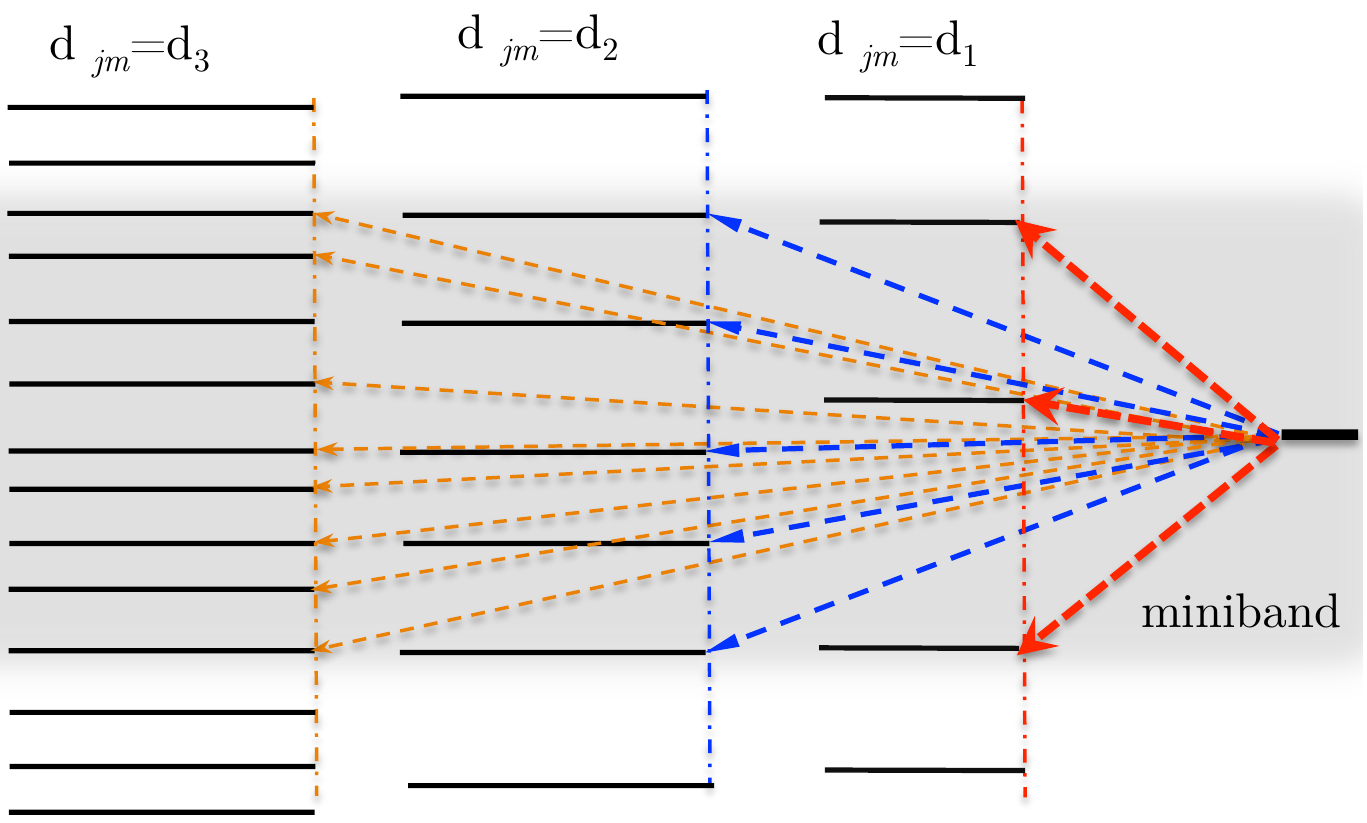}
  \caption{Cartoon of the energies of the marked states $\e_m$ within the impurity band.
 Energy levels are shown with solid black lines forming groups arranged vertically. All states $\ket{z_m}$ within one group  lie at the  same Hamming distance  $d_{jm}=d$ from  a given state $\ket{z_j}$ with $d$ increasing from right to left. The energy level $\e_j$   is depicted at the right side of the figure with thick black line. 
   Arrows depict the transitions away from the initial  state $\ket{\psi(0)}=\ket{z_j}$ into the marked states  $\ket{z_m}$  whose energy levels lie inside the miniband   of the width  $\Gamma_j$ centered at  $\e_j$, i.e., they  satisfy the condition  $|\e_j-\e_m|\apprle \Gamma_j$.  
  Miniband width is indicated with the gray shading area. Arrows of the same color  depict transitions within one decay channel, connecting  the state $\ket{z_j}$ to the states a Hamming distance $d$ away from it.  Smaller values of  $d$ correspond to bigger typical level spacings $\delta\e_j^d$ (\ref{eq:dep}) and fewer states in a miniband   $\Mr$ (\ref{eq:Mres}) within the  decay  channel  given by $d$.  
   \label{fig:FGR}}
\end{figure}

The amplitude $\psi(t,z_j)$ has a well-known form \cite{mahan2013many}
\begin{equation}
\psi(z_j,t)=\int_{-\infty}^{\infty}\frac{dz}{\pi}\,\frac{ \Sigma_{j}^{\prime\prime}(z) \,\exp(-i z t)}{( z-\Sigma_{j}^{\prime}(z)-\e_j)^2+(\Sigma_{j}^{\prime\prime}(z))^2}\;,\label{eq:psizj}
\end{equation}
where  we used a  short-hand notation 
\begin{equation}
\Sigma_j(z)=\Sr_j(z)-i\Si_j(z)\label{eq:S-def}
\end{equation}
for real and imaginary parts  of self-energy of the marked state $\ket{z_j}$
\begin{equation}
 \Sigma_{j}(z)=\sum_{m\neq j}\frac{\scH_{jm}^{2} }{z-\e_m+i\eta}\;,\label{eq:self}
\end{equation}
and we keep $z$ real. Calculating the above integral to the leading order in   $\scH_{j m}$ ($j\neq m$) we get
 \begin{equation}
\psi(z_j, t)\simeq \exp\left[-i(\e_j+\Delta\e_j )t-\frac{1}{2}\Gamma_j t\right]\;,\label{eq:exp-d}
 \end{equation}
 where
\begin{equation}
 \Delta\e_j\simeq \Sr_{j}(\e_j),\quad  \Gamma_j \simeq 2\Si_{j}(\e_j)\;.\label{eq:GD}
 \end{equation}
 
The quantity $\Gamma_j$ above is the total decay rate of the state $\ket{z_j}$ which is twice the imaginary part of the self-energy  $\Si_{j}$. The latter equals to the   "width" of the level $\e_j$ due to the decay. Expressions (\ref{eq:exp-d}),(\ref{eq:GD}) correspond to a well-known Born approximation for the   self-energy $\Si_{j}$. Using (\ref{eq:self}) we get,
\begin{equation}
\Si_{j}=\pi \sum_{m \in[1\ldots M]/j} \scH_{mj}^{2} \, \delta(\e_j-\e_m,\eta)\;.\label{eq:FGR0}
\end{equation}
where we defined a function,
\begin{equation}
\delta(\e,\eta)\equiv \frac{1}{\pi}\frac{ \eta}{\e^2+\eta^2}\;.\label{eq:deltaY}
\end{equation}
The matrix elements $\scH_{mj}^{2}$, see (\ref{eq:IBH}),(\ref{eq:V2}), depend only on the Hamming distance $d_{mj}$.
%\begin{gather}
%\Gamma_{j} 
 %=2\pi \,
 %\sum_{m\neq j}V^2(d_{mj})2\sin^2\phi(d_{mj})\delta(\epsilon_{j}-\epsilon_m,\eta)\;.
%\end{gather}
The dominant contribution in to the sum (\ref{eq:FGR0}) comes from the transitions to the states  with $|\epsilon_{j}-\epsilon_m| \apprle \eta$. If the number of such states is large   the sum can be replaced by the integral corresponding to the approximation where the Lorentzian $\delta(\epsilon_{j}-\epsilon_m,\eta)\approx \langle \delta(\epsilon_{j}-\epsilon_m,\eta)\rangle$ is replaced with its average over realizations of $\epsilon_m$. The average is independent of $\eta\ll W$ which therefore drops out from the PDF of  the transition rate $\Gamma$ and  the resulting  level width   $\Si=\Gamma/2$.  
 This case corresponds to  the leading order Born approximation described  in the next Section~\ref{sec:LeadingOrderBorn}.

  A more accurate treatment of $\delta(\epsilon_{j}-\epsilon_m,\eta)$  as a random variable results  in the form of the PDF of $\Gamma$   (and $\Si$) being explicitly dependent on $\eta$. The physical meaning of $\eta$ is the decay rate at the "children" sites $\epsilon_m, m\neq j$ which gives rise  to the  width $\Si$ or the energy level $\e_j$ at the parent site. In a large system the statistics of the decay rate for children and parents are expected to be the same. The crude approximation that captures this effect  is obtained by   substituting  $\eta$ with typical value of $\Si$. This corresponds to   self-consistent Born approximation described in  Sec.~\ref{sec:self-consist-B}. It  gives rise to a more accurate expression for  PDF of $\Si$ (and $\Gamma$) whose shape is rescaled compared to the leading order Born.
   Systematic analysis  is given by the cavity method described in Secs.~\ref{sec:Cavity}, \ref{sec:sol}.

\subsection{\label{sec:LeadingOrderBorn}Leading order Born approximation}

We can break down the decay rate  $\Gamma_j=2\Si_j$ into a sum over different decay channels 
\begin{equation}
\Si_j=\pi \sum_{d=1}^{n} V^2(d)(1-\cos2\phi(d)) \varrho_\eta^j(d)\;,\label{eq:chan}
\end{equation} 
\noindent
where each term in the sum   corresponds to  the transition rate from  the initial state $\ket{z_j}$  into  the subset of the marked states on a given Hamming distance $d$ from  $\ket{z_j}$ (see Fig.~\ref{fig:FGR}).  The factor $\varrho_\eta^j(d)$  in  (\ref{eq:chan}) is a spectral density of the marked states located at a distance $d$ from the state 
$\ket{z_j}$ within the window of energies $\eta$ around $\e_j$ 
 \begin{align}
 \varrho_\eta^j(d) = \sum_{m\neq j}\delta(\epsilon_{j}-\epsilon_m,\eta)\Delta(d-d_{jm})\;,\label{eq:red}
 \end{align}
 \noindent
where $\Delta(d)$ is a Kronecker delta and $\delta(\e,\eta)$ is defined in (\ref{eq:deltaY}).

We denote  as $M_{j}^{(d)}$ the number  of marked states that are separated by a Hamming distance $d$ from the state $\ket{z_j}$ (number of terms in the sum (\ref{eq:red}))
\begin{equation}
M_j^{(d)}=\sum_{m \neq j}\Delta(d-d_{jm})\;.\label{eq:Mdj}
\end{equation}
As discussed in Sec.~\ref{sec:stat-app} the elements of the set   $\{ M_{j}^{(d)}\}_{d=1}^{n}$  are   sampled from the multinomial distribution  with mean values 
\begin{equation}
\langle M_{j}^{(d)}\rangle= M p_d\;, \quad p_d\simeq 2^{-n}\binom{n}{d}\;,\label{eq:KKd}
\end{equation}\noindent
where coefficient $p_d$  defined in (\ref{eq:p_d}) is the probability  that a randomly chosen state is located on a Hamming distance $d\neq 0$ from $\ket{z_j}$. The mean separation %$\delta\e_j^d $
 between the adjacent energies $\e_m$ in the sum (\ref{eq:red})  equals
\begin{equation}
\frac{W}{M_{j}^{(d)} }\sim \de\,\frac{ 2^{n}}{\binom{n}{d}},\quad (M_{j}^{(d)}\geq 1)%\,e^{2n\left (\frac{1}{2}-\frac{d}{n} \right)^2}
%>\de\;,\quad \delta\e \equiv W/M
\;,\label{eq:dep}
\end{equation}
where $\delta\e=W/M$ is mean spacing between the marked state energies. %and we used the estimate (\ref{eq:Asmall}) for binomial coefficient .
A substantial contribution to the sum in  (\ref{eq:red}) comes from   the  terms 
corresponding to the marked states whose energy levels $\e_j$ lie within the width $\eta$   from the energy $\e_m$, i.e., they  satisfy the resonant condition  $|\e_j-\e_m|\apprle \eta$ as shown in Fig.~\ref{fig:FGR}.  

The contribution to a sum from each  resonance  is  $\sim 1/\eta$ and the
number of the resonances in a given decay channel  is $\Mr\sim M_{j}^{(d)}\eta/W$ (cf. Fig.~\ref{fig:FGR}). It is  shown in the Appendix \ref{sec:Just} that the dominant contribution  to the {\it typical} values of  $\Si_j$ (\ref{eq:chan}) comes from the values of $d$ that correspond to $\Mr\gg 1$. For them  the function $\delta(\e_j-\e_m,\eta)$ in Eq.~(\ref{eq:red}) changes  weakly between the adjacent values of $\e_m$,  and in the leading order Born approximation we estimate the  sum over $m$ in (\ref{eq:red})  by  replacing it with an integral. %The   error induced by such approximation will  be discussed below.
Then  the spectral density can be estimated as 
\begin{equation}
\varrho_\eta^j(d)\simeq M_{j}^{(d)}  p(\e_j)\;,\label{eq:rhod-est} 
\end{equation}
where we required  
 \begin{equation}
 \de\ll\eta\ll W\;.\label{eq:eta-cond}
 \end{equation}
and $p(\e)$ is PDF of the marked state energies $\e$ with the width $W$ (see  (\ref{eq:pA})).

We plug (\ref{eq:rhod-est}) into the expression (\ref{eq:chan}), obtaining  the following relation
\begin{align}
\Si_j =\pi \, p(\e_j)\sum_{d=1}^{n} M_{j}^{(d)}V^2(d)(1-\cos2\phi(d))\;,\label{eq:G-j-sum}
\end{align}
where the  sum  is dominated by values of $d$ corresponding to  large values $M_{j}^{(d)}\gg 1$ (see Appendix \ref{sec:Just}). The steep exponential decrease  with $d$
of  the matrix element $ V^2(d) \propto 1/\binom{n}{d}$ (\ref{eq:V2}) is canceled by equally steep growth with $d$ of the average number of states in the $d$ channel $\langle M_{j}^{(d)}\rangle\propto \binom{n}{d}$  (\ref{eq:KKd}). As a result, the binomial factors cancels out and the average quantity $ \langle M_{j}^{(d)}\rangle  V^2(d)$ changes only by  $\cO(n^{-1})$ when $d$  changed by 1.

The term involving $\cos2\phi(d)$ above  oscillates around 0 on the scale $d\sim 1$ (cf. Eq.~(\ref{eq:phi-largeB})). Therefore the contributions to the sum from the terms $\propto \langle M_{j}^{(d)}\rangle  \cos2\phi(d)$ average out.
 In what following we shall  neglect the cross-product of fluctuational and oscillatory parts  $( M_{j}^{(d)} -\langle M_{j}^{(d)}\rangle)\cos2\phi(d)$ and drop the second term  in the r.h.s of (\ref{eq:G-j-sum}) that contains $ \cos2\phi(d)$. %the oscillatory part.

Essentially, the above approximation corresponds to replacing the oscillatory part in the expression for the off-diagonal matrix elements  $\scH_{ij\neq i}=V(d_{ij})\sqrt{2}\sin\phi(d_{ij}) $ (\ref{eq:IBH}) as follows:
\begin{equation}
\scH_{i j} \rightarrow V(d_{ij}) \beta_{ij},\quad \beta_{ij}=\pm1,\quad i < j\;,\label{eq:Hbeta}
\end{equation}
where  $\beta_{ij}$ are instances of a dichotomous random variable that takes values $\pm1$ with probability $1/2$. This approximate model of the ensemble of $\scH_{i j}$ will be also used in cavity method calculation in Sec.~\ref{sec:sol}.

Using the expression (\ref{eq:pA}) for $p(\e)$   and also Eqs.~(\ref{eq:w-m}), (\ref{eq:V2}),(\ref{eq:typ}),  we obtain the relation between the  PDFs  of the random variables  
% \begin{equation}
%\Gamma \disteq   2\pi V_{\rm typ}^{2} M s_M\, p(\e)\;,\label{eq:Gdef}
%\end{equation}
%\vspace{-0.2in} %where
\begin{equation}
 \Si \disteq \,  \Sti  s_M,\quad  \Sti=\pi \frac{V_{\rm typ}^{2}}{W/M}\;,\label{eq:Gdef1}
\end{equation}
\begin{equation}
s_M=\frac{1}{M}\sum_{m=1}^{M}w_m\;.\label{eq:SM}
\end{equation}
Here  $w_m$ are random variables independently sampled from the probability distribution $g_\infty(w)$ given in (\ref{eq:g_0(w)}). The level widths $\Si_j$
  of  individual marked states for $1\leq j\leq M$ are  samples  of  the random variable  $\Si$. 

In Eq.~(\ref{eq:Gdef1})  we introduced  the characteristic value of the level width  $\Sti$. This equation relates 
  the PDF of $\Si$  (or the decay rate $\Gamma=2\Si$) to that of   $\e$ and  $M s_M$.   
  We note that the  resulting  expression for the  level width $\Si$ of a  marked state   formally corresponds to that given by FGR for the decay of the discrete level into the continuum \cite{mahan2013many}. The energies  of the marked states $\e_m$ into which a given marked state $\ket{z_j}$ decays form a miniband of the width $\Si_j$. 
The  decay occurs simultaneously in many channels  corresponding to different Hamming distances between the initial marked state and the states of the  miniband.

 The heavy-tailed PDF of the  random variable $s_M$ is studied in details in Appendix \ref{sec:CLT}. 
Using Generalized Central Limit Theorem (GCLT) for the  sums of a  large number of identical heavy-tailed random variables \cite{gnedenko1954limit,cizeau1994theory}
it can be represented in the form
\begin{equation}
s_M  \disteq  \sigma_M x+b_M
\end{equation}
where   $x$ obeys a  so-called Levy alpha-stable distribution  $L^{1,1}_{1}(x) $  \cite{cizeau1994theory} defined in the Appendix, Eq.~(\ref{eq:char}), and shown  in Fig.~\ref{fig:st-dist}.
  Scaling factor and shift are
\begin{equation}
\sigma_M=\sqrt{\frac{\pi}{4\log M}}\;,\label{eq:sigmaM}
\end{equation}
\vspace{-0.2in}
\begin{equation}
b_M\simeq \sigma_{M}^{-1}-\frac{2}{\pi} \sigma_{M}\log(\sigma_{M}^{-1}) +\frac{2}{\pi} (1-\gamma_{\rm Euler}) \sigma_M\;,\label{eq:bM}
\end{equation}
($\mathcal \gamma_{\rm Euler}\simeq 0.577$ is the  Euler constant). They display very weak logarithmic dependence on $M$ as compared with the main factor $\propto V_{\rm typ}^{2}/\de$ in (\ref{eq:Gdef1}). The width of the PDF  of $s_M$ is shrunk   by a  factor $(\log M)^{1/2}\gg1$ and  the location of its maximum is increased
by  a factor $(\log M)^{1/2}\gg1$  compared to   $L_{1}^{1,1}(x)$.

\begin{figure}[ht]
  \includegraphics[width= 3.35in]{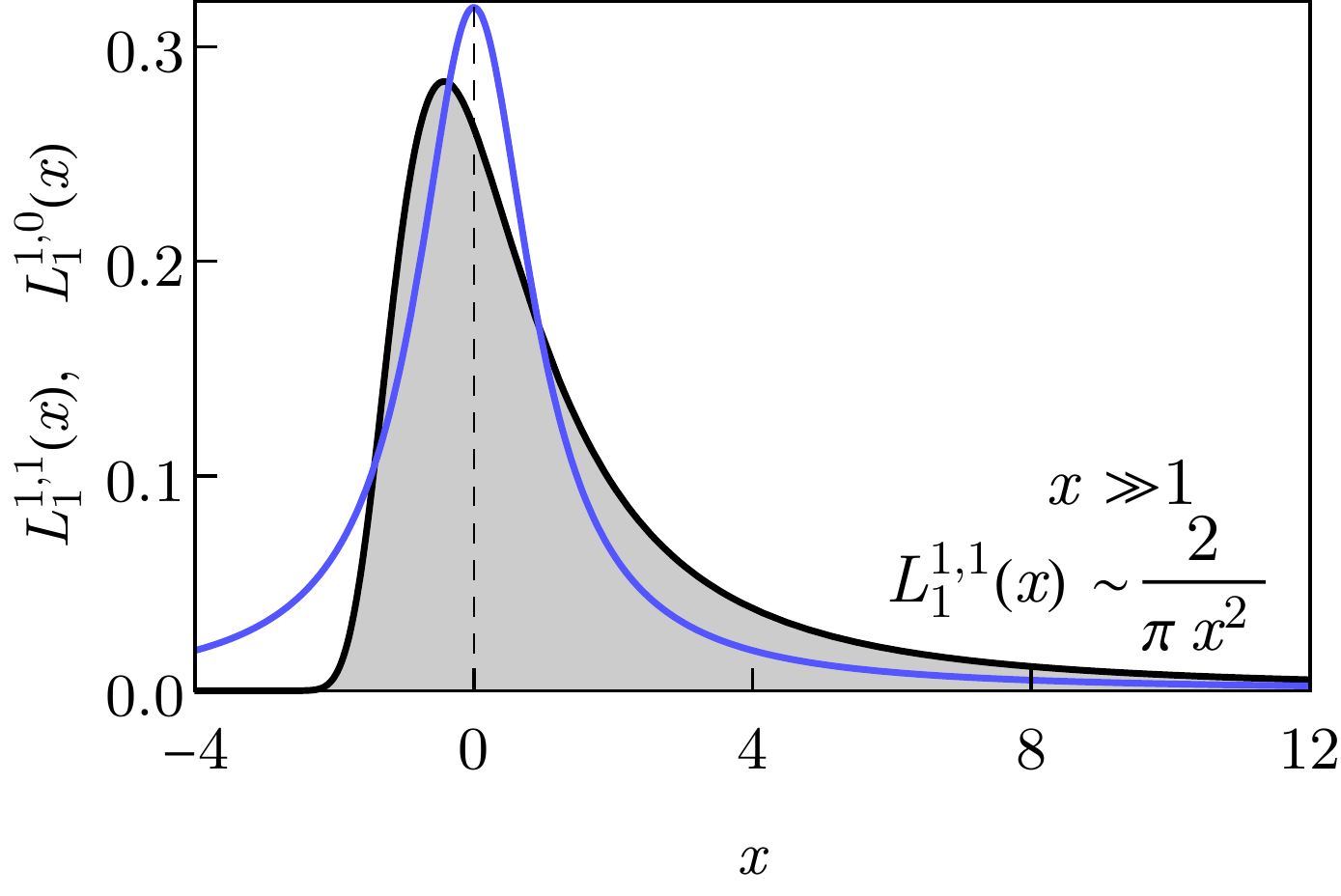}
  \caption{Black solid line shows the plot of the  Levy alpha-stable distribution  $L_{\alpha}^{C,\beta}(x)$  \cite{cizeau1994theory}   with tail index $\alpha=1$, asymmetry  parameter $\beta=1$  and  unit scale parameter $C=1$.  Inset shows asymptotic behavior of the distribution at large positive $x$.  At  $-x \gg 1$ the function decays steeply as a double exponential,  $\log L_{1}^{1,1}(x)\propto -
 \exp(-\frac{\pi}{2} x)$.  Blue line shows the Cauchy distribution $L_{1}^{1,0}(x)=\frac{1}{\pi(1+x^2)}$. 
 We follow here the definition introduced in   \cite{cizeau1994theory} and used in subsequent papers on Levi matrices in physics literature. In mathematical literature  \cite{WikiStable,voit2013statistical} a different definition  is usually used, corresponding to $f(x; \alpha, \beta,C^{1/\alpha},0)=L_{\alpha}^{C,\beta}(x)$. }\label{fig:st-dist}
\end{figure}

 The PDF of $s_M$ has polynomial tail. Therefore  decay rates  of marked states $\Gamma_j=2 \Si_j$ can take a range values that are much bigger than their typical values $2\Sti$ (\ref{eq:Gdef1}), up to   $M$ times bigger in the sample of the size $M$.  These atypically large decay rates   correspond to rare  clusters  of marked states that are located anomalously  close to each other. When clusters are formed by ${\cal O}(1)$ states the above  picture of the decay fails. 

%\subsection{\label{sec:self-consist-B0}Self-consistent  Born approximation}

\section{\label{sec:NumRes} Number of states in a miniband within Born approximation}

 Using the  expression (\ref{eq:Gdef1}) for the miniband  width  we can estimate the number of marked states $\Omega$ in a miniband corresponding to a given  state $\ket{z_j}$. As before,  we divide the states  into the groups of the sizes $\Mr$, each corresponding  to the transitions away from $\ket{z_j}$ with a fixed number of flipped bits $d$. The level width can  be written in the form  $\Si_j=\sum_{d=1}^{M}\Si_{j,d}$ where $\Si_{j,d}$ is  the  partial level-width  due to  the transitions  with flipping $d$ bits.
 Then, using (\ref{eq:G-j-sum})   and making use of the expression  (\ref{eq:KKd})  for the average values of $M_{j}^{(d)}$ 
 we obtain 
\begin{equation}
\Si_{j,d}\simeq \pi \frac{V_{\rm typ}^{2}}{\delta\e} \,\frac{1}{\sqrt{\pi n/2}}{M_{j}^{(d)}  \over \langle M_{j}^{(d)}\rangle} \;.\label{eq:Gamma-jd}
\end{equation}
The quantity  $M_{j}^{(d)}/\langle M_{j}^{(d)}\rangle\sim 1$ in (\ref{eq:Gamma-jd}). This results in the interesting phenomenon   due to cancellation mentioned din the previous section: while   the typical number of marked states  in  a decay channel   
varies  very steeply with $d$,  typical values of  partial decay rates $\Si_{j,d}$  in different channels do not.

The estimate for the typical number of states in the  miniband  at the distance $d$ from $\ket{z_j}$ is
 $\Mr\sim \Si_{j,d}/\de_{j}^{d}$,
\begin{equation}
\Mr\sim \Omega\, p_d,\quad \Omega\sim\frac{ \Si_j }{\de} \sim \left(\frac{V_{\rm typ}}{\de}\right)^2\;,\label{eq:Mres}
\end{equation}
where $p_d=2^{-n}\binom{n}{d}$ and $ \Omega$ is the total number of states in the miniband.
 
One can also write the partial decay rate as $\Gamma_{j}^{(d)}\sim V(d)\Omega_d$ where the  product $V(d)\Omega_d$ does not depend on $d$ (except from the prefactor). Of course, the analysis based on the decay rate does not apply for the  transition to the channels 
with  very few states. The condition
$\Omega_d\simeq 1$ leads to  $\Gamma_{j}^{(d)}\sim V(d)$ for  $d=\dminr$ corresponding to the typical Hamming  distance from $\ket{z_j}$ to the nearest  marked state in a miniband 
where the condition $V(d)\simeq \de_j^d$  is satisfied (see Eq.~(\ref{eq:dminr}) in Appendix).

The above  estimate gives the correct time scale $\sim1/ V(\dminr)$ over which the two states become hybridized. We note however that the total number of channels is  $n-2\dminr=\cO(n)$. As all $\Gamma_{j}^{(d)}$ are nearly   the same, each channel contributes a small fraction $\cO(1/n)$ to the total rate. Therefore   $V(\dminr)\sim\Gamma_j/n$ and  marked  state $\ket{z_j}$  decays into the large number of marked states  within a miniband  before it has a chance to hybridize with the nearest one  at a distance $\dminr$.
 This property is markedly different from  the situation at finite dimension  \cite{anderson1958absence}.
 
Using the scaling ansatz (\ref{eq:param}) we estimate the mean separation between the energies   of marked states as
\begin{equation}
\de=\frac{W}{M}=\lambda \cVt M^{\gamma/2-1}\;.\label{eq:de1}
\end{equation}
Using the Eqs.~(\ref{eq:Gdef1})  and (\ref{eq:Mres}) we obtain the estimates for typical values of the decay rates and number of marked states in a miniband 
\begin{equation}
\Gamma=2\Si \sim \cVt \,M^{1-\gamma/2},\quad \Omega\sim M^{2-\gamma}\;.\label{eq:GM}
\end{equation} 

We immediately observe that in the range of  $\gamma>2$ the number of marked states in a miniband  vanishes. It corresponds to a localized phase, consistent  with the fact that typical energy  spacing $\delta$ becomes  greater  than the typical tunneling matrix element $\cVt$ connecting the states. The number of states in a miniband   $\Omega$ cannot be greater than the total number of states $M$ in the IB. The expression  above does not apply  for $\gamma \leq 1$.
This regime corresponds to ergodic phase.  

In the region $2>\gamma>1$ the separation between adjacent eigenvalues of
$\scH$ is of the same order as $\de$. The typical number of marked states in a miniband  $\Omega$ corresponds to   the typical number of non-ergodic delocalized eigenstates of $\scH$ that form the miniband. 
\begin{equation}
W\gg \Gamma\gg \de=\frac{W}{M}\;.\label{eq:n-er}
\end{equation}
The number of states in a miniband scales as a fractional power of $M$ less than one.  This is a hallmark of non-ergodic delocalized phase.

\section{\label{sec:Cavity} Cavity method:  summary of the previous results}

 The cavity method has been actively used to study  Anderson Localization in  Levy matrices in the last several decades \cite{cizeau1994theory,burda2007free,rogers2008cavity,metz2010localization,tarquini2016level,facoetti2016non,monthus2016localization} starting from the seminal  work \cite{cizeau1994theory}.
  In the   present work we use cavity method to study the properties of minibands of delocalized non-ergodic states that were previously discovered in the studies of Rosenzweig-Porter  \cite{facoetti2016non,Kravtzov2015RzPr,facoetti2016non} and Regular Random Graph (RRG) \cite{altshuler2016nonergodic,altshuler2016multifractal} models.
  Initial studies suggested the existence of the mixed region with localized but non-ergodic states  \cite{cizeau1994theory}. However, recent numerical studies based on exact diagonalization using very large number of samples established that initially large crossover region between localized and extend states collapses in the limit of increasing matrix sizes \cite{tarquini2016level}. Multifractal properties of eigenstates  in the localized phase and at criticality  were studied in  \cite{monthus2016localization} using strong disorder perturbation theory.
 
Numerical solution of  cavity equations  to study localization transition in  Levi matrices with power-law distributions $P(\scH_{ij}^2)\propto 1/\scH_{ij}^{2(\alpha+1)}$ were obtained using population dynamics algorithm  \cite{metz2010localization}  utilizing the approach developed in  \cite{rogers2008cavity}. An alternative approach is based on the integral equation for the PDF of the diagonal elements of the resolvent \cite{cizeau1994theory,burda2007free}. It was obtained   in the limit where    imaginary part of the self-energy is vanishingly small \cite{cizeau1994theory,burda2007free,tarquini2016level}  (with the limit of infinite matrix size taken first). This allows one to derive  analytically the global density of states  \cite{cizeau1994theory,burda2007free} and the mobility edge $E^*(\alpha)$ which gives the 
 $\alpha$-dependence of the energy   $E^*$ separating extended and localized eigenvalues of $\scH$ \cite{tarquini2016level}. 

The cavity method proceeds as follows.  First,  we generate  a random $M\times M$ matrix  $\scH_{ij}$ (\ref{eq:IBH})  from the  ensemble  described in  Sec.~\ref{sec:ensembleH}. Then  we add  a new row (and a symmetric column) of independent numbers identically distributed as those in the old matrix  $\scH_{ij}$. This is done by generating a random energy $\e_0$ from the distribution $\frac{1}{W}p_A(\e/W)$; then  generating a random  bit-string $z_0$, computing  the array of Hamming distances $d_{j0}$ between $z_0$ and $z_j$ and the corresponding matrix elements 
$\scH_{j0}=\scH_{0j}$ for integer $j\in [1, M]$. As a result we obtain a new  $(M+1)\times (M+1)$ matrix $\scH^{+1}$, where $+1$ emphasizes that it has one more row and one more column than $\scH$. We will number elements of the new matrix by  indices running over the range $[0,M]$ where the   index 0 corresponds to the added marked state $\ket{z_0}$.  The cavity equations have the form    \cite{cizeau1994theory,burda2007free} \[\Sigma_{0}^{+1}(z)=\sum_{m=1}^{M}\scH_{0m}^{2}G_{m m}(z)\;,\]   where \[ G_{mm}(z)=(z-\e_m- \Sigma_{m})^{-1}\;. \]  It does not involve the non-diagonal matrix elements of the Green's  function $G_{mm^\prime}(z)$ when statistical average $\langle \scH_{0m}\rangle=0$. This is effectively our case as well (see Eq.(\ref{eq:Hmean})).

The main assumption of cavity method is that in the limit  $M\rightarrow \infty$ the difference between the PDFs of $\Sigma_{0}^{+1}(z)$ and $\Sigma_{0}(z)$ disappears. This results in a self-consistent
 equations for the self-energy.  Following \cite{abou1973selfconsistent}  we add  small imaginary parts to the diagonal matrix elements $\scH_{mm}=\e_m-i \eta$. It is a small ``fictitious'' quantity that is  still assumed to be much bigger than the marked state energy spacing 
  $\eta\gg W/M$. Results are  not expected to depend on the value of $\eta$ provided its scaling with $M$ is chosen appropriately, as will be discussed below.
  We separate the real and imaginary parts of the self-energy, $\Sigma_m(z)=\Sr_m(z)-i \Si_m(z)$ (cf. (\ref{eq:S-def})), obtaining
 %\begin{equation}
%\Sigma_m(z)=\Sr_m(z)-i \Si_m(z)\;.\label{eq:Sigma-m} 
%\end{equation}
 %\noindent
 \begin{subequations}\label{eq:XY0}
 \begin{align}
 \Sr_0  & \disteq  \pi\sum_{m=1}^{M}\scH_{0m}^{2}\, \delta(\Si_m+\eta,z-\e_m-\Sr_m)\;,\label{eq:Xc}\\
\Si_0  & \disteq  \pi\sum_{m=1}^{M}\scH_{0m}^{2}\, \delta(z-\e_m-\Sr_m,\Si_m+\eta)\;.\label{eq:Yc}\
\end{align}
\end{subequations}
where the function $\delta(x,y)\equiv \frac{1}{\pi}\frac{y}{x^2+y^2}$ was already introduced in (\ref{eq:deltaY}).

The self-consistent  Eqs.~(\ref{eq:XY0}) were  derived  by Abou-Chacra, Anderson and Thouless \cite{abou1973selfconsistent} for matrices on Bethe lattices and by Bouchaud and Cizeau  for Levy matrices \cite{cizeau1994theory}.
The solution of these equation was only found in the case when they can be linearized in $\Si_m$ \cite{abou1973selfconsistent,cizeau1994theory,tarquini2016level} giving the location of mobility edge $E^*(\alpha)$ as a function of the power $\alpha$
in the tail of the PDF of the matrix elements $P(\scH_{ij}^2)\propto 1/\scH_{ij}^{2(\alpha+1)}$.
Here we will provide a full solution of the nonlinear equations.

We will solve the self-consistent equations (\ref{eq:XY0}) under the assumption that pairs of variables $(\Sr_m,\Si_m)$ for each state $m\in [0,M]$ are taken from the {\it same} PDF 
$\scP(\Sr,\Si;z)$ defined over the domain $x\in(-\infty,\infty)$, $y\in [0,\infty)$. In what following for brevity  we omit the explicit dependence on the parameter $z$. Following \cite{abou1973selfconsistent} we introduce the characteristic  function $\scF(k_1,k_2)$ of the PDF $\scP(\Sr,\Si)$
\begin{equation}
 \scF(k_1,k_2)=\int_{-\infty}^{\infty}d\Sr\int_{0}^{\infty}d\Si\,\scP(\Sr,\Si)e^{i k_1 \Sr+i k_2 \Si}\;,\nonumber %\label{eq:F}
\end{equation}
that satisfies the equation $\scF_\eta(k_1,k_2)$=$\scG^M_\eta(k_1,k_2)$
%\begin{equation}
%F(k_1,k_2)=G^M(k_1,k_2)\;,\label{eq:GM}
%\end{equation}
where
%\begin{align}
%&\scG_\eta(k_1,k_2) =\int_{-\infty}^{\infty}d\Sr\int_{0}^{\infty}d\Si\, \scP(\Sr,\Si)\int_{-\infty}^{\infty}\frac{d\e}{W}\, p_A\(\frac{\e}{W}\)\nonumber \\
%\end{align}
%\begin{align}
%& \times \int_{0}^{\infty}dfP(f)  \,e^{i f k_1\,\delta(\eta+\Si, z-\e-\Sr)+ i f k_2 \,\delta(z-\e-\Sr,\eta+\Si)}\;,\nonumber \\
%& \times \int_{0}^{\infty}dfP(f)  \,e^{i f k_1\,\delta(\eta+\Si, z-\e-\Sr)+ i f k_2 \,\delta(z-\e-\Sr,\eta+\Si)}\nonumber %\label{eq:Gk1k2}
%\end{align}
\begin{align}
\scG(k_1,k_2) =%\end{align}
%\begin{align}
%& \times \int_{0}^{\infty}dfP(f)  \,e^{i f k_1\,\delta(\eta+\Si, z-\e-\Sr)+ i f k_2 \,\delta(z-\e-\Sr,\eta+\Si)}\;,\nonumber \\
\avg{  e^{i f k_1\,\delta(\eta+\Si, z-\e-\Sr)+ i f k_2 \,\delta(z-\e-\Sr,\eta+\Si)} }\nonumber %\label{eq:Gk1k2}
\end{align} 
Here   $f$=$\scH_{0m}^{2}$ and the average is performed with the joint PDF $\scP(\Sr,\Si)\frac{1}{W}\, p_A\(\frac{\e}{W}\)dfP(f) $.
%$\int_{-\infty}^{\infty}d\Sr\int_{0}^{\infty}d\Si\, \scP(\Sr,\Si)\int_{-\infty}^{\infty}\frac{d\e}{W}\, p_A\(\frac{\e}{W}\)\int_{0}^{\infty}dfP(f) $
The above relation  between $\scF(k_1,k_2)$ and $\scG(k_1,k_2)$ is actually an equation for  the PDF $\scP(\Sr,\Si)$ because
 both $\scG$ and $\scF$ depend on  $\scP$. 

%\subsection{ \label{sec:sol} Solution of self-consistent equations in non-ergodic delocalized phase}

\section{ \label{sec:sol}  Solution of cavity equations  in non-ergodic delocalized phase}

\subsection{\label{sec:Im}Analysis of the imaginary part of self-energy}

We note that the   exponent in the integrand of the above  expression for $\scG$  depends on  $\Sr$ and $\e-z$ only via their combination $\Sr+\e-z$.
In the non-ergodic delocalized phase the typical width of the PDF of  $\Sr$ is much more narrow than the width $W$ of $p(\e)$ (\ref{eq:pA}).  We will also consider small values of $|z|\ll W$. Therefore in the first approximation we will neglect  $\Sr$ and $z$ compared to $\e$.  Then $\scG(k_1,k_2)$ depends only on the marginalized PDF
\begin{equation}
\scP(\Si)=\int_{-\infty}^{\infty}d\Sr\, \scP(\Sr,\Si)\;.\label{eq:pY}
\end{equation}
Once this PDF is obtained, the PDF $\scP(\Sr,\Si)$ can be analyzed from its characteristic
 function $\scF(0,k_2)$. Inverting it we obtain  the self-consistent equation for $\scP(\Si)$ in the limit $M\rightarrow \infty$
\begin{equation}
\scP(\Si)=\frac{1}{2\pi}\int_{-\infty}^{\infty}dk e^{M \theta(k)-i k \Si}\;.\label{eq:PetaYm}
\end{equation}
\vspace{-0.2in}
\begin{equation}
 \theta(k)=\int^{\infty}_{0}df d\Si dh \, P(f) \scP(\Si)p_{\eta+\Si}(h) (e^{i k f h}-1)\;\nonumber % \label{eq:th-k}
\end{equation}
Here $\theta(k)=1-\scG_\eta(0,k)$ and the  domain of integration for all variables is $[0,\infty)$. The function $p_{\eta+Y}(h)$  above is a conditional PDF of a random variable \[h=\delta(\e,\eta+Y)\] 
with $Y$ fixed and $\delta(x,y)$ given in (\ref{eq:deltaY}). The explicit form of  the PDF $p_{\eta+Y}(h)$ is obtained  in  Sec.~\ref{sec:peh} of the Appendix, Eqs.~(\ref{eq:p_et}),(\ref{eq:p_et-1}).
% \ref{sec:peh}. Its explicit form is (cf. Eq.~\eqref{eq:p_et}) 
%\begin{equation}
%p_Y(h)=\frac{\sqrt{\Si}}{h^{3/2}\sqrt{1-\Si h}}\,p_A(\sqrt{\Si(1/h-\Si)})\;.\label{eq:p_yh}
%\end{equation}

\begin{figure}[htb]
  \includegraphics[width= 3.35in]{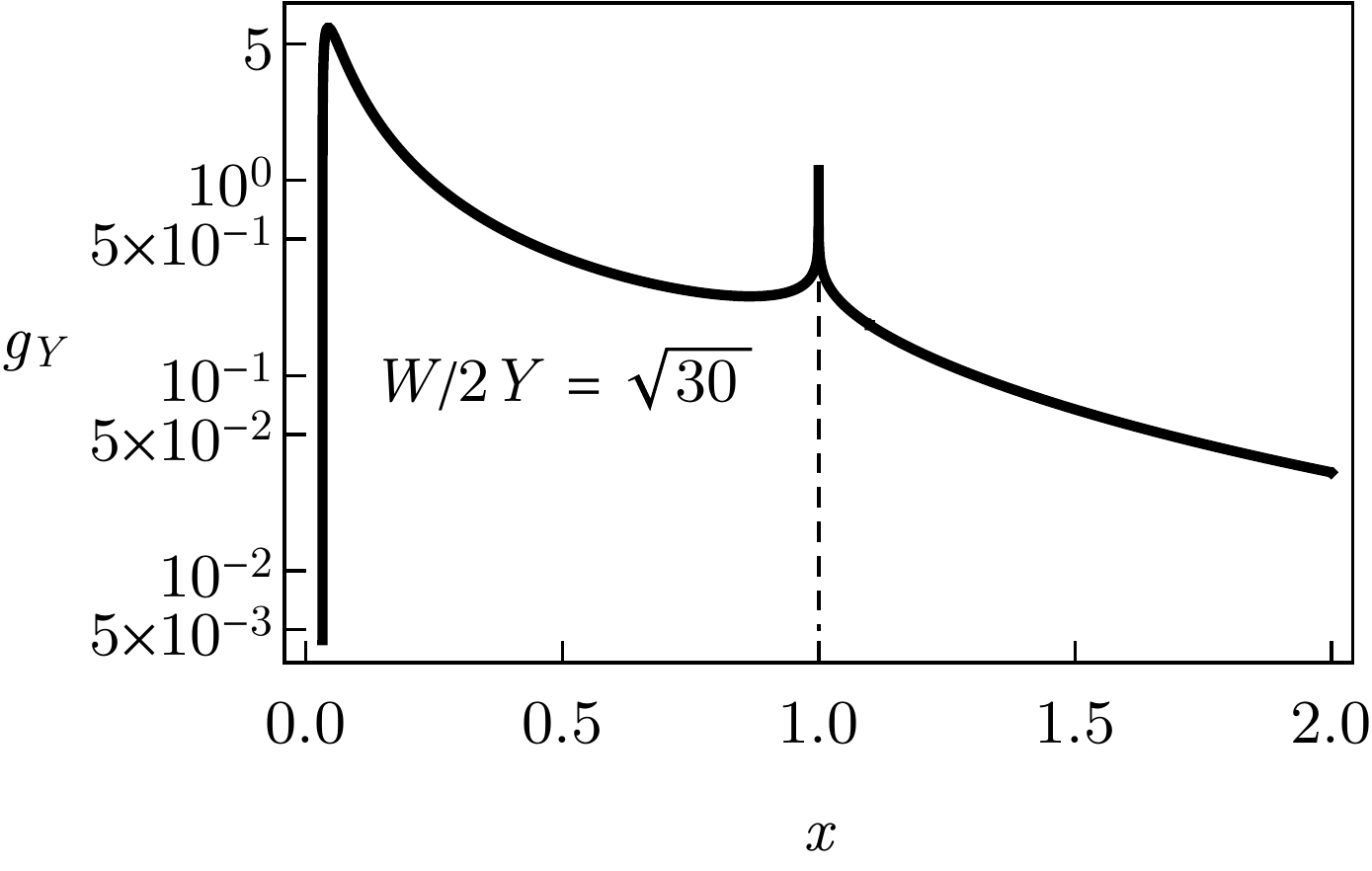}
  \caption{Plot of the PDF $g_Y(x)$ of the random variable $x$=$\frac{w Y^2}{(z-\e)^2+Y^2}$ where  random variables $\e$ and $w$ obey distributions  $W^{-1}p_A(\e/W)$ and  $g_\infty(w)$, respectively,  and  $W/(2 Y)=\sqrt{30}$.
  Detailed discussion of  $g_Y(x)$ is given in Appendix \ref{sec:getax} (see Eq.~(\ref{eq:pzP})). Its maximum is located at $x\sim (Y/W)^2$. The singularity at $x=1$ corresponds to $\e=z$. For large values of $x\gg1$
 the conditional PDF of  $\frac{Y^2}{(z-\e)^2+X^2}$  is  narrowly peaked around its mean value $\pi Y/W$ with  $|\e-z|\sim Y$,  giving rise to the relation in Eq.~(\ref{eq:pzPam}).
   \label{fig:pzP} }
\end{figure}

To achieve further progress we use  the approximation (\ref{eq:Hbeta})  and drop oscillatory  factors in the off-diagonal matrix elements $\scH_{0m}$. Then we have for the PDF
$P(f)=g_\infty(f/\cVt^2)/\cVt^2$ (\ref{eq:g_0(w)}) and in what follows we will use the rescaled variable $w=f/ \cVt^2$ for the squared matrix elements, in accordance with (\ref{eq:w-m}). Instead of the variable $h$ in (\ref{eq:PetaYm}) we will use the re-scaled variable
\begin{equation}
y=\sqrt{h(\eta+\Si)}\;,\label{eq:resc}
\end{equation}
 that  obeys the distribution 
\begin{equation}
\fp_{\eta+\Si}(y)=\frac{2(\eta+\Si)}{W}\frac{1}{ y^2\sqrt{1-y^2}}\;\label{eq:p-ym}
\end{equation}
(see details in Appendix \ref{sec:peh}, (\ref{eq:p-y})).
Then $ \theta(k)$ takes the form
\begin{equation}
\theta(k)=\int_{0}^{\infty}d\Si \scP(\Si)\,\phi_{\Si+\eta}\(\frac{k \cVt^2}{\Si+\eta}\)\;.\label{eq:theta1}
\end{equation}
Here $\phi_Y(u)$ is a characteristic function
\begin{equation}
\phi_Y(u)=\int_{0}^{\infty}dx \, g_Y(x)(e^{i u x}-1)\;.\label{eq:phiY}
\end{equation}
of the PDF  $g_Y(x)$ of the random variable $x= w y^2$ where  $w$ obeys $g_\infty(w)$ and $y$ obeys $\fp_Y(y)$ (\ref{eq:p-ym}).  Detailed study of $g_Y(x)$ is given in  Appendix \ref{sec:getax}. The PDF  $g_Y(x)$ depends on $Y$ via the ratio $Y/W$ and its plot is shown in  Fig.~\ref{fig:pzP}. It goes over into $g_\infty(y)$ for $Y\rightarrow \infty$.

We now make a key observation: in the limit of large $x\gg 1$ and for $W\gg Y$ the following relations holds for the PDF  $\phi_Y(u)$  and its characteristic function (see the corresponding Eqs.~\eqref{eq:phi0u} and (\ref{eq:pzPa}) in Appendix \ref{sec:getax})
\begin{equation}
g_Y(x)\simeq \frac{\pi Y}{W}g_\infty(x),\quad \phi_Y(u)\simeq \frac{\pi Y}{W}\phi_\infty(u)\;.\label{eq:pzPam}
\end{equation}
The reason for this can be explained as follows.
For   large  deviations of   $x=w Y^2/(\e^2+Y^2)$   the conditional PDF $p(\e|x)$ of the marked state energy $\e$  is  narrowly peaked in the range of values  $|\e|\sim Y$.
In contrast, typical  energy values are much bigger $\e\sim W$.  This narrowing  of the conditional PDF $p(\e|x)$
gives rise to a small factor $\pi Y/W$ in the  r.h.s. of (\ref{eq:pzPam}).

We observe that $\lim_{k\rightarrow\infty}\theta(k)=0$ and for $M\rightarrow\infty$ the integral  in (\ref{eq:PetaYm}) is dominated by $|k|\ll 1$.
We make an  assumption (whose validity becomes obvious below) that for small enough $k$ the integral in (\ref{eq:theta1}) is dominated by values of $\Si$ such that $k\cVt^2/(\Si+\eta)\ll 1$. Therefore we will use in (\ref{eq:theta1})
the approximate expression for the characteristic  function $\phi_{\Si+\eta}$ given by  Eq.~(\ref{eq:pzPam}). We rescale $\Si$ with the typical value of imaginary part of self-energy of marked states  $\Sti$ (\ref{eq:Gdef1}) obtained in FGR-based calculation in Sec.~\ref{sec:FGR}. Making a change of variables
\begin{equation}
\Si=\Sti s,\quad \scP(\Si)=\frac{1}{\Sti} \rho(s)\;,\label{eq:change}
\end{equation}
we rewrite the self-consistent equation~(\ref{eq:PetaYm}) in the limit $x\gg1, W\gg Y$ for the rescaled PDF $\rho(s)$ in the following form:
\begin{equation}
\rho(s)=\frac{1}{2\pi}\int_{-\infty}^{\infty}du\,e^{-i u s+ \Phi(u,\Omega)}\;,\label{eq:rho-et}
\end{equation}
\vspace{-0.2in}
\begin{equation}
 \Phi(u,\Omega)=\int_{-\infty}^{\infty}d\nu\, \rho\(\nu- \beta_\eta\)\,\Omega\,\nu\,\phi_{\infty}\(\frac{q}{\Omega\,\nu}\)\;,\nonumber
 \end{equation}
\begin{equation}
 \beta_\eta=\frac{\eta}{\Sti}\;,\label{eq:beta}
 \end{equation}
  and
\begin{equation}
\Omega=\frac{\pi\Sti}{\delta\e}=\(\frac{\pi \cVt}{W/M} \)^2\;,\label{eq:L}
\end{equation}
$\Sti$ and $\cVt$ are  defined in (\ref{eq:Gdef1}) and (\ref{eq:typ}), respectively. We observe that $\Omega$ corresponds to  the typical number of marked states in the mini-band  that we estimated in Sec.~\ref{sec:FGR} using  the Born approximation.

Assuming $\Omega\gg1$ (delocalized phase) we expand $\Omega\nu \phi_\infty(q/(\Omega\nu))$ in inverse powers of $\log \Omega$ using asymptotic form of the characteristic function $ \phi_\infty(u)$ at small argument studied in Appendix \ref{sec:phi0},   Eqs.~(\eqref{eq:Rephi0a}),(\ref{eq:Imphi0a}). Truncating the expansion at terms $\sim (\log M)^{-1/2}$ we get 
\begin{align}
 \Omega\, \Re\,\phi_\infty\left(\frac{q}{\Omega}\right) &\simeq -\frac{\pi |q|}{2\sqrt{\log \Omega}}\;,\label{eq:ReImphim}\\
\Omega\,\Im\,\phi_\infty\left(\frac{q}{\Omega}\right)& \simeq  2q  \left(\frac{\log \Omega}{\pi}\right)^{1/2}+ q \frac{1-{\mathcal C}-\log |q|}{(\pi \log \Omega)^{1/2}}\;,\nonumber
\end{align}
\noindent
where ${\mathcal C}\approx 0.577$ is the Euler constant.
It is clear from comparing  individual terms in  Eq.~(\ref{eq:ReImphim}) with the exponential in  Eq.(\ref{eq:rho-et}) that $q=\cO(\sqrt{\log \Omega})$. This justifies the order of truncation (see details in Appendix \ref{sec:CLT}, Eq.~\ref{eq:ReImphi}). 

We make change of variables in the integral in (\ref{eq:rho-et})
%\begin{equation}
$q=2 \sqrt{\log \Omega / \pi}\;t $ 
and obtain 
\begin{equation}
\rho(s)=\sigma_{\Omega}^{-1} L_{1}^{1,1}((s-\mu_\Omega)/\sigma_\Omega)\;,\label{eq:rhos}
\end{equation}
where quantity $\mu_\Omega$ satisfies the   equation
\begin{equation}
\mu_\Omega=b_\Omega+\frac{2\sigma_\Omega}{\pi}\int_{-\infty}^{\infty}ds\;\rho(s)\log|s+\beta_\eta |\;.\label{eq:muL}
\end{equation}
Above   $L^{1,1}_{1}(x) $ is Levy distribution    \cite{cizeau1994theory} defined in the Appendix, Eq.~(\ref{eq:char}), and shown  in Fig.~\ref{fig:st-dist}.  Coefficients $\sigma_\Omega =\sqrt{\pi/(4\log\Omega)}$  and $b_\Omega \simeq 1/\sigma_\Omega$ are  given   in Eqs.~(\ref{eq:sigmaM}),(\ref{eq:bM}) where the   parameter $M$ needs to be replaced by $\Omega$. 

We plug the above expression for $\rho(s)$ into \eqref{eq:muL} and express $\mu_\Omega$ in terms of a new variable $x$ 
\begin{equation}
\mu_\Omega \equiv b_\Omega-\frac{2\sigma_\Omega}{\pi}\log\sigma_{\Omega}^{-1}+\sigma_\Omega\,x\;.\label{eq:mu-x}
\end{equation}
Then this variable   satisfies the following  equation
\begin{equation}
x=\frac{2}{\pi}\int_{-\infty}^{\infty}ds\, L_{1}^{1,1}(s)\log |s+x+\zeta_\Omega|\;,\label{eq:x-eq}
\end{equation}
that  involves a scale-free Levy distribution and  a single parameter $\zeta_\Omega$
\begin{equation}
\zeta_\Omega=\frac{b_\Omega}{\sigma_\Omega}-\frac{2}{\pi}\log\(\frac{1}{\sigma_\Omega}\) +\frac{1}{\sigma_\Omega}\frac{\eta}{\Sti}\;\label{eq:zeta}
\end{equation}
where  we used an explicit form of $\beta_\eta$ (\ref{eq:beta}). We note that the  self-consistent equation for the   function $\rho(s)$ is now reduced to the simple transcendental  equation (\ref{eq:x-eq}).

Using explicit form of   $\sigma_\Omega$ and $b_\Omega$ (\ref{eq:sigmaM}),(\ref{eq:bM}) one can see that $\zeta_\Omega$ is large compared to unity   in the delocalized phase,  $\zeta_\Omega \simeq \sigma_{\Omega}^{-2}$ $\sim$ $ \log\Omega \gg 1$. With  this property   the equation for $x$ (\ref{eq:x-eq}) can be  solved by iteration using the asymptotic expansion of Levy distribution
at large arguments,  $L_{1}^{1,1}(\nu)\simeq (2/\pi )\nu^{-2}$ ($\nu\gg1$). To the leading order
\begin{equation}
x\simeq \frac{2}{\pi}\log\zeta_\Omega+\cO\(\frac{\log\zeta_\Omega}{\zeta_\Omega}\)\;.\label{eq:xsol-a}
\end{equation}
Then using (\ref{eq:mu-x}) the expression for $\mu_\Omega$ is
\begin{equation}
\mu_\Omega\simeq \frac{1}{\sigma_\Omega}+\frac{2\sigma_\Omega}{\pi}\log\(1+\frac{\eta\, \sigma_\Omega}{\Sti}\)+\frac{2\sigma_\Omega(1-\gamma_{\rm Euler})}{\pi}\label{eq:mu-Omega0}
\end{equation}
where we neglected terms $\sim\sigma^{3}_{\Omega}\log\Omega$ that are much smaller than the width $\sigma_\Omega$ of the distribution $\rho(s)=\sigma_{\Omega}^{-1}L_{1}^{1,1}((s-\mu_\Omega)/\sigma_\Omega)$.

We note that the dependence of  $\mu_\Omega$ (\ref{eq:mu-Omega0})  on the initial (fictitious) level broadening $\eta$ disappears when the later is chosen to be much smaller  than the mini-band width \cite{altshuler2016nonergodic, altshuler2016multifractal,facoetti2016non}, $W/M\ll \eta \ll \Sti \sigma_\Omega$.
 Using (\ref{eq:Gdef1}),  (\ref{eq:param}) the scaling behavior of $\eta$ with $M$ in the non-ergodic delocalized regime must satisfy the condition 
\begin{equation}
\eta=M^\kappa,\quad |\kappa|< 1-\frac{\gamma}{2},\qquad \gamma\in(1,2)\;.\label{eq:eta-scale}
\end{equation}

Finally  the expression for the distribution function of  the imaginary part of self-energy has the form
\begin{equation}
\scP(\Si)=\frac{1}{C} L_{1}^{1,1}\(\frac{\Si-\Styp}{C}\)\;,\label{eq:scPIm}
\end{equation}
\begin{equation}
\Styp=\mu_\Omega\Sti,\quad C=\sigma_\Omega \Sti\;.\label{eq:SC}
\end{equation}
Here $\Styp$ is a   shift  of the distribution and $C$ its scale parameter  (characteristic width). Also,
\begin{align}
\mu_\Omega & \simeq \frac{1}{\sigma_\Omega}+\frac{2\sigma_\Omega(1-\gamma_{\rm Euler})}{\pi}\;.\label{eq:mu-Omega}\\
\sigma_\Omega &=\sqrt{\frac{\pi}{4\log \Omega}}\;.\label{eq:sigmaOmega0}
\end{align}

Using the  scaling ansatz (\ref{eq:param}) for the width $W$ of the IB in terms of $M$ ,  the typical number of states in a mini-band 
(number of resonances) equals, 
\begin{equation}
\Omega=\(\frac{\pi}{\lambda}\)^2 M^{2-\gamma}\;.\label{eq:Omegac}
\end{equation}
 
Using  the same scaling ansatz   (\ref{eq:param})  and  the expressions for  $\sigma_\Omega$ (\ref{eq:sigmaM}) and $\mu_\Omega$ (\ref{eq:mu-Omega}) we obtain,
\begin{align}
\Styp  &\simeq \frac{2 \pi^{1/2} }{\lambda}\, \cVt M^{1-\gamma/2} (\log \Omega)^{1/2}\;,\label{eq:shift}\\
C& \simeq \frac{\pi^{3/2}}{2\lambda}\,\,\cVt M^{1-\gamma/2}   (\log \Omega)^{-1/2}\;.\label{eq:width}
\end{align}
%\begin{align}
%\Styp  \simeq \frac{\sqrt{8\pi(1-\gamma/2)}}{\lambda}\,\,& \cVt M^{1-\gamma/2} \sqrt{\log M}\;,\label{eq:shift}\\
%C \simeq \frac{\pi^{3/2}}{ \lambda \sqrt{8(1-\gamma/2)}}\,\,&\frac{\cVt M^{1-\gamma/2}}{   \sqrt{\log M}}\;.\label{eq:width}
%\end{align}

Here $\cVt $$\sim $$n^{1/2}2^{-n/2} e^{-n/(4 B_{\perp}^{2})}$  is given in (\ref{eq:typ}).
The  shift $\Styp$ corresponds to  the typical value of $\Si$.   One can see from the above that it is  $ \log \Omega\sim\log M\gg 1$ times bigger than the distribution width. We note in passing that distribution of $\Sigma^{\prime\prime}$ determines that of the miniband width $\Gamma=2\Sigma^{\prime\prime}$ (\ref{eq:GM}).

\subsubsection{\label{sec:cav-born}Comparison between the cavity method and leading-order Born approximation}

It is instructive to compare the above distribution of $\Si$ obtained using the cavity method with that obtained within the Born approximation \eqref{eq:Gdef1}-(\ref{eq:bM}).
In both cases the distribution of $\Si$ is given by the appropriately rescaled and shifted Levy alpha-stable distribution $L_{1}^{1,1}(x)$.
In both cases, the  scale parameter  $C$ (characteristic width)  of the disribution  has the form $C=\sigma_S\Sti$ with $\sigma_S=\sqrt{\pi/(4\log S)}$. In   the case of the Born approximation  $S=M$, corresponding to the total number of marked states, and in the case of cavity method $S=\Omega\ll M$, corresponding to the (much smaller) number of states in the mini-band. Using (\ref{eq:Omegac}) we estimate
\begin{equation}
\frac{\sigma_M}{\sigma_\Omega}=\sqrt{2-\gamma} <1 \qquad (W=\lambda M^{\gamma/2}) \;.\label{eq:OM}
\end{equation}
Therefore Born approximation underestimates the width  of the distribution of $\Si$. The ratio (\ref{eq:OM}) is especially pronounced near the localization transition $\gamma=2$.  Value of $\sigma_{\Omega}^{-1}$ shrinks to zero at the transition while that of 
$\sigma_{M}$  does not depend on the closeness to the transition point.

We note however that factors $\sigma_\Omega$ and $\sigma_M$ depend on $M$ only logarithmically.  At the same time, the leading order (power-law) dependence of the rescaling coefficient  on $M$ is given by the factor $\Sti\propto M^{1-\gamma/2}$,  and is identical in the cavity method and the Born approximation-based expressions. 

The situation is similar with the shift parameter $\Styp$  in the Levy distribution of $\Si$ corresponding its  typical value,  $\Styp $ $\simeq$ $ \Sti/\sigma_S$ with $S$=$M$ (Born approximation) and  $S$=$\Omega$ (cavity method). The  leading-order dependence of the shift on $M$ is the same in both cases and is given by $\Sti$. In both cases the shift is greater than the rescaling coefficient by a factor $\sim \log M$. However the Born approximation overestimates the shift by a factor  $(2-\gamma)^{-1/2}$ .

\subsubsection{\label{sec:self-consist-B}Comparison between the cavity method and self-consistent  Born approximation}

%
%The key approximation in Sec.~\ref{sec:FGR} was to assume that the partial sum in the expressions (\ref{eq:Gamma-d}),(\ref{eq:red}) for the decay rate $\Gamma_{j}^{(d)}$ can be replaced by an integral over energies of the marked states  
% located on the same   Hamming distance $d$ from the state $\ket{z_j}$.
%%is dominated by the terms with  $|\e_j-\e_m| \apprle \eta$ for marked states  $\ket{z_m}$
% %located on the same   Hamming distance $d$ from the state $\ket{z_j}$.
% We revisit  the decay rate equation (\ref{eq:FGR0}), and removing the oscillatory parts in the matrix elements (\ref{eq:Hbeta}) we write
%\begin{equation}
%\Si=\frac{\cVt^2}{\eta}\sum_{m=1}^{M}x_m,\quad x_m=\frac{w_m \eta^2}{\e_m^2+\eta^2}\;.\label{eq:Ysum-m}
%\end{equation}
%Here in the l.h.s. we  omitted  the subscript in  $\Si_j$ and  made our usual assumption that  $|\e_j|\ll W$, and the rescaling $V(d_{jm})^{2}=\cVt^2 w_m$. Random variables $x_m$  are sampled from the distribution $g_\eta(x)$ given in (\ref{eq:pzPam}) and plotted in  Fig.~\ref{fig:pzP}). 
%
%
%{\color{black}Unlike the situation in the cavity method where the level broadening $\eta$ in the effective  model (\ref{eq:tVen}) has a fictitious meaning and drops out from the final result, here we will make a self-consistent assumption and set it  equal to the characteristic  width of the miniband 
%\begin{equation}
%\eta=\Sti\;.\label{eq:eta-S}
%\end{equation}
%Then using GCLT for the sum in (\ref{eq:Ysum-m}) one can  obtain the PDF of $\Si$. The details are given in {Appendix}~\ref{sec:getax} and here we provide the result

The leading-order Born  approximation recovers    the typical shift $\Styp$ and the scale parameter  $C$  of the distribution of $\Si$ with exponential accuracy in $\log M$. However it gives an incorrect  dependence   of  the prefactor on $\log M$ in these coefficients.
The main approximation in Sec.~\ref{sec:LeadingOrderBorn} was to assume that the sum in the expression for the spectral density $\rho_{\eta}^{j}(d)$ 
(\ref{eq:red})  can be replaced by an integral. %As the result the parameter  $\eta$ dropped out from the PDF of $\Si$.
We revisit  the decay rate equation (\ref{eq:FGR0}) using the statistical ensemble (\ref{eq:Hbeta}) 
 %\begin{equation}
% \Si_j=\pi\sum_{m\neq j}V^2(d_{jm})\delta(\e_j-\e_m,\eta)
 %\end{equation}
 %The random variable $\Si$ can be expressed in terms of the rescaled random variables $w_m$ (\ref{eq:g_0(w)}) and $\eta \delta(\e_m,\eta)$ (the statistics of the latter is studied in the Appendix~\ref{sec:peh}) 
 \begin{equation}
\Si=\frac{\cVt^2}{\eta}\sum_{m=1}^{M}x_m,\quad x_m=\frac{w_m \eta^2}{\e_m^2+\eta^2}\;.\label{eq:Ysum-m}
\end{equation}
Here in the l.h.s. we  omitted  the subscript in  $\Si_j$ and  made the rescaling $V(d_{jm})^{2}=\cVt^2 w_m$. Random variables $x_m$  are sampled from the distribution $g_\eta(x)$ given in (\ref{eq:pzPam}) and plotted in  Fig.~\ref{fig:pzP}). 
Using GCLT for the sum in (\ref{eq:Ysum-m}) one can  obtain the PDF of $\Si$. The details are given in {Appendix}~\ref{sec:getax} and here we provide the result,
\begin{equation}
 \Si\disteq \tilde\Sigma_{{\rm typ}}^{\prime\prime} + x \,C,  \nonumber
 \end{equation}
 \begin{equation}
  \tilde\Sigma_{{\rm typ}}^{\prime\prime}  = b_{\Omega_\eta}\,\Sti , \quad C= \sigma_{\Omega_\eta}\,\Sti\;.\label{eq:SAB}
\end{equation}
Here  $x$ is a random variable that obeys Levy distribution $L_{1}^{1,1}(x)$, coefficient 
$\sigma_\Omega$  is given in (\ref{eq:sigmaOmega0}) and $b_\Omega$ is  given   in (\ref{eq:bM}) where one should replace $M$ with the number of marked states in a mini-band of width $\eta$
\begin{gather}
\Omega_\eta= \frac{\pi \eta}{\delta \epsilon}.
\end{gather}
Unlike the discussion  in the cavity method, the statistics of $\Sigma^{\prime\prime}$ explicitly depends on $\eta$. 
We  make a self-consistent assumption and set $\eta$  equal to the characteristic  width of the miniband 
\begin{equation}
\eta=\Sti\,\Longrightarrow\, \Omega_\eta=\frac{\pi \Sti}{\de}\;.\label{eq:eta-S}
\end{equation}
 We conclude that the   typical number of states in a miniband $\Omega_\eta=\Omega$  given by the self-consistent Born approximation is the same as that  given by the cavity method,  Eqs.~(\ref{eq:L}). Therefore using (\ref{eq:SAB})  one can  see that the width   $C$  of the  distribution of $\Si$  is also the same in both methods.
  The difference 
between the typical values of $\Si$  in the two methods is  
\[\Sigma_{\rm typ}^{\prime\prime}-\tilde\Sigma_{\rm typ}^{\prime\prime}=\frac{2}{\pi}C \log\sigma_{\Omega}^{-1}\ll \Sigma_{\rm typ}^{\prime\prime}\]
This error is much smaller than in the case  discussed in Sec.~\ref{sec:cav-born} (cf. Eq.~\eqref{eq:OM}) where the self-consistent condition is not used.
However it  exceeds the distribution width $C$ for sufficiently large   $M\gg1$ because in the non-ergodic delocalized phase
 $\log\sigma^{-1}_{\Omega}\sim\log\log M$.

\subsection{\label{sec:Re}Real part of self-energy}

In this section  we will find the marginalized probability distribution of real parts  of self-energy 
\begin{equation}
\cP(\Sr)=\int_{0}^{\infty}d\Si\,\scP(\Sr,\Si)\;.\label{eq:pX}
\end{equation}
We  consider the first equation in (\ref{eq:XY0}).  Following the arguments provided in Sec.~\ref{sec:Im} we neglect the terms $z-\Sigma^{\prime}_{m}$ in the r.h.s of the equation and   drop the oscillatory  factors in  $\scH_{0m}$ using the  probability distribution $P(f)=g_\infty(f/\cVt^2)/\cVt^2$ (\ref{eq:g_0(w)}) instead. Then Eq.~(\ref{eq:Xc}) takes the form
\begin{equation}
\Sr  \disteq \sum_{m=1}^{M} r_m\;.\label{eq:SReR}
\end{equation} 
Here $r_m$ are instances of a random variable $R$ such that
\begin{equation}
r= f \frac{\e}{\e^{2}+ (\Si)^{2}}\;,\label{eq:R}
\end{equation}
where  $\e,\,f,\,\Sigma^{\prime\prime}$ are random variables independently sampled from the distributions $p(\e),\,P(f)$ and $\scP(\Si)$, respectively. Using GCLT, in the asymptotic limit of $M\rightarrow\infty$ the sum in (\ref{eq:SReR}) is determined by the tail of the probability distribution  of $r$ at $|r| \rightarrow \infty$. This analysis is very similar to the one already discussed in Sec.~\ref{sec:FGR},\ref{sec:Im} and in Appendix \ref{sec:CLT}. Here we omit details of the calculations and simply provide the result.
The tail of the PDF of $r$ in the limit $|r|\rightarrow\infty$ has the form
\begin{equation}
\rho=\frac{r}{2\Sti/(\pi M)},\quad {\rm PDF}(\rho)\simeq \frac{1}{\rho^2} \sqrt{\frac{\log(\rho)}{\pi}}\;.\label{eq:PDFR}
\end{equation}
($\rho\gg 1$). The distribution function $\cP(\Sr)$ of the sum in (\ref{eq:SReR}) is the  Cauchy distribution
\begin{equation}
\cP(\Sr)=\frac{1}{\pi} \frac{\Str }{(\Str)^2+(\Sr)^2},\quad \Str=\frac{\Sti}{ \sigma_M}\;.\label{eq:SRe-dist}
\end{equation}
Here the expression for $\sigma_M\sim 1/\sqrt{\log M}$ is given in (\ref{eq:sigmaM}). Cauchy distribution has the form very similar to the stable distribution $L_{1}^{1,1}(x)$ that describes the fluctuations of the $\Si$ (\ref{eq:scPIm}) up to the shift and rescaling coefficients. Both distributions are displayed in Fig.~\ref{fig:st-dist}. The tail of the Cauchy distribution differs  from that of $L_{1}^{1,1}(x)$ by a factor of 2. Unlike that of $\Si$ the  distribution of 
$\Sr$  is symmetric for impurity states with energies near the center of the band. 
The typical value of  $\Sr$ is greater than that of $\Si$ by a constant factor
\begin{equation}
\frac{\Str}{\Styp} =\frac{1}{\sqrt{2-\gamma}} \qquad (W=\lambda M^{\gamma/2})\;.\label{eq:Re/Im}
 \end{equation}
The width  of the distribution of $\Sr$ is the same as its typical value while the width  $C$ of the distribution of $\Si$ is   smaller by a factor $\sim  1/\log M$ (cf. Eqs.~(\ref{eq:shift}),(\ref{eq:width})).  These relations between the distributions of $\Sr$ and $\Si$ have implications for the complexity of the population transfer as discussed below. We also note that the real and imaginary parts of self-energy of  a given marked state are correlated with each other because according to Eqs. \eqref{eq:Xc},\eqref{eq:Yc} the values of  $\Sr_{j}$ and $\Si_j$ depend on the same set of parameters ($\scH_{jm}$, $\e_m$, etc). In this work we will not study their correlations.

\subsection{\label{sec:Dynamic correlations} Dynamic correlations}

For states close to the center of the band of marked states the typical value of the mini-band width can be connected to the average of the dynamical correlator, with the delta function regularized by a finite scale $\eta, \Sigma''_{\rm typ}\gg\eta\gg \delta \epsilon$, $\delta(x) \rightarrow \delta_{\eta} (x)\equiv \tfrac{1}{\pi} \tfrac{\eta}{x^2+\eta^2}$, 
\begin{gather}
\frac{1}{\Sigma''_{\rm typ}+\eta }=\frac{1}{\pi }\int _{-\infty }^{\infty }d\omega  \frac{ \eta}{\eta ^2+\omega ^2}   p (\omega),
\end{gather}
which can be inverted to obtain,
\begin{gather}
p ( \omega) \approx 
 \left[
 \begin{array}{ll}
 \frac{1}{\Sigma''_{\rm typ}}, & \omega\leq \omega_{\textrm{Th}} \\
  \frac{1}{\Sigma''_{\rm typ}}\left(\frac{\omega _{\textrm{Th}}}{\left| \omega \right| }\right)^2, & \omega>\omega_{\textrm{Th}} 
 \end{array}
 \right.
\end{gather}
where we introduced the Thouless energy,
\begin{gather}
\omega _{\textrm{Th}}=\frac{1}{2} \pi \Sigma^{\prime\prime}_{\rm typ}.
\end{gather}
The typical value of the mini-band width was obtained in Eq.~(\ref{eq:shift}). From the comparison of the respective Fig.~\ref{fig:ds_vs_a} we conclude that the scaling of the typical population transfer time $1/\omega_{\rm Th}$ and the scaling of the value of the dynamical correlator $K(\omega)$ are consistent in numerical and analytical calculations, subject only to a small correction in the scaling exponent $\varepsilon=0.05$.

\section{\label{sec:compl} Complexity of the Population Transfer protocol}

After the system is prepared at a given marked state $\ket{z_j}$ at $t=0$ the probability for the population to be transferred to other marked states is $1-\psi^{2}(z_j, t)$. At the initial stage 
the survival probability decays exponentially \eqref{eq:exp-d} with the mean decay time $1/\Gamma_{j}=1/(2\Si_j)$.

The  initial marked state decays into the eigenstates $\ket{\psi_\beta}$ of the IB Hamiltonian $\scH$ with typical energies $E_\beta$ inside the narrow interval  corresponding to the miniband associated with $\ket{z_j}$. It has a width  $\Si_j$ and is centered around $\scH_{jj}=\e_j$.  Typical classical energies $\e$ of the bit-strings measured at the end of PT protocol will obey the probability distribution $\cP(\e-\e_j-\Sr_j)$ with $\cP$ given in (\ref{eq:SRe-dist}). The  success of PT protocol  is  to find a bit-string distinct from $z_j$ at a time $t$ with energy inside that window  $\Delta \cE_{\rm cl}$ around $\e_j$.  The expected time to succeed in PT equals
\begin{equation}
t_{\rm PT}^{j}= \frac{1}{2\Si_j p_{\Delta\cE}},\quad p_{\Delta\cE}=\int_{0}^{\Delta\cE_{\rm cl}}\cP\(\e-\Sr_j-\frac{\Delta\cE_{\rm cl}}{2} \)d\e\;.\nonumber\label{eq:ptj}
\end{equation}
Here $p_{\Delta\cE}$ is the probability of detecting a bit-string inside the target window  $\Delta \cE_{\rm cl}$ under the condition that initial state has decayed.
Let us assume that the PT window  is as wide as the typical miniband  width, $\Delta \cE_{\rm cl}=\Styp$.  In this case $p_{\rm mb}$ differs from 1 only by a constant factor that does not depend on $M$ (cf. \eqref{eq:Re/Im}). Therefore we will detect the bit-string inside the PT window with finite  probability as long as we waited long  enough for the  transition away from the initial marked state to occur. Because the initial state  $\ket{z_j}$ is picked at random we can estimate typical  time  to success  of PT   $\pt\sim 1/ \Styp$  corresponding to the inverse typical  width of the miniband.   All of the states in a miniband are populated at (roughly) the same time $\pt$ because    transition rate to a subset of states on a distance $d$ away from $\ket{z_j}$ depends on $d$ very weakly (see Eq.~(\ref{eq:Gamma-jd}) and related discussion in Sec.~\ref{sec:NumRes})).
 
 From a computational perspective it is of interest to characterize the PT  by the relation between the  the typical success time   of PT  $\pt$ and the number of states $\Omega$ over which the population is spread  during PT
\begin{equation}
\pt\sim\frac{1}{\cVt\,\sqrt{\Omega \log\Omega}}\sim \left(\frac{2^{n}}{n \Omega \log\Omega} \right)^{1/2} e^{2\theta n}\;.\label{eq:Gr-rel}
\end{equation}
where we set $\Delta\cE_{\rm cl}\sim \Sti$ (see discussion above).  We note that the time $t_{\rm G}$ for the Grover algorithm for unstructured quantum search to find $\Omega$ items in a database of the size $2^n$ is $t_{\rm G}\sim (2^{n}/\Omega)^{1/2}$. 
PT time $\pt$ scales worse than Grover time $t_{\rm G}$ by an additional exponential factor $e^{2\theta n}\simeq e^{ \frac{n}{2B_{\perp}^{2}}}$   (\ref{eq:g-exp}). The  scaling exponent $2 \theta$  can be made arbitrarily small at large transverse  fields $1\ll \B=\cO(n^0)$.
 
One can expect that the distributions of eigenvalues and eigenvectors inside the mini-band  are very similar to those    in the ergodic case, albeit with the appropriately rescaled effective dimension $\Omega$ of the Hilbert  space~\cite{Kravtzov2015RzPr}. For example,  the energy spectrum of the  mini-bands in the non-ergodic delocalized phase of Rosenzweig-Porter (RP) model corresponds to  the Gaussian Orthogonal ensemble.
There, according to  the semicircle law~\cite{tao2012topics},  the typical spectral width of the mini-band  ($\sim 1/\pt$)  is proportional to the square root of the number of states $\Omega$ in it. 
Therefore the Grover scaling  (\ref{eq:Gr-rel}) for PT is consistent with semicircle law in the Gaussian random matrix models  that allow for non-ergodic delocalized phase such as RP model.

 However in the case of Levy matrices the distribution of eigenvalues has polynomial tails \cite{cizeau1994theory},  their spectrum is not bounded and semi-circle law does not apply. As mentioned above, this leads to a broad distribution of PT rates.  There exist statistically significant clusters of states of a relatively small size that will be populated faster than typical case because the corresponding classical bit-strings  are located closer  to each  in Hamming distance than the  typical inter-state separation. At first glance,  this tendency is counter to the Grover scaling \eqref{eq:Gr-rel}. We note however that fluctuations of $\Sr$ and $\Si$ are correlated with each other. Faster  decay of a marked state will also correspond to bigger self-energy shift which will reduce the likelihood of finding a marked state with its energy inside the target window $\Delta\cE_{\rm cl}\sim\Sti$.
 
However the Grover  scaling still survives in a typical case corresponding to PT away from a randomly selected bit-string. For Levy matrices  \cite{cizeau1994theory}  it reflects  the fact that  the {\it typical} width $\Styp$ of the curve of the global density of states along the energy axis must scale as a square root of the corresponding typical number of states (area under the curve).

\section{\label{sec:Grover} Comparison with the analogue  Grover search}

\subsection{\label{sec:Grover-sym} Grover search starting from a fully symmetrized  state}

So far we have studied the PT protocol with the Hamiltonian (\ref{eq:H}) $H=H_D+H_{\rm cl}$ that  starts from a  given marked state of an IB model  $H_{\rm cl}$ (\ref{eq:H_ib})  and aims at finding a different  marked state inside a given window of energies using a transverse field Hamiltonian $H_D=-\B \sum_{m=1}^{n} \sigma_{m}^{x}$ (\ref{eq:H}) as a driver. 

We  consider here a different protocol inspired by the Hamiltonian version of Grover algorithm proposed in \cite{farhi1998analog}. The new protocol 
finds  marked states in the IB model $H_{\rm cl}$ starting from the  ground state of $H_D$   which is a
fully symmetric state $\ket{S}=2^{-n/2}\sum_{j=1}^{n}\ket{z}$  in a computational basis. This protocol can be implemented   by adjusting  the value of transverse field $\B$ $\approx$ 1 so that  the ground state energy of the driver is set near the center of the IB. Then we can replace the full driver with the projector on its ground state, $H_D\rightarrow -n\B\ket{S}\bra{S}$. 
The quantum evolution is guided by the  Hamiltonian: 
\begin{equation}
H_{\rm G}=-n\B \ket{S}\bra{S}+\sum_{j=1}^{M}\cE(z_j)\ket{z_j}\bra{z_j}\;.\label{eq:HG}
\end{equation}
With the   initial condition  $\ket{\psi(0)}=\ket{S}$. In the case where all impurity energies are equal to each other, $\{\cE(z_j)=-n\}_{j=1}^{M}$, and $\B=1$ the Hamiltonian $H_{\rm G}$  is a generalization of the analog version of Grover search \cite{farhi1998analog} for the case of  $M$ target states. The system performs Rabi oscillations between the initial state $\ket{S}$ and  the state which is an equal superposition of all marked (solution)  states. Time to solution is the half-period of the oscillations, the "Grover time" $t_{\rm G}$
\begin{equation}
t_{\rm G}=\frac{\pi}{2n\B} \sqrt{ \frac{2^n}{M}}\;.\label{eq:Grover}
\end{equation}
Hamiltonian versions of Grover search with transverse field driver whose ground state were tuned at resonance with that of the solution state were considered in \cite{farhi2000quantum,childs2002quantum}.

Robustness of the Grover algorithms to phase noise was considered previously in the case of a single marked  state \cite{long2002phase,shenvi2003effects}. Here we investigate the role of 
systematic phase errors in quantum oracle for the case of multiple  solutions by assuming that  marked state energies   take distinct values $\cE(z_j)=-n+\e_j$
randomly   distributed over some  narrow  range  $W$. We will also investigate the systematic error in the Grover diffusion operator\cite{grover1997quantum}. In the Hamiltonian formulation \cite{farhi1998analog}
this corresponds to the deviation from unity of the parameter $\B$  that controls the weight of the driver in (\ref{eq:Grover}). We will define
\begin{equation}
\B=1-\frac{\e_0}{n}\;,\label{eq:e0}
\end{equation}
where  $\e_0$ is the driver error. 

We denote the computational basis states as $\ket{j}\equiv \ket{z_j}$ with $j\in[1,N],\,\,N=2^n$ and assume that marked states correspond to the range  $j\in[1,M]$. We also introduce the state 
 %\begin{equation}
$ \ket{0}=\frac{1}{\sqrt{N-M}}\sum_{j=M+1}^{N}\ket{j}$
% \end{equation}
that  is  orthogonal  to all  the marked states. The subset of  basis vectors $\scS=\{\ket{j}\}_{j=0}^{M}$ spans the $M+1$ dimensional subspace  with the remaining set $\scS_{\perp}$ of  basis vectors spanning the  orthogonal $N-M-1$ dimensional subspace. One can show that  $H_G$ does not have matrix elements that couple  $\scS $ with $\scS_{\perp}$.

%We define the deviation of the transverse field from unity
Assuming that $N\gg M$ one can  consider the  decay of the  state $\ket{0}$ instead of the state $\ket{S}$. 
We use (\ref{eq:e0}) and omit constant terms and small corrections $\cO(M/N)$ in $H_G$.  The  non-zero matrix elements  $H_{G}^{ij}=\bra{i}H_G\ket{j}$ 
in this subspace $\scS$ have the form 
%\begin{equation}
%H_{G}=\sum_{j=0}^{M}\e_j\ket{j}\bra{j}+V\sum_{j=1}^{M}(\ket{j}\bra{0}+h.c.),\;\; V=-\frac{n}{\sqrt{N}}\nonumber
%\end{equation}
\begin{equation}
H_{G}^{jj}=\e_j,\quad  H_{G}^{j0}= -(1-\delta_{j0})V,\quad V=n 2^{-n/2}\;,\label{eq:HijG}
\end{equation}
where  $j\in[0,M]$ and $H_{G}^{j0}=H_{G}^{0j}$.
On a time scale $t\ll1/\de=M/W$ much smaller than the inverse spacing of the  energies $\e_j$
the quantum evolution with  initial condition $\ket{\psi(0)}=\ket{0}$ corresponds to the decay of the discrete state with energy $\e_0$ into the  continuum \cite{mahan2013many} with the finite spectral width $W$ \cite{kogan2006analytic}. It is a similar problem to that  discussed in the Sec.~\ref{sec:FGR}.

%We also note that in the analogue Grover search  with many marked states whose energies  are distributed in  the range  $W\simt_{\rm G}^{-1}\sim 2^{-n} \sqrt{M}$ the time of the algorithm
%can be chosen in the broad range  such that $\exp(-t/t_{\rm G})=\cO(1)$. This is a consequence of the ffact that the atc that the  ich that anywhere on the scale does not have to be adjusted near $t_{\rm G}$ as in the usual Grover search corresponding  to $W=0$.

\subsubsection{Sensitivity to  systematic oracle phase error}

We first consider the case of  relatively large oracle errors (wide energy band $W$) 
\begin{equation}
V\sqrt{M}\ll W \ll V M\;,\label{eq:Grover-dec}
\end{equation}
and modest driver errors 
\begin{equation}
\e_0=n(1-\B) \apprle W\;.\label{eq:Bconst}
\end{equation}
In this case, following the results of the Sec.~\ref{sec:FGR} on the  solution of the Fano-Andreson model  \cite{kogan2006analytic} we obtain an exponential decay of the initial amplitude
(cf. (\ref{eq:exp-d}))
\begin{equation}
\psi_{0}(t) \simeq \exp\left[ -\Sigma_{0}^{\prime\prime} t-i\e_0 t-i \Sigma^{\prime}_{0}(\e_0+i 0^+)t \right]\;.\label{eq:cont-dec}
\end{equation}
where $\Sigma_{0}(z)=\Sigma_{0}^{\prime}(z)+i \Sigma_{0}^{\prime\prime}(z)$  is a self-energy
and 
\begin{equation}
\Sigma_{0}(z)=V^2\sum_{m=1}^{M}\frac{1}{z-\e_m},\quad \Sigma_{0}^{\prime\prime}\equiv\frac{1}{2}\Gamma_0=\frac{\pi V^2}{W/M}\;.\label{eq:S0r}
%\end{equation}
\end{equation}
The state $\ket{0}$  undergoes an exponential decay  with the  rate $\Gamma_0=2\Sigma_{0}^{\prime\prime}$. After the characteristic  time $\pt\sim 1/\Gamma_0$  the population is transferred into a  subset of the marked states with energies inside the window $|\e_j-\e_0|\simeq \Sigma_{0}^{\prime\prime}\ll W$.

 The number of marked states (solutions)  to which the population is transferred is $\Omega\sim \Sigma^{\prime\prime}_{0}/\de$.
 The relation between  $\pt$ and  $\Omega$ is
 \begin{equation}
\pt \sim \frac{1}{V\sqrt{\Omega}},\quad \Omega\sim  \left(\frac{V}{W/M}\right)^2\;,\label{eq:GG} \end{equation}
 the same  as in the Grover algorithm (\ref{eq:Grover}). It also recovers the scaling with $\Omega$ and $n$, up to a factor $\exp(-n/(2\B^2))$, for the  time of PT  considered in the rest of the paper that uses  transverse field as a driver  and starts from any  marked state instead of a fully-symmetric state.

 To characterize the effect of oracle errors we introduce the scaling  ansatz  for the marked states bandwidth $W\sim 2^{-n/2} M^{\gamma/2}$ similar to that in   (\ref{eq:param}).  We observe that the number $\Omega$ of solution states populated over the time $\pt$  cannot be greater than $M$ by construction.  For  $W\apprle V\sqrt{M}$ (or $\gamma<1$) the value of $\Omega\simeq  M$ and the scaling of the transfer time $\pt$ with $M$ is the same as $t_{\rm G}$ in the ideal  Grover  algorithm (\ref{eq:Grover}). In the region given by  (\ref{eq:Grover-dec}) (or $2>\gamma>1$)  the algorithm performance is degraded because $\Omega\ll M$. For  {\color{black}  $W\gg V M$ (or $\gamma>2$) }% $W\ll V$ (or $\gamma>2$) 
 the algorithm fails to find even one solution.

\subsubsection{Sensitivity to  the systematic driver error}

We now consider the sensitivity of the algorithm to an error in  the  weight of the driver Hamiltonian, i.e., to the nonzero value of the parameter $e_0=n(1-\B)$ (\ref{eq:e0}).  We assume that 
$\e_0\gg W$ while the spread of the marked state energies
%, $W\lesssim V\sqrt{M}$
{\color{black}  the condition (\ref{eq:Grover-dec})}, so that absent driver errors, PT time  would follow a Grover-like scaling law (\ref{eq:GG}).

%This case is bounded by the condition (\ref{eq:Grover-dec}).  
In this case the state $\ket{0}$ is coupled non-resonantly to a continuum with narrow bandwidth.  The   expression for the population transfer  to the marked states can be obtained  from the time-dependent perturbation theory in the parameter $\e_0/W$
\begin{equation}
\sum_{m=1}^{M}|\psi_{m}(t)|^2=\frac{2 M V^2}{\e_0^2}\(1-\cos(\e_0 t)\frac{\sin( Wt/2)}{W t/2}\right)\;.\nonumber
\end{equation}
Maximum transfer occurs at the time $t_0=\pi/\e_0$ with the total transferred probability $p_0=4M V^2/\e_0^2 $. Typical time $\pt\simeq t_0/p_0$ to achieve the successful population transfer to marked states involves repeating the experiment $1/p_0$ times 
\begin{equation}
\pt=\frac{1}{\Gamma_{0}}\,\frac{\pi^2\e_0}{W}\;,\label{eq:offres}
\end{equation}
where $\Gamma_0$ is given in (\ref{eq:S0r}) and the first multiple in r.h.s gives the typical transfer  time in the absence of driver errors. The later leads to an increase of the transfer time by a large factor $\e_0/W$.

For  the maximum possible bandwidth  $W$ when nearly all states are populated, $W$ $\sim$ $\Gamma_0$$\sim$$V\sqrt{M}$,   the time of population transfer (\ref{eq:offres}) is 
%\begin{equation}
%\pt \sim t_{\rm G}\frac{1}{\e_0 t_{\rm G}} \quad (\e_0 \gg t_{\rm G}^{-1}\sim V\sqrt{M})\;.\label{eq:tptG}
%\end{equation}
{\color{black}
\begin{equation}
\pt \sim t_{\rm G} \left(t_{\rm G} \e_0 \right)  \quad (\e_0 \gg t_{\rm G}^{-1}\sim V\sqrt{M})\;.\label{eq:tptG}
\end{equation}
}
As expected, when  the driver error exceeds inverse  Grover time $1/t_{\rm G}$ the performance of analogue Grover  algorithms  (\ref{eq:HG})  degrades  relative  to $t_{\rm G}$. 
 This is a direct consequence of the fact that the quantum evolution  begins from fully symmetric state which is a ground state  of the driver Hamiltonian whose energy is tuned at resonance with the marked states. In this case  the transverse field Hamiltonian driver effectively corresponds to  the projector (\ref{eq:HG}). Because the ground state is not degenerate, the resonance region is exponentially narrow ($\sim 2^{-n/2}\sqrt{M}$). This  results in the   exponential  sensitivity of the Grover algorithm performance  to the value  of driver weight. This critical behavior was studied in the work on quantum spatial search \cite{childs2004spatial} for the case of one marked state.

In contrast, in the PT protocol considered earlier in the paper there was no need to fine-tune  the value of $\B$ other than making  it  large, $\B\gg1$. 
This happened  because   the effective coupling between the marked states described by the down-folded Hamiltonian $\scH$ (\ref{eq:IBH}) was not due to any one particular eigenstate of the driver (such as the state $\ket{S}$ for the Grover case). Instead this coupling was 
formed due to an exponentially large (in $n$)  number of  non-resonant,  virtual  transitions between the marked states and highly  exited states  of the transverse field Hamiltonian $H_D$.  This resulted in a significant  improvement in robustness for the proposed PT relative to the analogue Grover  algorithm.

\subsection{\label{sec:Grover-marked} Grover search starting from a marked state}

We now consider an implementation of the analogue Grover search that starts from the marked state similar to the PT protocol considered in previous Sections. The transition amplitude $U_{ij}(t)=\bra{i} \exp(-i H_{\rm G} t)\ket{j}$ between the two marked states  can be written in the form
%\begin{equation}
%U_{ji}(t)=V^2\sum_{\beta=0}^{M} \frac{e^{-i\lambda_\beta t}}{Z(\lambda_\beta) (\lambda_\beta-e_i)(\lambda_\beta-e_j)}\;,\label{eq:Uij}
%\end{equation}
\begin{equation}
U_{ji}(t)=\sum_{\lambda} e^{-i\lambda t} {\psi_\lambda}(i){\psi_\lambda}(j)\;.\label{eq:Uij}
\end{equation}
Here  $\psi_\lambda(j)=\braket{j}{\psi_\lambda}$ are amplitudes of the eigenstates of $H_{\rm G}$ in  the $M+1$ dimensional subspace and $\lambda$ are
the corresponding eigenvalues  that obey the equation 
\begin{equation}
\lambda=\e_0+\sum_{j=1}^{M}\frac{V^2}{\lambda-e_j},\quad \psi_\lambda(j)=\frac{V}{\lambda-\e_j}\frac{1}{\sqrt{Z_\lambda}}\;. \label{eq:Geig}
\end{equation}
Here
\begin{equation}
Z(\lambda)=1+\sum_{m=0}^{M}\frac{V^2}{(\lambda-\e_m)^2}\;.\label{eq:Z}
\end{equation}
Instead of providing a detailed analysis of the above solution we  provide  an order of magnitude estimate to extract the relevant scaling behavior. 
We again assume that   the spread of the marked state energies, $W=t_{\rm G}^{-1}=\cO(V \sqrt{M})$ corresponds to the inverse of the Grover time  $t_{\rm G}$ needed  to find any one of the solutions with equal probability.  The typical separation between the adjacent vales of $\e_j$ is  $\de=W/M\sim V/\sqrt{M}$. 

It follows from (\ref{eq:Geig}) that in the {\it ordered} array obtained by  combining  together the sets of  energies $\{e_j\}_{j=0}^{M}$  and eigenvalues $\{\lambda_m\}_{m=0}^{M}$ their values  appear alternatively and sequentially, e.g., $\e_{j-1}<\lambda_{j}<e_{j}<\lambda_{j+1}$. The typical separation between the adjacent elements   in the array is $|\lambda_{j}-\e_j|\sim \de$. We observe that for a given value of $\lambda$ the  sum in the expression for $Z(\lambda)$ (\ref{eq:Z}) is dominated by  the small, $\cO(1)$,  number of  terms  with  $|\e_m-\lambda |\sim \de$, each term of the order of $M$.  Indeed, there are  $\cO(M)$ remaining terms   corresponding to   $|\e_m-\lambda |\sim W$. The  magnitude of those terms is $V^2/W^2\sim 1/M$ and their aggregated contribution to the sum is $\cO(1)$. Therefore we can estimate  $Z(\lambda)=\cO(M)$ and for the amplitudes we have
\begin{equation}
\psi_\lambda(m) \sim  \frac{V}{\lambda-\e_m}\frac{1}{\sqrt{M}}, \quad m=i,j\;.\label{eq:c-est}
\end{equation}
For a given initial state $\ket{i}$ at time $t$ we pick the final state $\ket{j}$ within the energy window $\e_j-\e_j\sim \Delta = 1/t$ around $\e_i$. The sum in the expression (\ref{eq:Uij}) for the transition amplitude $U_{ji}(t)$ is dominated by the number of terms $\Omega= \Delta/\de\sim\Delta\sqrt{M}/V$ corresponding to the  eigenvalues $\lambda$ inside  the same  window of energies. For those terms  $\lambda-\e_i,\;  \e_j-\lambda\sim\Delta$ giving the estimate for the amplitudes $\psi_\lambda(i),\;\psi_\lambda(j)\sim 1/\Omega$ (cf. (\ref{eq:c-est})). The magnitude of the sum in (\ref{eq:Uij}) can be estimated as $|U_{ij}(t)|\sim\Omega|\psi_\lambda(i)\psi_\lambda(j)|\sim1/\Omega$. On the other hand, because ordered values of $\lambda$ and $\e_m$ alternate in sequence the probability $|U_{ij}(t)|^2$ is distributed over $\Omega$ marked states and $|U_{ij}(t)|\sim\Omega^{-1/2}$. By equating the  above two estimates for $|U_{ij}(t)|$ we immediately obtain $\Omega\sim 1$ and therefore
\begin{equation}
\Delta =\frac{1}{t} \sim\de\sim  \frac{V}{\sqrt{M}},\quad (\Omega\sim 1).\label{eq:DO-est}
\end{equation}

In the case when there are only a few marked states ($M\sim 1$ and $W\sim V$)  the  
probability is initially localized on a given marked state $\ket{i}$ and then it spreads  over to others states  separated in energy by $V$ during the time  $t_{\rm G}\sim 1/V\sim 2^{n/2}$. In this case the algorithm 
 time scales with $n$ identically to that of the analogue Grover search that starts at the fully symmetric state $\ket{S}$. Similar performance is achieved by the PT protocol using transverse field $\B\gg1$ and discussed
 in previous sections. 

The difference from analogue Grover search starting at $\ket{S}$ from the above PT protocol using a transverse field becomes  dramatic for large number of marked states $M\gg1$. Both analogue Grover search and the PT protocol  benefit from the increase in $M$: the algorithmic time shrinks $\propto 1/\sqrt{M}$ and the number of marked (solution) states $\Omega$  in the number of states in the final superposition increases  with $M$.

In contrast, the quantum search with $H_{\rm G}$ starting form the marked state $\ket{i}$ does not create massive superpositions of solution states when $M$ increases. Instead it involves  a very few others states that are adjacent in energy, $|\e_j-\e_i|\sim V/\sqrt{M}$.  The time of the algorithm increases with $M$ (\ref{eq:DO-est}). This happens because unlike the Hamiltonian $H$ with a transverse field (\ref{eq:H}), the Hamiltonian $H_{\rm G}$ is integrable. The wave-function remains localized near the initial marked state.

\section{\label{sec:conc}Conclusion}

We analyze
the computational role of coherent multiqubit tunneling {\color{black}that gives rise to  %when a large
%umber of tunneling paths interfere constructively.  This gives rise to
bands of nonergodic delocalized quantum  states as a coherent pathway
for population transfer (PT) between computational states with close 
energies}.  In this regime PT cannot be efficiently simulated by QMC.  

We consider optimization problems with an energy function $\cE(z)$ defined over the set of $2^n$ $n$-bit-strings $z$. 
We define a computational primitive  with the objective to find   bit-strings  $z_j\neq z_i$  inside some narrow energy window $\Delta \cE_{\cl}$ around the energy of the initial bit-string $z_i$.
The problem is hard for sufficiently low   starting   energy $\cE(z_i)$ in the region  proliferated by  deep local minima   that are separated by  large Hamming distances.

We propose to solve this problem  using the  following quantum population transfer (PT) protocol: prepare the system in a computational state $\ket{z_j}$ with classical energy $\cE(z_j)$, then evolve it with the transverse-field quantum spin Hamiltonian. Classical energies $\cE(z)$ are encoded in the problem Hamiltonian diagonal in the basis of states $\ket{z}$ similar to quantum annealing (QA) approaches \cite{kadowaki1998quantum,farhi2001quantum,brooke1999quantum}. A key difference  from  QA or analogue quantum search Hamiltonians   \cite{farhi1998analog,childs2004spatial} is that the transverse field is kept constant throughout  the algorithm and  is not fine-tuned to any particular value.   At the final moment of PT  we
projectively measure in the computational basis and check if the
outcome $z$ is a ``solution'', i.e.,  $z \ne z_j$, and the energy
$\cE(z)$ is inside the window $\Delta \cE_{\cl}$.

  In this paper we  analyzed PT dynamics in  Impurity Band (IB) model with a ``bimodal'' energy function: $\cE(z)=0$  for all states  except for  $M$ ``marked''  states  $\ket{z_j}$ picked at random with energies forming a narrow band of the width $W$ separated by a large gap $\cO(n)$ from the rest of the states.  This landscape is  similar to that in analogue Grover   search   \cite{farhi1998analog,farhi2000quantum} with multiple target states and a distribution of oracle values for the targets. The best known classical algorithm for finding another marked state has cost $O(2^n/M)$. 

The transverse field gives rise to tunneling between a pair of marked states  corresponding to a  sum over a large number of virtual transitions connecting the two marked states via the states with $\cE(z)=0$.  As a result the PT dynamics  is described by the down-folded
$M\times M$ Hamiltonian $\scH$ that is dense in the space of the marked states $\ket{z_j}$. Its  off-diagonal matrix elements  $\scH_{ij}=V(d_{ij})\cos\phi(d_{ij})$ depend only on the Hamming distance $d$ and are obtained using  WKB method. The distribution of matrix elements $\scH_{ij}$  has a heavy tail decaying as a cubic power for $V(d)\gg V_{\rm typ}$. This is a remarkable result of the competition between the very steep decay of the off-diagonal tunneling matrix element with the Hamming distance $d$, and the steep increase in the number of marked states $M_d\propto\binom{n}{d}$ at distance $d$. We emphasize that such polynomial tail in the distribution of matrix elements is only possible either in infinite dimension or in presence of long-range interactions (e.g, dipolar glass). 

The dispersion of the diagonal elements  $\scH_{jj}=\cE(z_j)$ is expected to be large, $W\sim V_{\rm typ} M^{\gamma/2} \gg V_{\rm typ} $ with $\gamma\in\left[1, 2\right]$. Therefore we call  $\scH_{ij}$ a Preferred Basis Levi matrix (PBLM), a generalization of the Levi matrix from the random matrix theory. We demonstrate two localization transitions in the PBLM ensemble whose locations are determined by the strong hierarchy of elements of the PBLM  $\scH_{ij}$. In the range $1<\gamma<2$ there  exist  minibands of non-ergodic delocalized eigenstates of $\scH$. Their  width is proportional to $1/\pt \ll W$.   Each miniband associated with a support set $\scS$ over the marked states.   If  $\gamma>2$ then  $W$ exceeds the largest matrix element of  $\scH_{ij}$ and the support set is empty, all  eigenstates  are localized. If $\gamma<1$  then  $W$ is smaller than the typical largest element in a  row of $\scH_{ij}$ and the support set extends to all marked states -- all eigenstates   are ``ergodic''.

We find the distribution of the miniband width $\Gamma=1/\pt$ analytically by solving the non-linear cavity equations for an ensemble of PBLMs. Unlike previous analyses focused on linearized cavity equations near the Anderson transition, we find the solution of the fully non-linear cavity equations in the non-ergodic delocalized phase. 

The distribution of miniband widths $\Gamma$  obeys alpha-stable Levi law with tail index 1. The typical value of $\Gamma$ and its  characteristic variance  exceeds the typical matrix element of $\scH$ by  a factor $\Omega^{1/2}$ where $\Omega=(M V_{\rm typ}/W)^2$ is a size of the support set in a typical miniband.

We demonstrate that quantum PT finds another state within a target window of energies $\Omega$ in time $\pt \propto 2^{n/2} \Omega^{-1/2} \exp( n/(2B_\perp^2))$. The scaling exponent of $\pt$ with $n$ differs from that in Grover's algorithm by a factor $\propto B_\perp^{-2}$, which can be made small with large transverse fields $n \gg B_\perp^2 \gg 1$. 

Crucial distinctions between this case and the Hamiltonian in the  analogue version of Grover's algorithm \cite{farhi1998analog}  for the case of multiple target states are the
non-integrability of our model, and the delocalized nature of the
eigenstates within the energy band $W$.  Furthermore, analogue 
Grover's algorithm for multiple targets is  exponentially sensitive in $n$ to the
weight of the driver Hamiltonian,  and cannot be initialized with a
computational basis state. 

The model  (\ref{eq:H}) considered in the paper belongs to the class of n-local infinite range spin glasses similar to   quantum Random Energy Model in transverse field \cite{PhysRevB.41.4858}. However the key feature of our analysis  --transport via miniband of non-ergodic delocalized states at the tail of the density of states dominated by deep local minima -- 
is ubiquitous to a  broad class  of quantum spin glass models (\ref{eq:H}), such  as transverse field Sherrington Kirkpatrick, p-spin model \cite{kirkpatrick1987p}, K-Satisfiability, etc.

In the above models one can identify    two distinct energy scales. The first scale is the  typical change in classical energy  corresponding to one bit flip: $\cE_{\rm flip} \apprge \B$. The second scale is the typical width of  non-ergodic minibands $\Gamma < \Delta \cE_{\rm cl}$, which decreases   exponentially with $n$.
The tunneling transitions between the states inside the  miniband  require  a large number of spin flips, and therefore 
$\cE_{\rm flip} \gg \Gamma$. Starting from the initial state $\ket{z_i}$ inside the strip of energies $\Delta \cE_{\rm cl}$,  the quantum evolution   is confined within the corresponding miniband.  The quantum  PT can be described by an effective down-folded Hamiltonian $\scH_{ij}$ 
defined over a subset of computational basis states whose classical energies lie within the energy strip $\Delta \cE_{\rm cl}$ at the tail of the density  of states. 

We note that once a computational  problem contains a structure, the associated   minibands   can be organized in a   more complex way than in the IB model  considered in our paper. E.g., the population transfer can proceed via the tree of resonances \cite{altshuler2016nonergodic,altshuler2016multifractal}. In the structured problems the typical  tunneling matrix elements $\cH_{ij}$  can be exponentially greater in $n$ than those in the Grover algorithm  and than  the transition rates in  the classical local search algorithms.
Extensions of our approach  for the analysis of  the computational complexity of Population Transfer for generic spin glass models  presents a promising direction for  the future research.

\begin{acknowledgments}
Authors  are grateful to Edward Farhi, Lev Ioffe, Vladimir Kravtsov, Christopher Laumann, and  Antonello Scardicchio for the fruitful discussions of this work. K.K. acknowledges support by NASA Academic Mission Services, contract number NNA16BD14C.  This research is based upon work supported in part by the AFRL Information Directorate under grant F4HBKC4162G001 and the Office of the Director of National Intelligence (ODNI) and the Intelligence Advanced Research Projects Activity (IARPA), via IAA 145483. The views and conclusions contained herein are those of the authors and should not be interpreted as necessarily representing the official policies or endorsements, either expressed or implied, of ODNI, IARPA, AFRL, or the U.S. Government. The U.S. Government is authorized to reproduce and distribute reprints for Governmental  purpose notwithstanding any copyright annotation thereon.

\end{acknowledgments}.

\newpage
\bibliographystyle{IEEEtran}

\bibliography{mbdl-database}

\appendix%*

%\section*{Appendix}

%%%%%%%%%%%%%%%%%%%%%%%%APPENDIX%%%%%%%%%%%%%%%%%%%%%%%%%%%%%%%%%

\section{\label{sec:dfH} Matrix elements of the downfolded  Hamiltonian and the normalization condition for its eigenvectors}
We introduce eigenstates $\ket{x}$ of the transverse field (driver) Hamiltonian 
\begin{equation}
H_D=- \B \sum_{j=0}^n \sigma_j^x=\sum_{x\in\stn}H_D^x \ket{x}\bra{x},\label{eq:V-app}
\end{equation}
Here
\begin{equation}
\ket{x}=\ket{x^1}\otimes\ldots\otimes\ket{x^n}\;,\label{eq:eigen-x-app}
\end{equation}
where $ \ket{x^k}$ is the state of $k$th qubit  such that $\sigma_x \ket{x^k}= (1-2x_k)\ket{x^k}$ and $x$-bits take values $x^k =0,1$. Also 
\begin{equation}
H_{D}^{x} =-\B \,(n-2 h_x),\quad h_x=  \sum_{k=1}^{n} x^k,\label{eq:Vx-app}
\end{equation}
where $h_x$ is a Hamming weight of the bit-string $x$ and $-\B \,(n-2 h),\,h\in(0,n)$ are eigenvalues of  $H_D$.

We expand  the eigenstates $\ket{\psi}$ of  the system Hamiltonian $H$ (\ref{eq:H})  into the basis of the 
 eigenstates $\ket{x}$
\begin{equation}
\ket{\psi}=\sum_{x\in \stn}\Psi(x)\ket{x}\;.
\end{equation}
We write the  Schrodinger equation $H\ket{\psi}=E\ket{\psi}$ in the form
\begin{equation}
H_D \ket{\psi}+\sum_{j=1}^{M}\cE(z_j)\ket{z_j}\psi(z_j) =E \ket{\psi}\;,\label{eq:Hst-app}
\end{equation}
where $\psi(z_j)=\braket{z_{j}}{\psi}$. 
 Then we multiply it  from the  left by $\bra x$ and obtain $\Psi(x)$ in terms of $\psi(z_j)$
\begin{align}
\Psi(x) =\frac{\sum _{j=1}^M \cE(z_j)\upsilon_{x,j} \psi(z_j)}{ E- H_{D}^{x} }\;.\label{eq:psix-app}
\end{align}
In Eq.~(\ref{eq:psix-app}) the coefficients  $\upsilon_{x,j} =\braket{x}{z_j}$ equal
\begin{align}
\upsilon_{x,j} =2^{-n/2}(-1)^{x\cdot z_j},\quad x\cdot z_j \equiv  \sum_{k=1}^{n}x^k z_j^k,\label{eq:upsilon-xj-app}
\end{align}
and  $z_j^k=0,1$.

We now multiply Eq.~(\ref{eq:Hst-app}) from the left by $\bra{z_j}$ where  $j \in(1,M)$ enumerates marked states and obtain
\begin{equation}
\sum_{x\in \stn } H_{D}^{x}\, \Psi(x)\upsilon_{x,j}=(E-\cE(z_j))\braket{z_j}{\psi}\;.
\end{equation}
Plugging here the expression for $\Psi(x)$ (\ref{eq:psix-app}) the matrix eigenvalue problem (\ref{eq:Hst-app})  we obtain 
\begin{equation}
\cE(z_i)\psi(z_i)-\sum_{j=1}^{M} \cE(z_j) c_{i j}(E) \,\psi(z_j)=E \psi(z_i) \label{eq:Hc0-app}
\end{equation}
where
\begin{equation}
c_{i j}(E)=\sum_{x\in \stn}\upsilon_{x,i}\upsilon_{x,j}\frac{H_{D}^{x}}{E- H_{D}^{x}},\label{eq:c_ij-app}
\end{equation}
Because $H_{D}^{x} $ depends on a bit-string $x$ only via its Hamming weight $\sum_{j=1}^{n} x^j $ one can perform the partial summation in (\ref{eq:c_ij-app}) getting
\begin{equation}
c_{i j}(E)\equiv c(E,|z_i-z_j|),\quad |z_i-z_j|=\sum_{k=1}^{n}|z_i^k-z_j^k |, \label{eq:c_ij_d-app}
\end{equation}
where the function $c(E,d)$ has he form
\begin{equation}
 c(E,d)=\sum_{k=0}^{n-d}\sum_{l=0}^{d}\binom{n}{k}\binom{n-d}{l}\frac{(-1)^l \,2^{-n}}{1+\frac{E}{\B(n-2 k-2 l)}}.\label{eq:CEd-app}
\end{equation}
Above $|z_i-z_j|$ denotes the Hamming distance between the bit-strings $z_i$ and $z_j$.
We introduce the rescaling 
\begin{equation}
\psi(z_i)=\frac{A_i}{\sqrt{\cE(z_i)}},\quad i\in[1..M]\;. \label{Ak}
\end{equation}
Then Eq.~ (\ref{eq:Hc0-app}) can be written in the form
\begin{equation}
\sum_{j=1}^{M}\cH_{ij}(E)A_j=E A_i,\label{eq:Hc1-app}
\end{equation}
where  $\cH_{ij}$ is a symmetric $M\times M$ matrix
\begin{equation}
\cH_{ij}(E)=\delta_{kj} \cE(z_i) +\sqrt{\cE(z_i)\cE(z_j)} c(E, {d_{ij}}),\label{eq:Hc2-app}
\end{equation}
indices $k,j=1$:$M$ and $\delta_{kj} $ is Kronecker delta. This is a nonlinear eigenproblem given in the main  text, Eq.~(\ref{eq:Hc2}).

We note that  the projections of the eigenvectors  of  $H$ onto  the  marked state subspace are not, in general, normalized nor they are orthogonal.
Let us consider the eigenstate $\ket{\psi_\beta}$ and the corresponding eigenvalue $E_\beta$ of $H$. We calculate the corresponding amplitude $\Psi_\beta(x)$  using Eq.~(\ref{eq:psix-app}) and plug it into the  normalization condition
\begin{equation}
\sum_{x\in \stn}\Psi_\beta^2(x)=1\;,
\end{equation}
obtaining after  partial summation 
 \begin{equation}
 \sum_{i,j=1}^{M}\cE_i\cE_j \,r(E_\beta,d_{ij})\psi_\beta(z_i)\psi_\beta(z_j)=1\;,\label{eq:EEr-app}
 \end{equation}
 where the coefficient $r(E,d)$ equals
 \begin{equation}
 r(E,d)=2^{-n}\sum_{k=0}^{n-d}\sum_{l=0}^{d}\frac{ (-1)^{k} \binom{d}{k}\binom{n-d}{l} }{ \B(n-2(k+l)+E)^2}\;.
 \end{equation}
 It can be written in the form
 \begin{equation}
 r(E,d)=\frac{\partial}{\partial E}\left(\frac{c(E,d)-\delta_{d,0}}{E}\right)\;,
 \end{equation}
 where $\delta_{d,0}$ is the Kronnecker delta. We use (\ref{eq:Hc2-app}) and write
 \begin{equation}
 r(E,d_{ij})=\frac{1}{\sqrt{\cE_i\cE_j}}\frac{\partial \cH_{ij}(E)}{\partial E}\;.
\end{equation} 
 We now define  the coefficients $Q_{ij}(E)$ such that
  \begin{equation}
\frac{1}{ Q_{ij}(E)}=\cE_i\cE_j\, r(E,d_{ij})=\sqrt{\cE_i\cE_j}\,\frac{\partial \cH_{ij}(E)}{\partial E}\;.\label{eq:Qij-app}
\end{equation}
Then Eq.(\ref{eq:EEr-app}) takes the form
\begin{equation}
\sum_{i,j}\frac{1}{Q_{ij}(E)}\psi_\beta(z_i)\psi_\beta(z_j)=1\;.\label{eq:normA-app}
\end{equation}
 The above equations   (\ref{eq:Qij-app}) and  (\ref{eq:normA-app})  correspond to  Eqs.~(\ref{eq:normA}) and (\ref{eq:Qjk}) of the main text.

\section{\label{sec:det-wkb} Details of the WKB analysis of the coupling coefficients}

%\subsection{\label{sec:n/2}Coupling coefficient for  $d=n/2$}

%\subsection{\label{sec:c10-large} Coupling coefficient for large values of $d={\cal O}(n)$}

 In the main text we expressed the  coupling coefficient $c(E,d)$  in terms of the  off-diagonal matrix elements of the resolvent (\ref{eq:GE}) of the transverse field Hamiltonian  $H_D$ between the states that belong to  a maximum total spin subspace $S=n/2$. The results are given in 
  the expressions (\ref{eq:c-g}), 
 (\ref{eq:3term}) from the main text repeated below for convenience
 \begin{equation}
c(E,d)=\delta_{d,0}-\frac{E}{\sqrt{\binom{n}{d}}} G_{\frac{n}{2} -d,\frac{n}{2}}(E)    \;. \label{eq:c-g-app}
\end{equation}
Here the resolvent $G_{\frac{n}{2} -d,\frac{n}{2}}(E) $ obeys the inhomogeneous  equation
 \begin{equation}
\delta_{m,\frac{n}{2}}+\sum_{s=\pm 1}u(m-s/2)  G_{{m+s},\frac{n}{2}}=E G_{m,\frac{n}{2}},\label{eq:3term-app}
\end{equation}
\vspace{-0.1in}
\begin{equation}
u(m)=-\B\sqrt{L^2-m^2},\quad L=\frac{n+1}{2}.\label{eq:u-app}
\end{equation}

We will solve the above equations  for the  case where the  energy $E$ of the resolvent is not far from the center of the Impurity Band 
\begin{equation}
E=-n+\Delta,\quad \Delta=\cO(n^0)\;.\label{eq:EnD}
\end{equation}
The WKB solution to Eq.~(\ref{eq:3term-app})   is sought in the exponential form 
\begin{equation}
G_{m,\frac{n}{2}}\propto \exp \left(i \int^{m} dk\, p(k)  \right).\label{eq:wkb-form}
\end{equation}
It is assumed that $\int^{m}_{0} dk\, p(k) ={\cal O}(n)$ and $|p(k)|={\cal O}(n^{0})$ so that $G_{m,\frac{n}{2}}$ is varying  steeply with  $m$ changing  by 1.  However  $|p^{\prime}(m)|={\cal O}(1/n)$ and $p(m)$ is varying very slowly with $m$  due to the similar property of the coefficients $u(m)/L$ in the Eq.~(\ref{eq:3term-app}). This property is at the root of WKB approximation \cite{braun1993discrete}.
The quantity $p$ corresponds to the \lq\lq momentum" of the 
 effective mechanical  system with coordinate $m$, energy $E$ and Hamiltonian function
 $u(m)\cos p$. The function $p=p(E,m)$ is obtained from the equation
 \begin{equation}
u(m)\cos p=E.\label{eq:Ep}
 \end{equation}
 This equation also defines the  curve on the $(m,E)$ plane with  $p=0$ shown in Fig.~\ref{fig:eff-pot}. Points on that curve are turning points of the classical motion   with  energy $E$.
 
 For  not too small transverse fields \begin{equation}
 \B>\frac{2L}{|E|}\simeq 1\;,\label{eq:largeB}
 \end{equation}
the Eq.~(\ref{eq:Ep})  has two types of WKB solutions that correspond to   real or  imaginary momentum $p(m)$ depending on the value of $m$ relative to the  turning points $m=\pm m_0(E)$ given below
\footnote{
The expression (\ref{eq:m0-app}) for $m_0(E)$ should be evaluated for
the energy  $E=E^{0)}\simeq -n-B_{\perp}^{2}$ corresponding to the  eigenvalues of impurity band (\ref{eq:E0}),(\ref{eq:Delta0as})
In this case the ${\cal O}(n^{0})$ corrections in   the r.h.s. of  (\ref{eq:m0-app}) vanishes. 
\begin{equation}
m_0(E^{(0)})\simeq \frac{n}{2}\sqrt{1-B_{\perp}^{-2}}+\frac{1-3\Bs}{4n\sqrt{1-B_{\perp}^{-2}}}\;.\nonumber
\end{equation}
The terms  ${\cal O}(1/n)$ above and can be neglected in the exponents of the WKB solutions (\ref{eq:f-wkb}) and (\ref{eq:f-wkb-1}).}
   \begin{equation}
  m_0=\sqrt{L^2-\left(\frac{E}{2\B}\right)^2}\;.    \label{eq:m0-app}
\end{equation}

%\textcolor{red}{To the leading order in $n$ the value of $m_0$ above is $E$-independent for $E={\cal O}(n)$. }\footnote{ \textcolor{red}{

  %To connect to the discussion elsewhere in the paper we shall  describe the WKB amplitudes  $f_m=f_{n/2-d}$ in terms of the Hamming distance $d$  (cf. (\ref{eq:m})).

 In the region
 \begin{equation}
 n/2+m_0> d> n/2-m_0,\label{eq:allowed}
 \end{equation}
 the amplitude    $G_{\frac{n}{2}-d,\frac{n}{2}}$ (\ref{eq:wkb-form}) is rapidly oscillating with $d$ and can be written in the form
%\begin{align}
 %f_{m} =B(E)\frac{ \sin\left (S(m_0(E),m) -\frac{\pi}{4}  \right )}{ (m_0^2(E)-m^2)^{1/4} } , \label{eq:f-wkb}
 %\end{align}
 \begin{align}
G_{\frac{n}{2}-d,\frac{n}{2}} =-\scC(E)\frac{ \sin\phi(E,d)}{ [m_0^2(E)-(n/2-d)^2]^{1/4} } , \label{eq:f-wkb}
 \end{align}
 where
 \begin{equation}
\phi(E,d)=\int_{n/2-d}^{m_0}dk\,\arcsin\left(\sqrt{\frac{m_0^2-k^2}{L^2-k^2}}\right)-\frac{\pi}{4}\;,\label{eq:phi-m}
\end{equation}
% \begin{equation}
% S(m_0,m)=\int_{m}^{m_0}dk\,\arcsin\left(\sqrt{\frac{m_0^2-k^2}{L^2-k^2}}\right).\label{eq:Swkb}
% \end{equation}
 is a phase of WKB solution and  $\scC(E)$ is the constant of integration that will be discussed  below.

 On the other hand, in the two regions 
 \begin{equation}
 d\in[0,n/2-m_0]\cup [n/2+m_0,n] \label{eq:forbid}
 \end{equation}
 the resolvent  $G_{\frac{n}{2}-d,\frac{n}{2}}$  is decreasing exponentially with $d$.
 For example,   in the  left  region 
\begin{equation}
G_{\frac{n}{2}-d,\frac{n}{2}}= \frac{\scC(E)}{2} \frac{e^{ |{\rm Im}\, \phi(E,d)|}}{[(n/2-d)^2-m_0^2(E)]^{1/4}}\;.\label{eq:f-wkb-1}
\end{equation}
%\begin{equation}
%f_m=-B(E) \frac{\exp\left(  |S(m, m_0(E))|\right)}{( m^2-m_0^2(E))^{1/4}  },\quad  \frac{n}{2}-m \gg1.\label{eq:f-wkb-1}
%\end{equation}
We omit here for brevity the expression in the right  region (\ref{eq:forbid}).

\subsection{Determination of the integration constant in WKB solution}
Within the  WKB approach the  integration constant   $\scC(E)$  can be obtained by matching the exponential asymptotic  (\ref{eq:f-wkb-1}) with the 
solution obtained near the boundary of the interval $d=0$. However as discussed in Sec.~\ref{sec:tails} of the main text,  for the  relevant range of the model parameters
the properties of the typical sample in  the ensemble of the IB Hamiltonians $\scH$ depend only on $G_{\frac{n}{2}-d,\frac{n}{2}}$ in  the  region of its oscillatory behavior (\ref{eq:allowed}) away from the boundaries of the interval $d=0,\,n$. To avoid the analysis in the region of no consequence for us  we determine $\scC(E)$ by equating  
the above WKB asymptotic for  $G_{\frac{n}{2}-d,\frac{n}{2}}$ at the center of the interval $d=n/2$ with  expression for $G_{0,\frac{n}{2}}$  at that point obtained  in a different way.% using the following integral arepresentation of the double sum in Eq.~(\ref{eq:CEd}) of the main text.

Using  Eq.~(\ref{eq:CEd}) we  write $c(E,n/2)$ in the integral form
\begin{equation}
c\left(E,\frac{n}{2}\right)= \frac{i E}{2^n  \B} \int_{0}^{\infty}d\tau (1-e^{4i\tau})^{n/2} e^{i(E/\B-n+i o)\tau}\nonumber
\end{equation}
($o\rightarrow +0$). The integral can be expressed in terms of the Gamma function $\Gamma(x)$. In the region of not too small  transverse fields (\ref{eq:largeB})
it has the form
\begin{equation}
c\left(E,\frac{n}{2}\right)=\frac{2^{1-n} \, \pi a (a^2-1)^{-1}\Gamma \left( \frac{n}{2} \right )  }{ \sin\left( \frac{\pi(a-1)n}{4a}\right) \Gamma\left(\frac{(a+1)n}{4a}\right)\Gamma\left(\frac{(a-1)n}{4a} \right)}.\label{eq:n2}
\end{equation}
where 
\begin{equation}
a=-\frac{n \B}{E} >1\;.\label{eq:a}
\end{equation}
Using Sterling  formulae for Gamma function we obtain in the limit $n\gg 1$, $a=\cO(n^0)$
\begin{align}
&c\left(E,\frac{n}{2}\right) =\frac{ \sqrt{n\pi} }{2 a\sin\left( \frac{\pi(a-1)n}{4a}\right)}2^{-n/2} e^{-n \theta(a)},\label{eq:c-n-2}\\
%& \theta=\frac{2\arccoth(\B)+\B \ln(1-B_{\perp}^{-2})}{4\B}\;.\label{eq:theta}
&\theta(a) =\frac{2\arctanh\left(\frac{1}{a}\right)+a\ln\(1-a^{-2}\)}{4a}\;.\label{eq:kappa}
\end{align}
For  large transverse fields $a\gg 1$ and we have  $\theta\simeq a^2/4$.

Using Eq.~(\ref{eq:c-g-app}) we obtain the asymptotic of the Green function at the zone center
\begin{equation}
G_{0,\frac{n}{2}}(E)=\(\frac{\pi}{8 n^3}\)^{1/4}\frac{\exp(-n \theta(a))}{\sqrt{\Bs-1}\,\sin\phi(n/2,E)}\;.\label{eq:G0int}
\end{equation}
Here we used the equality for the phase WKB $\phi(E,n/2)$ (\ref{eq:phi-m}) at the zone center
\begin{equation}
\phi(E,n/2)=\pi\frac{(a-1)n}{4a}\;.\label{eq:phi-n2}
\end{equation}
On the other  hand, from the WKB expression (\ref{eq:f-wkb})
we get
 \begin{align}
G_{0,\frac{n}{2}} =-\scC(E) \(\frac{2}{n}\)^{1/2}\,\frac{ \sin\phi(E,n/2)}{ (1-B_{\perp}^{-2})^{1/4} }\;. \label{eq:G0wkb}
 \end{align}
By comparing the Eqs.~(\ref{eq:G0int}) and (\ref{eq:G0wkb}) we finally obtain the constant of integration $\scC(E)$
\begin{equation}
\scC(E)=-\frac{\pi^{1/4}}{(32n \Bs(\Bs-1)^{1/4}}\frac{\exp(-n\theta(a))}{(\sin\phi(E,n/2))^2}.\label{eq:Cwkb}
\end{equation}

One can use   (\ref{eq:Cwkb}) in (\ref{eq:f-wkb}) and (\ref{eq:c-g})  to obtain the  expression for $c(E,d)$  in the region (\ref{eq:allowed}). Before providing  the result 
we observe that for energies $E$  not too far from the Impurity Band center (cf. Eq.~(\ref{eq:EnD})) the expression for $n\theta(a)$ can be expanded in powers of $1/n$
\begin{equation}
n\theta(a)\simeq n\theta(\B)-\frac{E+n}{2\B}\arccoth\B+\cO(n^{-1})\;.
\end{equation}
where $E+n\equiv \Delta=\cO(n^0)$.

Finally the expression for the coupling coefficient has the form
\begin{align}
c(E,d) =&\sqrt{A(E,d/n)} \,\,\frac{n^{\frac{1}{4} } \,e^{-n\theta(\B)}}{ \sqrt{\binom{n}{d}}}   \label{eq:c-app} \\
\times& \sqrt{2} \, \sin\phi(E,d)\;,\nonumber
%\cV(E,d) &=  \sqrt{\frac{A(E)}{\upsilon(d/n)\,\binom{n}{d}}} \;.\nonumber %\label{eq:c}
\end{align}
where  the WKB phase $\phi(E,d)$ is given in (\ref{eq:phi-m}) and the coefficient $A(E,\rho)$ equals
\begin{equation}
A(E,\rho)=\sqrt{\frac{\pi}{32 }} \frac{e^{\frac{E+n}{\B}\arccoth\B}}{(B_{\perp}^{2}-1)\upsilon(\rho)\sin^4(\phi(E,n/2))}\;,\label{eq:A-app}
\end{equation}
\begin{equation}
\upsilon(\rho)=\left(1-\frac{(1-2\rho)^2}{1-B_{\perp}^{-2}}\right)^{1/2}\;,\label{eq:upsilon-app}
\end{equation}
It is related to  $A(\rho)$ in the Eq.~(\ref{eq:A}) of the main text as follows: $A(\rho)= A(E^{(0)},\rho)$.
The phase  $\phi( E,n/2)$ in (\ref{eq:A-app}) has an explicit form
\begin{equation}
\phi( E,n/2)=\frac{\pi}{4}\left(n(1-B_{\perp}^{-1})+\frac{n+E}{\B}\right)\;.\label{eq:phi0-app}
\end{equation}
%The coefficient $\theta(E)$  that appears in  (\ref{eq:c}) is given in (\ref{eq:kappa}).

\subsection{Limit of large transverse fields $\bm{\B\gg1}$}
In the limit of large transverse fields the tuning point $m_0$ (\ref{eq:m0-app}) is very close to the boundary of the interval $m=L$ so that one has a small parameter 
\begin{equation}
\sqrt{\frac{L-m_0}{L}}=\frac{1}{\sqrt{8}}\frac{|E|}{L\B}\ll 1
\end{equation}
In this case the expression for the WKB phase takes a simple  form
\begin{equation}
\phi(E,d)=\frac{\pi d}{2}-\frac{\pi n}{4}\frac{\chi(E,d/n)}{\B}\;,\label{eq:phi-largeB-app}
\end{equation}
\vspace{-0.2in}
\begin{equation}
\chi(E,\rho)= \(1-\frac{\Delta}{n} \) \(1-\frac{2}{\pi}\tan^{-1}\frac{1-2\rho}{\sqrt{1-(1-2\rho)^2}}\)\nonumber \label{eq:chi}
\end{equation}
where $\Delta=E+n=\cO(n^0)$ and values of $d$ are not too close to the interval boundaries
\begin{equation}
n-d,\, d\gg L-m_0\sim \frac{n}{\Bs}\;.\label{eq:d-far}
\end{equation}
We note that for large transverse fields $\B\gg1$ the phase is a sum of the two terms. First term changes rapidly with $d$ with the slope $\pi/2$ and second term changes very little (by an amount $\cO(n^{-1})$) when $d$ is changed  by 1.

We note that unlike the study of the WKB eigenfunctions  where one has to  select the WKB solution that  decays into the classically forbidden region (\ref{eq:forbid}), 
the Green function  $G_{n/2-d,n/2}(E)$ corresponds to the  solution that {\it increases} exponentially with  $m=n/2-d>m_0$.  
Using the oscillating (\ref{eq:f-wkb}) and exponentially growing  (\ref{eq:f-wkb-1})  WKB solutions   one can obtain the  coefficient $c(E,d)$ from the relation (\ref{eq:c-g}).
% \begin{equation}
%c(E,d)=-\frac{E}{\sqrt{\binom{n}{d}}}f_{\frac{n}{2}-d}\quad (d\neq 0) .\label{eq:f} 
%\end{equation}
This  will provide an asymptotic  WKB form of $c(E,d)$   almost everywhere on the interval $d\in [0,n]$ except  for the small vicinities  of the turning  points, $| n/2-m_0(E)-d |={\cal O}(n^{0})$ and end points, $n-d, d={\cal O}(n^{0})$.
In Fig.~\ref{fig:c01} we plot the comparison between the coefficients $c(E,d)$ computed based on exact expression (\ref{eq:CEd}) and the results of asymptotic  WKB analysis using Eqs.~ (\ref{eq:f-wkb}),(\ref{eq:f-wkb-1}).

\section{\label{sec:lin-H-app}Linearization of the down-folded Hamiltonian near the center of the Impurity Band}

We divide   the  Hamiltonian ${\cal H}(E)$ for a given $E$  on two parts, accordingly
\begin{equation}
\cH_{ij}(E)=\cH^{(0)}_{ij}(E)+\cH_{ij}^{(1)}(E)\;,
\end{equation}
where we defined
\begin{equation}
\cH^{(0)}_{ij}(E)=n(c(E,0)-1)\,\delta_{ij},\label{eq:H0}
\end{equation}
\vspace{-0.25in}
\begin{equation}
\cH_{ij}^{(1)}(E)=\delta_{ij}(1-c(E,0))\epsilon_{i}+n c(E,d_{ij})(1-\delta_{ij}).\label{eq:H1}
\end{equation}
We write similar expansions for energies and amplitudes
\begin{align}
E\approx E^{(0)}+E^{(1)},\quad \psi(z_j)\approx  \psi^{(0)}(z_j)+\psi^{(1)}(z_j),\label{eq:expan} 
\end{align}
and get
\begin{align}
\cH(E)  \approx \cH^{(0)}(E^{(0)})+\frac{\cH^{(0)}(E^{(0)})}{\partial E}E^{(1)}+\cH^{(1)}(E^{(0)}),\nonumber
\end{align}
where the parts of the Hamiltonian $\cH^{(0,1)}$ are given above.
We plug the above expansions  into the system of equations  (\ref{eq:Hc1})  $\sum_{j=1}^{M}\cH_{ij}(E)\cA_j$=$E \cA_i$, and  use   (\ref{Ak}) to express   $\cA_{j}^{(0)}=n^{1/2}\psi^{(0)}(z_j)$. 
Equating terms of the same order in $\epsilon_j$ and $c(E,d_{ij}),\,i\neq j$, we obtain the equation for eigenstates and eigenvalues in  zeroth order
\begin{equation}
n[c(E^{(0)},0)-1]\psi^{(0)}(z_j)=E^{(0)}\psi^{(0)}(z_j)\;,\label{eq:0th}
\end{equation}
$j\in[1..M]$, and in the first order
 \begin{align}
a\, \epsilon_i \psi^{(0)}(z_j)+b\sum_{j\neq i=1}^{M}n\, c(E^{(0)},d_{i j})\psi^{(0)}(z_j)\nonumber \\
=E^{(1)} \psi^{(0)}(z_j)\label{eq:1st}
\end{align}
Above index $j$ enumerates marked states. Also the coefficients $a,b$ equal
\begin{equation}
a=b (1-c(E^{(0)},0),\quad b^{-1}=1-n\frac{\partial c(E^{(0)},0)}{\partial E}\;.\label{eq:ab}
\end{equation}

Similarly to the above we find from Eqs.~(\ref{eq:normA}),(\ref{eq:Qjk}) the zeroth-order approximation to the the total probabilistic weight  of an eigenfunctions $\ket{\psi}$  over the marked state  subspace $Q_{jk}^{(0)}=\delta_{jk}Q$ where
\begin{equation}
Q_{jk}^{(0)}=\delta_{jk}Q,\quad \frac{1}{Q} =n^2 \frac{\partial}{\partial E}\left(\frac{c(E,0)-1}{E}\right)_{E=E^{(0)}}.\label{eq:Qij}
\end{equation}

 \subsubsection{\label{sec:lin-0th} Zeroth-order of the perturbation theory}

Eq.~(\ref{eq:0th}) admits the solution corresponding to the $M$-fold  degenerate energy level that originates from the band of the  marked  states, $E^{(0)}\rightarrow -n$
in the limit of $\B\rightarrow 0$. The  corresponding $M$ eigenstates $\psi_\beta(z_j)$ ($\beta\in[1..M]$) have support  over the  part of  computational basis   corresponding  to  
marked  states:   $\psi_{z_j}^{\beta}\neq 0,\,\,j\in(1,M)$.  
Using $c(E,0)$ from (\ref{eq:c0a}) the explicit form of the equation (\ref{eq:0th}) for eigenvalue in zeroth order is given in the main text, Eqs.~(\ref{eq:E0}),(\ref{eq:Delta0a}) which we repeat here for convenience. 
\begin{equation}
E^{(0)} =-n-\Delta_0\;,\label{eq:E0-app}
\end{equation}
\vspace{-0.23in}
 \begin{equation}
\Delta_0=n 2^{-n}\sum_{d=0}^{n}\binom{n}{d}\frac{\B(n-2d)}{n+\Delta_0-\B(n-2d) }\;.\label{eq:Delta0a-app}
\end{equation} 
 Here    $\Delta_0$ is  the root of the  above  transcendental equation that satisfies the condition  $\lim_{\B\rightarrow 0}\Delta_0=0$. In general,  
  the sum (\ref{eq:Delta0a}) is dominated by the region of   values of $d$ such that $|d-n/2|={\cal O}(n^{1/2})$. We    obtain   $\Delta_0$  in a form of a series expansion in powers of $n^{-1}$ 
\begin{equation}
\Delta_0\simeq -B_{\perp}^{2}-\frac{B_{\perp}^{4}}{n}+{\cal O}(n^{-2}), \label{eq:Delta0as-app}
\end{equation}

Similarly, using  $c(E,0)$ from (\ref{eq:CEd}) in the equation (\ref{eq:Qij}) for the zeroth-order total weight over the marked state subspace we obtain
\begin{equation}
\sum_{k=1}^{M}|\psi_{z_k}^{(0)}|^2=Q, \nonumber
\end{equation}
\vspace{-0.25in}
\noindent
\begin{equation}
\frac{1}{Q}=\frac{1}{2^n}\sum_{d=0}^{n}\binom{n}{d}\frac{1}{(\B(n-2d)-n-\Delta_0)^2}.\label{eq:Q-app}
\end{equation} 
Using (\ref{eq:Delta0as}) and employing   similar approximations to that from the above we get an asymptotical expression in large $n$ limit
\begin{equation}
Q\simeq 1-\frac{B_{\perp}^{2}}{n}-\frac{3B_{\perp}^{4}}{n^2}+{\cal O}(n^{-3}). \label{eq:Qas-app}
\end{equation}
We recall that in our study $n$ is asymptotically large and  we  always  assume  that the transverse field $\B={\cal O}(n^{0})$ (but can be parametrically large, $\B \gg 1$).

The denominator in Eqs.~(\ref{eq:Delta0a-app}),(\ref{eq:Q-app}) corresponding to $d=m$ will become zero  at \lq\lq resonant" transverse field value $\B=B_{\perp\, m}$   which is a root of the equation (\ref{eq:Gamma-res}) in the main text.
%\begin{equation}
%B_{\perp\, m}=\frac{n+\Delta_0(B_{\perp\, m})}{n-2m},\label{eq:Gamma-res}
%\end{equation}
  In the range of $\B$ under consideration $n/2-m\gg n^{1/2}$. 

Near the $m$th resonance the term with $d=m$ in the sum (\ref{eq:Delta0a}) becomes anomalously large due to a small denominator despite the factor $p_m$ being very small.
We keep this term  (\ref{eq:Delta0a}) along with the   terms corresponding to $|n/2-d|\sim n^{1/2}$ and obtain 
\begin{equation}
\Delta_0 \simeq\frac{\delta_B}{2}  \pm\sqrt{ \frac{\delta^2_B}{4} +n^2 p_m}\;,\label{eq:Delta-r}
\end{equation}
where we  introduced  rescaled transverse field difference from its value at resonance
\begin{equation}
\delta_B=n\frac{\B-B_{\perp\, m}}{B_{\perp\, m}^{(0)}}\;,\label{eq:xi}
\end{equation}
\noindent
where $B_{\perp\, m}^{(0)}=n/(n-2m)$.

Clearly, in the resonance region  $\delta_B\sim  n \,p_{m}^{1/2}$ and $|\B-B_{\perp\, m}|\sim \Delta B_{\perp\, m}$ where $\Delta B_{\perp\, m}\sim  2^{-n/2}\binom{n}{m}\,B_{\perp\, m}^{(0)}$.
There   the weight factor $Q$ is decreasing dramatically (cf.   Fig.~\ref{fig:Q-res}) and   the above perturbation theory breaks down.  The width of the resonant regions  $\Delta B_{\perp\, m}$   (\ref{eq:resGamma})   remains exponentially small in $n$  for   $n/2-m\gg n^{1/2}$.

 In this study we will only focus on the off-resonance case, assuming the condition
\begin{equation}
 \Delta B_{\perp\, m} \ll  |B_{\perp\, m}-\B |\sim | B_{\perp\, m+1}-\B | ={\cal O}(\B)\;.\nonumber%\label{eq:offres}
\end{equation}
%where  the integer  $m$ is chosen so that $B_{\perp\, m}<\B<B_{\perp\, m+1}$ .   

\subsubsection{\label{sec:lin-1st} First order of the perturbation theory}

  \iffalse
{\color{red}{
\begin{itemize}
\item In the fist order of perturbation theory we obtained effective Hamiltonian linearized in the vicinity of shifted band.
\item give the expression for this Hamiltonian: only introduce terms that depend on n and explain. Temove $\theta$ symbol given assymtotics.
Give cartoon of the of the levels  of the driver and impurity band.
\end{itemize}
}}
\fi
The first order equation (\ref{eq:1st}) determines the correct zeroth order eigenstates $\{\psi_\beta(z_j)\}_{\beta=1}^{M}$ and removes the degeneracy of the energy levels. To evaluate the coefficients $a,b$ in (\ref{eq:1st}) we calculate $c(E,0)$ away from resonance using the same approach as that in the evaluation of the sum
in (\ref{eq:Delta0a})
\begin{equation}
c(E,0)\simeq -\frac{n B_{\perp}^{2}}{E^2}+{\cal O}\left(\frac{n^2 B_{\perp}^{4}}{E^4}\right )\;.\label{eq:c0a}
\end{equation}
The coefficients    $a,b\simeq 1+{\cal O}(B_{\perp}^{2}/n)$ and in what following will be  replaced  by unity. Then Eq.~ (\ref{eq:1st}) corresponds to the effective Hamiltonian $\scH$ with the  matrix elements, $\scH_{ii}$=$\e_i$ and $\scH_{ij\neq i}=n c(E^{(0)},d_{ij})$ where coupling coefficients $c$ are  given in (\ref{eq:c}). 
%\begin{equation}
%\scH=\sum_{j=1}^{M} \epsilon_j\ket{z_j}\bra{z_j}+\sum_{k\neq j=1}^{M} n c(E^{(0)},d_{jk})\ket{z_j}\bra{z_k}\label{eq:IBH}
%\end{equation}
Using Eqs.~(\ref{eq:E0}),(\ref{eq:Delta0a}) for zeroth-order energy $E^{(0)}$,
the matrix $\scH_{ij}$  can be written in the form (\ref{eq:IBH}).

%%%%% new appendixes

\section{\label{sec:stat-app}Statistical independence of matrix elements}
% identical phrasing to the text!
In this paper the IB  Hamiltonian $\scH_{ij}$ is determined by the symmetric matrix of Hamming distances $d_{ij}$ between the bit-strings corresponding to the marked  states sampled without replacement   from the set of all possible  $2^n$ bit-strings.
% below is old section 
 Instead of this  ensemble one can consider   a different one,  where each of the $M$ bit-strings   is sampled {\it with replacement} from the full set $\stn$. In this ensemble Hamming distances  $d_{ij}$ for distinct pairs $i,j$  are statistically independent allowing for much simpler statistical averaging. Indeed, for a given row $i$ of the  matrix  $d_{ij}$  the  joint probability distribution of the  two distinct off-diagonal matrix elements can be estimated as,
\begin{equation}
p_{d_{i j_1},d_{i j_2}}-p_{d_{i j_1}} p_{d_{i j_2}} \propto  \frac{1}{2^n}  \Delta(d_{i j_1}-d_{i j_2}) p_{d_{i j_1}}.\label{eq:pd1d2}
\end{equation}
Here $\Delta(d)$ denotes the Kronecker delta,   $j_1\neq j_2\neq i$ and $p_d$ as before corresponds to the modified binomial distribution,
  \begin{equation}
p_d =\frac{1}{Z} \,2^{-n}\binom{n}{d},\quad Z=\sum_{d=1}^{n}2^{-n}\binom{n}{d},\; \label{eq:p_d-app}
\end{equation}
 (also   $\sum_{d_1,d_2=1}^{n}p_{d_1,d_2}=1$).
 One can see that  the statistical correlation between a pair of Hamming distances $d_{i j_1},d_{i j_2}$  is exponentially small (in $n$) and can be neglected.
  
 Such an ensemble allows for multiple copies of the same bit-string to be sampled. However this effect is not statistically significant for  modest values of $M$ 
 \begin{equation}
 1\ll M  \ll 2^{n/2}\;.\label{eq:Mcond}
 \end{equation}
 This can be seen by comparing the number of ways to perform unordered sampling of $M$ elements from the group of $2^n$ elements with and without replacement. Using Stirling's formula we write the former number as,
 \begin{equation}
 \binom{2^n+M-1}{M}\simeq \binom{2^n}{M}\, \exp\left(\frac{M^2}{2^n}\right)(1+\varepsilon), \label{eq:repl}
 \end{equation}
where the latter number is given by $ \binom{2^n}{M}$ with $\varepsilon \sim M 2^{-3n/4}\ll 1$.  It is clear that when condition (\ref{eq:Mcond}) is satisfied the two ensembles are statistically equivalent because repetitions  can be neglected.

 \section{\label{sec:GerschgorinTh} Bound on the largest eigenvalue of $\cV_{ij}$ from Gerschgorin  circle theorem}
 
One can use the above estimates of the typical largest matrix elements of the matrix $\cV_{ij}$    to consider  the  bounds on its   eigenvalues  given by the  Gerschgorin  circle theorem \cite{weisstein2003gershgorin}.
  For the  case of real  eigenvalues the theorem states that every eigenvalue   lies within at least one of the intervals $[\cV_{ii}- R_i,\cV_{ii} +R_i ]$ where $i\in[1..M] $ and $R_i=\sum_{j\neq i}|\cV_{ij}|$ is a sum of absolute values of the off-diagonal elements  in the $i$th row.  
  For a  randomly chosen row the value of  $R_i$ can be estimated as follows 
\begin{equation} 
R_i \simeq M\sum_{d=1}^{n}  p_d |V(d)|\;,
\end{equation} 
where $p_d$ is defined  in (\ref{eq:p_d}). From Eq.~(\ref{eq:V2}) one can see that the above sum is dominated by the terms satisfying $|n/2-d|\ll n$. Using Stirling's approximation  we get 
%\begin{equation}
$R_i \sim M 2^{-n/2} e^{-n\theta}$.  %\;.\label{eq:Rest} 
%\end{equation}
For typical diagonal matrix elements $|\cV_{ii}|=|\epsilon_{i}|\apprle W$. 
 Therefore from the Gerschgorin  theorem we conclude the eigenvalues $E_{\beta}^{(1)}$ of  $\scH$ satisfy the following bound 
 \begin{equation}
| E_{\beta}^{(1)}| \leq  \max \left \{W,  M 2^{-n/2} e^{-n\theta}\right \}\;.\label{eq:gersh}
\end{equation}
One can see that Gerschgorin bound in our case precisely corresponds to the typical maximum  element in the matrix $\scH_{ij}$.

\section{\label{sec:MeanValueSigmaHeff} Mean value and standard deviation of the off-diagonal matrix elements $\scH_{ij}$}
 
 The mean value of the off-diagonal matrix element 
\begin{equation}
\langle \scH_{ij}\rangle=n\sum_{d=0}^{n}p_d c(E,d)\simeq \frac{n}{2^n}\frac{ \B}{ \B -  1}.\label{eq:Hmean}
\end{equation}
is much smaller than its  standard deviation
\begin{equation}
\langle ( \scH_{ij}-\langle  \scH_{ij}\rangle)^2\rangle^{1/2}\simeq \B \sqrt{\frac{n}{2^n}}. \label{eq:Hsd}
\end{equation}
This is related to the symmetry $p_d=p_{n-d}$ and a rapid oscillation of   $c(E^{(0)},d)$ with $d$ (cf. (\ref{eq:c}), (\ref{eq:phi-m}) and Fig.~\ref{fig:c01}). 
%The expressions above are given in the  leading order in $n\gg1$ and assuming  $\B={\cal O}(n^{0})$.

We note from (\ref{eq:typ}),(\ref{eq:Hsd}) that the standard deviation is exponentially larger than the typical value 
\[ \langle (\scH_{ij}-\langle \scH_{ij}\rangle)^2\rangle^{1/2}\sim  V_{\rm typ}\,e^{n \theta}.\]
 This can be understood by looking at the  values of $d_{ij}$ that dominate the  variance of $\scH_{ij}$.
 We write
\begin{equation}
\langle (\scH_{ij})^{2} \rangle = n^2\sum_{d=0}^{n} c^2(E^{(0)},d) p_d.\label{eq:H2}
\end{equation}
%The sum in the r.h.s is dominated by the values of $d={\cal O}(n^{0})$. The reason for this is the following.
It follows from the Eqs.~ (\ref{eq:c}) and   (\ref{eq:p_d}) that   for   $d\in(n/2-m_0,n/2+m_0)$ the coefficient $c^2(E,d)\propto 1/\binom{n}{d}$ decreases exponentially with $d$, while
the distribution $p_d \propto \binom{n}{d}$ increases exponentially with $d$. The binomial factors cancel out and  the expression under the summation in (\ref{eq:H2})  %$c^2(E,d) p_d\propto 2^{-n}e^{-n\theta}$ 
contains very slowly-varying with $d$ (non-oscillatory) part. However for  $d\in(0,n/2-m_0)$    the coefficient $c(E,d)$ grows exponentially  faster than $1/\binom{n}{d}$ with decreasing $d$ (see Eqs.~(\ref{eq:f-wkb-1}), (\ref{eq:c-g})).   Therefore the  variance (\ref{eq:H2}) is dominated by  non-extensive values of $d={\cal O}(n^{0})$
that are  much smaller than the smallest Hamming distance $d_{\rm min}={\cal O}(n)$  (\ref{eq:dmin}) in a  randomly chosen  row of $d_{ij}$. Therefore the variance of $\scH_{ij}$ is not a good statistical characteristic of the PDF  of $\scH_{ij}$. It  is dominated by the extremely rare atypical instances of the ensemble.

\section{\label{sec:distV2}PDF of the squared off-diagonal matrix elements of impurity band Hamiltonian
}

In this section we provide the details of the derivation of the PDF for the non-oscillatory parts of the (squared) off-diagonal matrix elements $\cV_{ij}^{2}$ of the IB Hamiltonian. As discussed in the main text,  in the  asymptotical limit of large $n\gg 1$ one can make an approximation that  $n$  is a continuous variable and  we replace the summation over $d$ in (\ref{eq:V2dist}) by an integral and Kronecker delta 
 $\delta(x)$ by Dirac delta. This results in the  Eq.~(\ref{eq:PVm}) displayed below for convenience
 \begin{equation}
 P(\cV_{ij}^{2})= \int_{0}^{n} p_x  \, \delta(V^2(x)-\cV_{ij}^{2})dx\;.\label{eq:PV}
 \end{equation}
It was discussed in the main text (see also below)  that the condition for this validity of this approximation is
\begin{equation}
\frac{1}{n}\log_2 M\ll 1\;.\label{eq:M-small}
\end{equation}
It corresponds to the number of marked states $M$ that is not very large. For example, it can still scale exponentially with $n$ so that $M=2^{\mu n}$, $\mu={\cal O}(n^{0})$,
but the coefficient $\mu$ in the exponent needs to be small $\mu\ll1$.

 The  expression (\ref{eq:PV}) is obtained  using the analytical continuation $p_x$ of the binomial distribution  $p_d$ (\ref{eq:p_d})  from the integer domain $d\in[0,n]$  onto  the interval of a real axis $x\in[0,n]$ in terms of the Beta function $B(x,y)$
\begin{equation}
p_x=2^{-n}\binom{n}{x}=\frac{2^{-n}}{(n+1)B(x+1,n+1-x)}\;,\label{eq:B}
\end{equation}
 and the resulting identity
\begin{equation}
\int_{0}^{n}dx \,p_x=1.\label{eq:intpd}
\end{equation}

\begin{figure}[htb]
  \includegraphics[width= 3.35in]{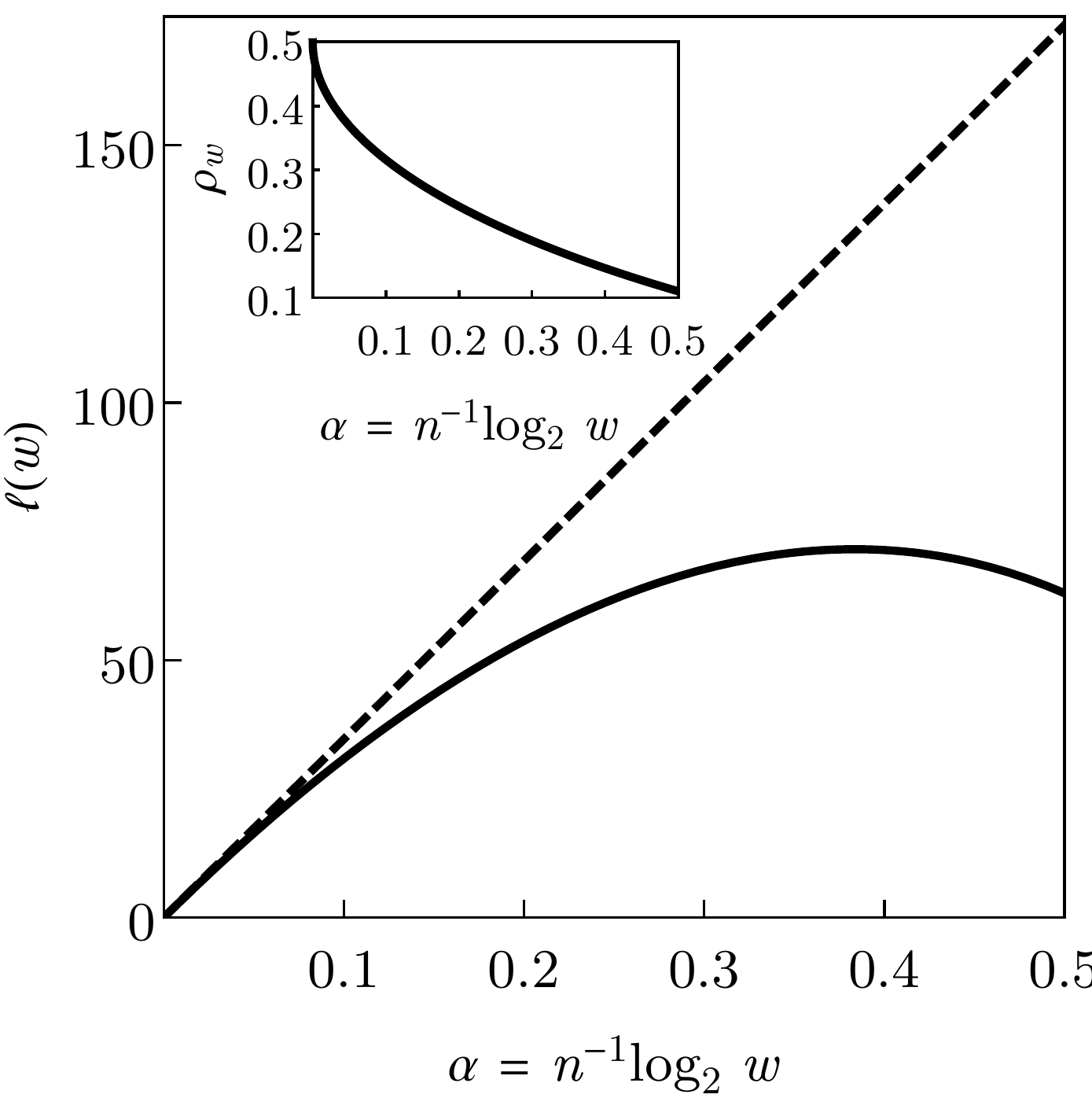}
  \caption{Solid line shows the dependence of $\ell(w)$  on $\alpha=\frac{1}{n}\log_{2}w$ from Eq.~(\ref{eq:ell}).  Dashed line  shows the tangent to the solid curve at the point $\alpha=0$ ($w=1$) . This line corresponds to  $\ell(w)\simeq \sqrt{\log w}$, in accordance with (\ref{eq:ell-small}).
  Inset shows the dependence  of the root $\rho_w$ of the equation (\ref{eq:qw})  on $\alpha=\log_2 w^{1/n}$. Small $\alpha\ll 1$ corresponds to Hamming distances $\rho_w\approx 1/2$. Near that point the dependence  of $\rho_w$ on $\alpha$ follows (\ref{eq:small-r}).
    \label{fig:w}}
\end{figure}

In what following we will study the   rescaled quantities 
\begin{equation}
w_{ij}\equiv \frac{\cV_{ij}^{2}}{V_{\rm typ}^{2}}=\left(\frac{2}{\pi n}\right)^{1/2}\frac{1}{p_{d_{ij}}}\;,\label{eq:w}
\end{equation}
\noindent
where  $i\neq j$,  $ V_{\rm typ}$  is  given in (\ref{eq:typ}) and $p_d=2^{-n}\binom{n}{d}$. Using Stirling's approximation in binomial coefficient
\begin{subequations}
\label{eq:binS}
\begin{equation}
p_{x}\equiv p_B(x/n),\quad p_B(\rho)=\frac{e^{-n \cA(\rho)}}{\sqrt{2\pi n\rho(1-\rho)}}\,.%\label{eq:binomS}
\end{equation}
\vspace{-0.2in}
\begin{equation}
\cA(\rho)=\rho\log\rho+(1-\rho)\log(1-\rho)+\log 2\;.\label{eq:AA} 
\end{equation}
\end{subequations}
we get from Eq.~(\ref{eq:V2}) for $\cV_{ij}=V(d_{ij})$
\begin{equation}
\pzw(\rho) \equiv\frac{V^2(n \rho)}{V_{\rm typ}^{2}}\simeq \frac{\sqrt{4\rho(1-\rho)}}{\upsilon(\rho)}\,e^{n \cA(\rho)}\,,\label{eq:q}
\end{equation}
where   $\upsilon(\rho)$ is given in (\ref{eq:upsilon}).
 Eq.~(\ref{eq:w}) takes the form
\begin{equation}
w_{ij}=\pzw(d_{ij}/n)\;.\label{eq:pzw}
\end{equation}
Then the expression for the PDF for $w_{ij}$
\begin{equation}
g(w_{ij})=V_{\rm typ}^{2}\,P(V_{\rm typ}^{2}\,w_{ij})\;,\label{eq:PDFw}
\end{equation}
can be written in the form (cf. (\ref{eq:PV}))
\begin{equation}
g(w)=2n\int_{0}^{1/2}p_B(\rho)\delta(w-\pzw(\rho))d\rho\;.\label{eq:gint}
\end{equation}
We note that the domain of $g(w)$ is bounded from below by $w=1$ and from above by $w={\cal O}(2^n)$.  Taking the integral in (\ref{eq:gint})  we get
\begin{equation}
g(w)=2n\frac{p_B(\rho_w)}{\left | \frac{d\pzw(\rho)}{d\rho}\right |_{\rho=\rho_w}}\;,\label{eq:g-der}
\end{equation}
where   the rescaled Hamming distance $\rho_w$  is a root of the transcendental equation
\begin{equation}
\pzw(\rho_w)=w\;.\label{eq:qw}
\end{equation}
In the leading order in $n\gg 1$ this equation gives
\begin{equation}
\cA(\rho_w)=\frac{1}{n}\log w\;,  \label{eq:Aw}
\end{equation}
where $\cA(\rho)$ is given in (\ref{eq:AA}).
Also  using Eqs.~(\ref{eq:binS}),(\ref{eq:q}) in (\ref{eq:g-der}) we  get 
 \begin{equation}
g(w)=\frac{1}{w^2\sqrt{\pi \ell(w)}}\;,\label{eq:g-ell}
\end{equation}
where 
\begin{equation}
\ell(w)=\frac{n}{8}\upsilon^2(\rho_w)\,|\log(\rho^{-1}_{w}-1)|^2\;\label{eq:ell}
\end{equation}
Here   the dependence  of  $\ell(w)$ on   $w$ is shown in Fig.~\ref{fig:w}. In the entire range the dependence is logarithmically  slow. 

We note that the (\ref{eq:Aw}) is a valid approximation to (\ref{eq:qw}) for $\rho-\rho_0\gg 1/n$ where $\rho_0$ is a zero of $\upsilon(\rho)$
\begin{equation}
\upsilon(\rho_0)=0,\quad \rho_0=\frac{1}{1-\sqrt{1-B_{\perp}^{-2}}}
\end{equation}
It corresponds to Hamming distance $n \rho_0 = n/2-m_0$  (\ref{eq:m00}) which lies at the boundary of the interval  (\ref{eq:allowed}) where the WKB solution (\ref{eq:c}),(\ref{eq:V2}) applies (see the discussion in the Sec.~\ref{sec:cd}). It is assumed that $n \rho_0$ is smaller than the typical smallest Hamming distance $\dmin$ in a randomly selected row
\begin{equation}
\dmin- n \rho_0\gg 1.\label{eq:rho0cond}
\end{equation}
Using the asymptotic expression (\ref{eq:binS})  for the binomial distribution in Eqs.~(\ref{eq:dmin})  we get the equation for $\dmin$ in the form
\begin{equation}
\cA(\dmin/n)=\frac{1}{n}\log M
\end{equation}
The function $\cA(\rho)$ is decreasing with $\rho$ for $\rho(0,1/2)$. Therefore Eq.~(\ref{eq:rho0cond}) leads to  the condition $ \cA(\rho_0)-\cA(\dmin/n)\gg 1/n$, or
\begin{equation}
 \cA(\rho_0)-\frac{1}{n}\log M\gg \frac{1}{n}\;.
\end{equation}
Using explicit forms of $\cA(\rho)$ and $\rho_0$ we get in the limit of $\B\gg 1$
\begin{align}
\log 2-\frac{1}{n}\log M  &>\frac{2\log\B+2\log 2+1}{4B_{\perp}^{2}}+\varepsilon\nonumber \\
0<\varepsilon &={\cal O}(B_{\perp}^{-4})\;.\label{eq:dmM}
\end{align}
This is the condition for (\ref{eq:rho0cond}). Clearly it corresponds to  a much weaker constraint on the values of $M$  than the condition $\frac{1}{n}\log M<\frac{1}{2}\log 2$
provided by the requirement of a statistical independence of matrix elements of $\cV_{ij}$ (cf. (\ref{eq:Mcond})).

\subsubsection{\label{sec:small-M} Case of $\,\frac{1}{n}\log_2 M \ll 1$}

 The rescaled Hamming distance $\rho_w$  depends on $w$ via the logarithmic factor $\alpha=\frac{1}{n}\log_{2}w$. This dependence is shown in the 
  inset to the Fig.~\ref{fig:w}.  In this section we consider  
  \begin{equation}
  \alpha=\frac{1}{n}\log_{2} w\ll 1\;.\label{eq:g-small}
  \end{equation}
  Then we get
  \begin{equation}
  p_B(\rho)\simeq \left(\frac{2}{\pi n}\right)^{1/2} e^{-n \cA(\rho)}\;,\label{eq:Asmall}
  \end{equation}
  \vspace{-0.2in}
  \begin{equation}
  \cA(\rho)\simeq 2\left(\frac{1}{2}-\rho\right)^2\;.\nonumber
  \end{equation}
  Then using (\ref{eq:Aw}) we get
  \begin{align}
  \rho_w \simeq \frac{1}{2}-\left(\frac{\alpha}{2}\right)^{1/2}\;.\label{eq:small-r}
  \end{align}
  \vspace{-0.2in}
  \begin{align}
  \ell(w)\simeq \log w\;,\label{eq:ell-small}
  \end{align}
  and finally,
  \begin{align}
g(w) \simeq g_\infty(w)=\frac{1}{w^2 \sqrt{\pi \log w}}\;.
\label{eq:gl}
\end{align}
The subscript here indicates that, unlike $g(w)$, the PDF $g_\infty(w)$ has the upper boundary of its domain  equal to infinity.
It is of interest to calculate  for a given $w$ the magnitude of the relative changes of   $V^2(d)$ and of the binomial coefficient $p_d$ when Hamming distance is chaining by 1 (and $\rho_w$ is changing by $1/n$). We define as in Eqs.~(\ref{eq:pzw}),(\ref{eq:q}) $w=V^2(d)/V_{\rm typ}^{2}$ and obtain
\begin{align}
&\frac{V^2(d+1)-V^2(d)}{V^2(d)} \simeq \frac{p_{d+1}-p_{d}}{p_{d}}\label{eq:dd1}\\
&\simeq 4\left(\frac{1}{2}-\rho_w \right)=\sqrt{8\alpha}\ll 1.
\end{align}
Here we used Eqs.~(\ref{eq:small-r}) and (\ref{eq:g-small}).  The above inequality justifies using the continues approximation (\ref{eq:PV}) in (\ref{eq:V2dist}).

 In a randomly chosen row of $w_{ij}$  the PDF that  the  largest element equals $w$ is
 \begin{equation}
 {\rm PDF}(\max_{m}w_m=w)\simeq \frac{M e^{-M w\log w}}{w\log w},\quad M\gg 1\;.\label{eq:wmax}
 \end{equation}
Typical largest  element in a row $\max_{1<j<i}w_{ij} \sim M$ in agrement with the results obtained earlier,  cf. Eqs.~(\ref{eq:V2}), (\ref{eq:dmin}) and (\ref{eq:w}). 
Therefore in order to ensure that  $\alpha\ll 1$ for all matrix elements in a typical row of $w_{ij}$ we require that $\log_2 M \ll n$ 
\begin{equation}
1\leq w\lesssim  M,\quad \frac{1}{n}\log_2 M\ll 1\;.\label{eq:range-r-app}
\end{equation}

The typical value of the smallest element in a randomly selected row of the rescaled matrix of Hamming distances $d_{ij}/n$ equals
\begin{equation}
\rho_{\rm min}=\frac{\dmin}{n}=\frac{1}{2}-\sqrt{\frac{\log_2 M}{2n}}\;.\label{eq:rho-min}
\end{equation}
We note that in the case we consider 
\begin{equation}
n \gg n/2-\dmin = {\cal O}(n)\;,\label{eq:dmin-close}
\end{equation}
minimum value $\dmin$ is close to $n/2$ but is still separated by extensive distance from it.

In this paper we use the  expression for  the  matrix elements of the IB Hamiltonian  $\scH_{ij}$ (\ref{eq:IBH}) that only applies in the region $|n/2-d_{ij}|<m_0$
 where  $m_0$  is given in (\ref{eq:m00}).  The elements in a typical row of the matrix $d_{ij}$ belong to this region if the condition $|n/2-\dmin|<m_0$ is fulfilled.
 Using (\ref{eq:rho-min}) we can re-write this as an inequality for $M$ 
 \begin{equation}
 M<2^{\frac{n}{2}(1-B_{\perp}^{-2})}\;.\label{eq:cond-ossc}
 \end{equation}
 This inequality is satisfied under the condition (\ref{eq:range-r-app}).

\section{\label{sec:phi0} Characteristic function of the PDF of  the squared off-diagonal matrix elements of impurity band Hamiltonian}

Here we compute the characteristic  function of the PDF  $g_\infty(w)$ (\ref{eq:gl}) (also given in Eq.~(\ref{eq:g_0(w)}) of the main text). It is defined as follows
\begin{equation}
 \phi_\infty(u)=\int_{1}^{\infty}dw\, g_\infty(w)(e^{i u w}-1)\;.  \label{eq:phi-inf}  
\end{equation}
We will be interested in the asymptotic limi of the above expression at small $|u|\ll1$. It is convenient  to calculate separately real  and imaginary parts of $\phi_\infty(u)$. 

%\subsubsection{Real part of the characteristic function}
For real part we have
\begin{equation}
-\frac{\sqrt{\pi}}{2}\,{\rm Re}[\phi_\infty(u)]=\int_{1}^{\infty}\frac{1}{x^2\sqrt{\log x}}\,\sin^2\left(\frac{u x}{2}\right)\;.\label{eq;Rephi0}
\end{equation}
Because $ \phi_\infty(-u)= \phi_\infty^*(u)$ we can assume that $u>0$ and break the  interval of integration  above on two parts
\begin{equation}
x\in [1,X/u] \cup [X/u,\infty),\quad u\ll X\ll 1\;.\label{eq:Xin}
\end{equation}
We write
\begin{equation}
-\frac{\sqrt{\pi}}{2}\,{\rm Re}[\phi_\infty(u)]=R_1(u)+R_2(u)\;.\label{eq:R1R2}
\end{equation}
Here
\begin{equation}
R_1(u)=\int_{1}^{X/u}\frac{1}{x^2\sqrt{\log x}} \sin^2\left(\frac{u x}{2}\right)\;,\label{eq:R1}
\end{equation}
\begin{equation}
R_2(u)=\int_{X/u}^{\infty}\frac{1}{x^2\sqrt{\log x}} \sin^2\left(\frac{u x}{2}\right)\;.\label{eq:R2}
\end{equation}
Using (\ref{eq:Xin}) asymptotic expansion of $R_1(u)$ has the form
\begin{equation}
R_1(u)\simeq \frac{u \,X}{4(\log(1/u))^{1/2}}+\frac{u \,X\log(1/X)}{8 (\log(1/u))^{3/2}} +\ldots\;\label{eq:R1a}
\end{equation}
Also after some tedious calculations we obtain
\begin{align}
R_2(u) &\simeq\frac{u}{(\log(1/u))^{1/2}}\left(\frac{\pi}{4}-\frac{X}{4}\)\nonumber \\
+&\frac{u}{2(\log(1/u))^{3/2}}\frac{\pi (\gamma_{\rm Euler}-1)}{4}\;.\label{eq:R2a}
\end{align} 
where
\begin{equation}
\gamma_{\rm Euler}\simeq 0.577\label{eq:Euler}
\end{equation} 
is the  Euler constant.

%\subsubsection{Imaginary part of the characteristic function}
Similarly to the above we also  break the interval of integration in the imaginary part of $\phi_\infty(u)$ on two parts given in (\ref{eq:Xin})
\begin{equation}
\Im[\phi_\infty(u)]=I_1(u)+I_2(u)\;.\label{eq:Imphi0}
\end{equation}
where
\begin{equation}
I_1(u)=\int_{1}^{X/u}\frac{\sin u x}{x^2\sqrt{\pi\log x}}\;,\label{eq:I1}
\end{equation}
\begin{equation}
I_2(u)=\int^{\infty}_{X/u}\frac{\sin u x}{x^2\sqrt{\pi\log x} }\;.\label{eq:I2}
\end{equation}
Expanding the integrand (\ref{eq:I1}) in $u$ and using condition  (\ref{eq:Xin})  we get
\begin{equation}
I_1(u)\simeq \frac{2u\sqrt{\log \frac{1}{|u|} }}{\sqrt{\pi}}-\frac{u \log \frac{1}{X}}{\sqrt{\pi\log \frac{1}{|u|}}}+\cO\(\frac{u \log^2X}{\log^{3/2}|u|}\)\;.\nonumber
\end{equation}
Performing similar asymptotic expansion in $I_2(u)$ we obtain
\begin{equation}
I_2(u)\simeq \frac{u(1-\gamma_{\rm Euler}-\log X)}{\sqrt{\pi\log\frac{1}{|u|}}}+\cO\(\frac{u\,\log^2X}{\log^{3/2}u}\)\;.\nonumber
\end{equation}

Finally, we combine together Eqs.~(\ref{eq:R1a}),(\ref{eq:R2a}) into Eq.~(\ref{eq:R1R2}) to obtain first two terms in the asymptotic expansion  of ${\rm Re}[\phi_\infty(u)]$  in powers of $1/\log u\ll 1$
\begin{equation}
{\rm Re}[\phi_\infty(u)]=-\frac{|u| \sqrt{\pi}}{2\sqrt{\log |u|^{-1}}}\(1-\frac{1- \gamma_{\rm Euler} }{2\log |u|^{-1}}\)\;.\label{eq:Rephi0a}
\end{equation}
We also combine together the above expressions for  $I_1$ and $I_2$ to obtain a similar asymptotic expansion of  ${\rm Im}[\phi_\infty(u)]$  
\begin{equation}
\Im[\phi_\infty(u)]\simeq \frac{2u\sqrt{\log \frac{1}{|u|} }}{\sqrt{\pi}}+\frac{u(1-  \gamma_{\rm Euler})}{\sqrt{\pi\log\frac{1}{|u|}}}\;.\label{eq:Imphi0a}
\end{equation}
Note that in both cases the terms involving $X$ cancels out confirming the validity of the matching procedures.

\section{\label{sec:CLT} Generalized Central Limit Theorem for the sum of  ${\bm M}$ random variables $\bm {w_m}$ that obey   the distribution  ${\bm g}_{\bm \infty}{\bm {(w)}}$ }

In this section we will study the asymptotic PDF  for the sum the independent identically  distributed random variables  in  Eq.~(\ref{eq:SM}) sampled from the probability distribution (\ref{eq:gl}).  We note that the variance of the random variables does not exist.
The  PDFs with polynomial tails at infinity are known as Pareto (heavy-tailed) distributions.
According to the Generalized Central Limit Theorem (GCLT), the  PDF  of the  sum of $M$ Pareto variables  for $M\rightarrow \infty$ approaches its asymptotic form   given by the stable law  \cite{gnedenko1954limit}.This general property coincides  with the  usual  Central Limit Theorem for the case when random variables in a  sum have finite variances. In this case the limiting  PDF has Gaussian form. 

We note that  the PDF given by Eq.~(\ref{eq:gl})  is not strictly polynomial at $w\rightarrow\infty$  because of the additional logarithmic factor. We will  derive the asymptotic form of the sum (\ref{eq:SM}) of random variables (\ref{eq:gl}) explicitly and compare with the standard GCLT result  without the logarithmic factor.

We are interested in the PDF of the random variable $s_M$ such that (cf. \eqref{eq:g_0(w)},\eqref{eq:SM})
\begin{equation}
 s_M =\frac{1}{M}\sum_{i=1}^{M}w_i,\quad g_\infty(w)=  \frac{1}{w^2 \sqrt{\pi \log w}}\;.\label{eq:sumOmega}
\end{equation}
Here  $w_i$ are i.i.d random variables sampled from $g_\infty(w)$ and we are interested in the asymptotic limit $M\gg 1$.
 
Using the convolution property of a sum of statistically independent random variables we get for the PDF   of $s_M$ 
\begin{equation}
\PDF(s_M)=\frac{1}{2\pi} \int_{-\infty}^{\infty}dq \, [\varphi_\infty(q/M)]^M\, e^{-i  q s_M}\;,\label{eq:bP}
\end{equation}
where %$\phi_\infty(u)=\int_{1}^{\infty}g_\infty(w) e^{i u w}dw$.   We  write
\begin{equation}
\varphi_\infty(u)=1+\phi_\infty(u)\;,\nonumber
\end{equation}
and $ \phi_\infty(u)$ is given in (\ref{eq:phi-inf}).
The limit $M\gg 1$ corresponds to $|u|\ll 1$. We note that
\begin{equation}
\lim_{u\rightarrow 0}\phi_\infty(u)=0
\end{equation}
Taking into account that  $\phi(u)$ is small in the above limit we write
\begin{equation}
\PDF(s_M)\simeq \frac{1}{2\pi}\int_{-\infty}^{\infty}dq\, \exp\left[-i q s_M+M\phi_\infty(q/M)\right]\;.\label{eq:PM1}
\end{equation}

Quantity  $M \phi_\infty(q/M)$ can be expanded in inverse powers of $\log M$$\gg$1  using asymptotic form of the characteristic function at small argument given in 
Eqs.~\eqref{eq:Rephi0a},(\ref{eq:Imphi0a}). First few terms of expansion have the form
\begin{align}
M\, \Re\,\phi_\infty\left(\frac{q}{M}\right)&\simeq -\frac{\pi |q|}{2\sqrt{\log M}}+\frac{ \sqrt{\pi} |q|  \(1-{\gamma_{\rm Euler}}-\log|q|\)    }{4(\log M)^{3/2}}\;,
\nonumber \\
M\,\Im\,\phi_\infty\left(\frac{q}{M}\right) &\simeq  2q  \left(\frac{\log M}{\pi}\right)^{1/2}+ q \frac{1-{\gamma_{\rm Euler}}-\log |q|}{(\pi \log M)^{1/2}}\nonumber\\
&+\frac{q\log|q|(1-{\gamma_{\rm Euler}} ) }{2\sqrt{\pi}(\log M)^{3/2}}\;.
\label{eq:ReImphi}
\end{align}
\noindent
where $\gamma_{\rm Euler}$ is the Euler constant.

It is clear from comparing  individual terms in  Eq.~(\ref{eq:ReImphi}) with the exponential in the integrand in Eq.(\ref{eq:PM1}) that $q=\cO(\sqrt{\log M})$. 
Therefore we can drop in Eqs.~\eqref{eq:ReImphi} terms $\cO(1/(\log M)^{3/2})$.
We make the change of variables in the integral in (\ref{eq:PM1})
\begin{equation}
q=2 \sqrt{\frac{\log M}{\pi}}\;t\;,\label{eq:qt}
\end{equation}
and obtain
%We plug the above expressions for $M\phi_\infty(q/M)$ to (\ref{eq:PM1}) and after rescaling (\ref{eq:qS}) obtain in the limit of $M\gg 1$ the probability distribution $\bp_0^M(S)$
\begin{equation}
\PDF(s_M)=\frac{1}{\sigma_M}L_{1}^{1,1}\left(\frac{s_M-b_M}{\sigma_M}\right)\;,\label{eq:SbM}
\end{equation}
\begin{equation}
L_{1}^{1,1}(x)\equiv\frac{1}{2\pi} \int_{-\infty}^{\infty}dt\,e^{-i t x-|t|-\frac{2 i}{\pi}t\log|t|}\;.\label{eq:char}
\end{equation}
Function $L_{1}^{1,1}(x)$ above  is  a so-called Levy alpha-stable distribution  \cite{cizeau1994theory,WikiStable,voit2013statistical}  shown  in Fig.~\ref{fig:st-dist}. The distribution  is defined by its characteristic function. Parameters $b_M$ and  $\sigma_M$ in (\ref{eq:SbM}) are typical values that  characterize the shift of the maximum of the $\PDF(s_M)$ from the origin and its  overall scale, respectively. They are given in 
 Eqs.~(\ref{eq:bM}) and (\ref{eq:sigmaM}) of the main text and we also provide them for convenience  below
\begin{equation}
\sigma_M=\frac{\pi}{2}\frac{1}{(\pi\log M)^{1/2}}\;,\label{eq:sigmaMa}
\end{equation}
\vspace{-0.2in}
\begin{equation}
b_M\simeq \sigma^{-1}_{M}-\frac{2}{\pi} \sigma_M\log(\sigma^{-1}_{M}) +\frac{2}{\pi} (1-{\gamma_{\rm Euler}}) \sigma_M\;,\label{eq:bMa}
\end{equation}
where $\gamma_{\rm Euler}$  is the  Euler constant.

It is instructive to compare the above expressions   with the result for the sum 
 of random  variables  that obey a standard Pareto distribution 
 (i.e.,  without the logarithmic factor present in $g_\infty(w)$)
\begin{equation}
s_M^0=\frac{1}{M}\sum_{i=1}^{M}w_i,\quad w_i \sim g_0(w)=w^{-2}\;.\label{eq:Sg0}
\end{equation}
The PDF of $s_M^0$ has the same form as the PDF of $s_M$ given in (\ref{eq:SbM})  but the expressions for 
the shift $b_M^0$ and  the overall scale $\sigma_M^0$ are   different
\begin{equation}
\sigma_M^0=\frac{\pi}{2},\quad b_M^0=\log M+1-{\gamma_{\rm Euler}}+\log \left(\frac{\pi}{2}  \right)
\end{equation}
One can see that
\begin{equation}
\frac{\sigma_M^0}{\sigma_M}\sim \frac{b_M^0}{b_M}\sim  (\log M)^{1/2}\gg 1\;.\label{eq:ratio}
\end{equation}
The rescaling  factor $(\log M)^{1/2}$ between the PDFs of $s_M$ and $s_M^0$  can be explained  by a similar logarithmic factor in the ratio $g_0(w)/g_\infty(w)\sim (\log w)^{1/2}$, taking into account the fact that typical of $w\sim M$.

\section{\label{sec:Just} Justification of replacing sum with integral in Eq.( \ref{eq:red}).}

We note that the number of marked states $ \Mr$ in a miniband (\ref{eq:Mres}) on a Hamming distance $d$ from a given marked  state $\ket{z_j}$ decreases  rapidly when $d$.   There is a  typical minimum Hamming distance  $d \simeq \dminr$ such that 
\begin{equation}
d_{\rm min}^{\rm res}={\rm argmin}(\Omega_d)=\cO(1)\;.\label{eq:dminr}
\end{equation}
There will be  no states in the  miniband located at the Hamming distances
$d$ from the state $\ket{z_j}$ that lie inside the intervals
 $d\in[1,\dminr)\cup (n-\dminr,n]$. For those values of $d$    we have $\Gamma_{j}^{(d)}=0$.
Using  (\ref{eq:Asmall}) we get
\begin{equation}
\dminr\simeq \frac{n}{2}-\sqrt{\frac{n}{2}\log \frac{2A\, \Omega}{\pi n^2}}\;.\label{eq:dminr-a}
\end{equation}
where $A=A(E^{(0)},1/2)$ (\ref{eq:A}). 

On the one hand we assume throughout the paper that  the number of marked states in a miniband $\Omega\gg1$ is sufficiently large so that
the number $n-2\dminr$ of dominant  terms in the sum    (\ref{eq:chan}) is much bigger than  1.
For example, using the  scaling ansatz (\ref{eq:param}) we have $\Omega\sim M^{2-\gamma}$ (\ref{eq:GM}). Then  assuming that $\gamma<2$ and $1>\frac{1}{n}\log_2 M={\cal O}(n^0)$ we can see that 
 the  second term in the r.h.s of  (\ref{eq:dminr-a}) is of the order of $n$ and  therefore the number $n-2\dminr=\cO(n)$.

On the other hand we note that the number $\Omega_d$ (\ref{eq:Mres}) of marked states in a miniband  on a Hamming distance $d$ from a given marked  state $\ket{z_j}$  is large ($M_{j}^{(d)} > \Mr\gg1$)  for almost all $d$, aside from  $\cO(n^0)$ values of $d$ near the boundaries of the interval $d\in[\dminr, n-\dminr]$.  

We recall that  all terms  in a sum  (\ref{eq:chan}) are nearly equal to each other and therefore the relative  contributions to $\Gamma_j$ from the boundary terms is $\cO(1/n)$ and can be neglected in a leading order estimates of the typical quantities. For $d$  away from the interval boundaries  the function $\delta_\eta(\e_j-\e_m)$ in Eq.~(\ref{eq:red}) changes little between the adjacent values of $\e_m$ (by an amount $\sim 1/\Mr\gg 1$). This provides the justification for us to replace  the sum over $m$ in (\ref{eq:red}) by an integral.

\section{\label{sec:peh} PDF  of  the  random variable $\bm{h=\frac{\bm{\eta}}{(z-\e)^2+\eta^2}}$}

Consider the PDF  $p_\eta(h;z)$  introduced in the Eq.~(\ref{eq:PetaYm})
\begin{equation}
p_{\eta}(h;z)=\int_{-\infty}^{\infty}\frac{1}{W} p_A(\e/W) \bm{\delta}[\,h-\delta(z-\e,\eta)\,] d\e\;,\label{eq:PetaY}
\end{equation}
Here the function of two arguments $\delta(x,y)$ is defined in (\ref{eq:deltaY}) and  $\bm{\delta}[x]$ is Dirac delta-function denoted here with bold font to distinguish from the above function. We also used the relation (\ref{eq:pA}) for the PDF of marked state energies.
Solving equation
\begin{equation}
h=\frac{\eta}{(z-\e)^2+\eta^2}\;,\label{eq:h}
\end{equation}
for $\e$ we get
\begin{equation}
\e_{\pm}=z\pm\sqrt{\eta(h^{-1}-\eta)}
\end{equation}
From here and from (\ref{eq:PetaY}) we get
\begin{equation}
p_\eta(h; z)=\frac{\sqrt{\eta}}{2 h^{3/2}\sqrt{1-\eta h}}(\varphi_+(h;z)+\varphi_-(h;z))
\end{equation}
\vspace{-0.2in}
\begin{equation}
\varphi_\pm(h;z)=W^{-1}p_A(z\pm\sqrt{\eta(1/h-\eta)})
\end{equation}
For $|z|\ll W$ we get $p_\eta(h;\eta)\simeq p_\eta(h;0)$
\begin{equation}
p_\eta(h;0)=\frac{\sqrt{\eta}}{h^{3/2}\sqrt{1-\eta h}}\,p_A(\sqrt{\eta(1/h-\eta)})\;.\label{eq:p_et}
\end{equation}

 \begin{figure}[htb]
  \includegraphics[width= 3.35in]{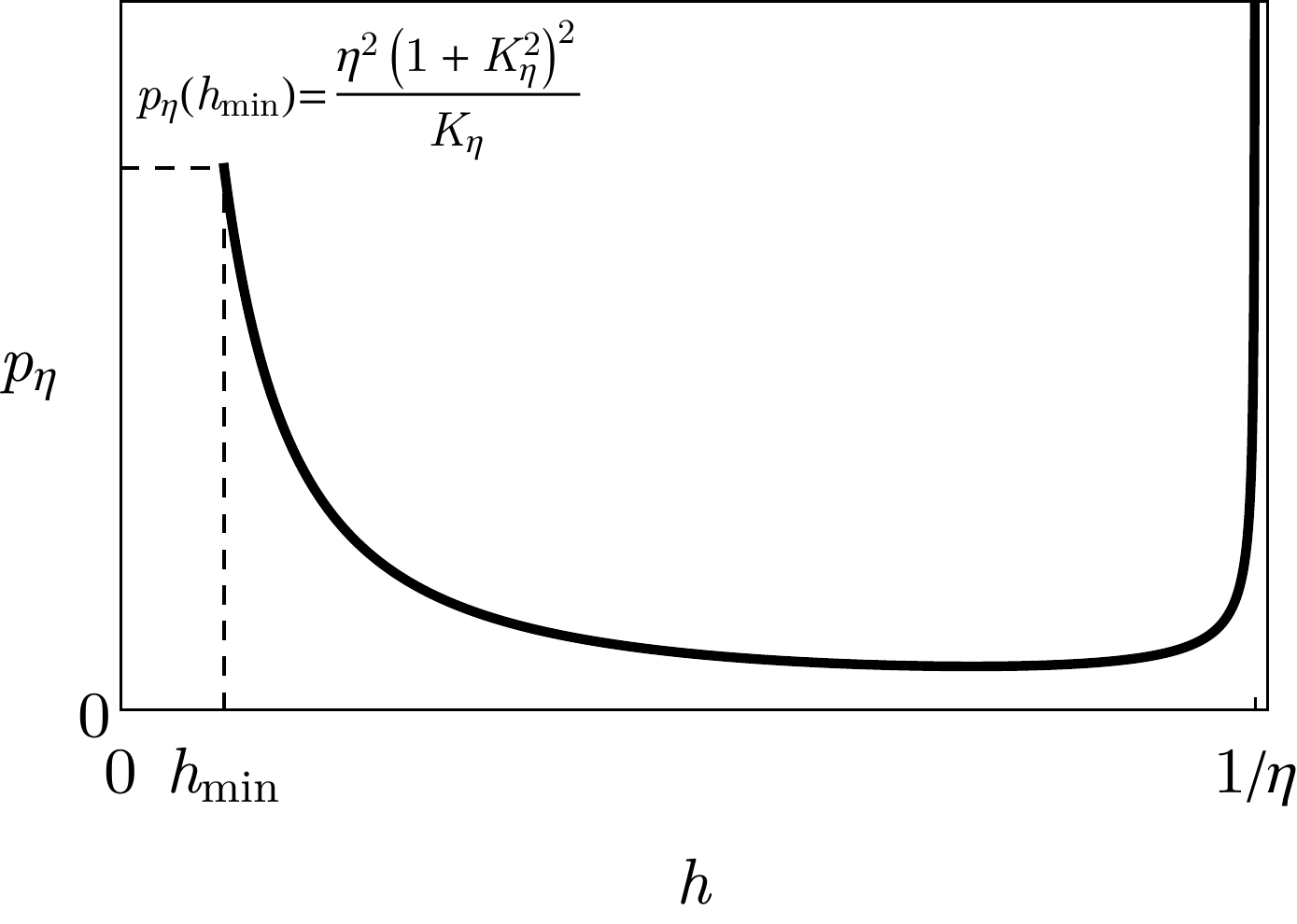}
  \caption{Plot of the PDF of $p_\eta(h;0)\equiv p_\eta(h)$ given in (\ref{eq:p_et}).      \label{fig:p-et}}
\end{figure}

\paragraph{Uniform Distribution}
For the case of uniform distribution
\begin{equation}
p_A(\e)=\frac{1}{W}\theta(W/2-\e)\;,\label{eq:unif}
\end{equation}
where $\theta(x)$ is Heaviside theta-function we have
\begin{equation}
p_\eta(h;0)=\frac{1}{h^{3/2}\sqrt{\eta^{-1}-h}}\;.\label{eq:p_et-1}
\end{equation}
Domain of values of $h$ is $h\in[h_{\rm min}, h_{\rm max}]$ where
\begin{equation}
h_{\rm min}=\frac{1}{\eta(1+K^2_\eta)},\quad h_{\rm max}=\frac{1}{\eta}\;.\label{eq:hminmax}
\end{equation}
\vspace{-0.2in}
\begin{equation}
K_\eta \equiv \frac{W}{2\eta}\;.\label{eq:K}
\end{equation}
And the value of the PDF on the lower boundary is
\begin{equation}
p_\eta(h_{\rm min})=\eta^2\,\frac{(1+K^2_\eta)^2}{K_\eta}\;.\label{eq:p-hmin}
\end{equation}

In the case of delocalized non-ergodic states  (\ref{eq:n-er})
\begin{equation}
M\gg K_\eta\gg 1\;.\label{eq:MK1}
\end{equation}

The PDF $p_\eta(h;0)\equiv p_\eta(h)$ is plotted in Fig.~\ref{fig:p-et}. 
 The PDF reaches the local  maximum on the lower  boundary $h_{\rm min}$ corresponding to  values of
marked state energies $\e\simeq W$ located at the edges   of the IB. In the region $h\sim 1/\eta$ the probability density reaches very small values, $p_\eta(h,z)\sim \eta^2$, corresponding to the energies of marked  states  $|\e-z|\simeq \eta$. Maximum value of $h=1/\eta$  corresponds to exact resonance $\e=z$. The PDF    $p_\eta(h;0)$ has an integrable singularity at this point.

It is of interest to consider the PDF of the sum of random variables $h_m$ over all marked states 
\begin{equation}
s^h_M=\frac{1}{M}\sum_{m=1}^{M}h_m,\quad h_m=\frac{\eta}{(z-\e_m)^2+\eta^2}\;.\label{eq:hm}
\end{equation}
In the non ergodic  phase $W\gg \eta$ mean value of $h_m$ is much smaller than its standard deviation
\begin{equation}
\langle h_m\rangle=\frac{\sum_{\sigma=\pm 1}{\rm arccot}\left(\frac{2\eta}{W-2 \sigma z}\right)}{W}
\simeq\frac{\pi}{W}\;,\label{eq:hav}
\end{equation}
\begin{equation}
\langle h^2_m\rangle\simeq\frac{\pi}{2W\eta} \gg \langle h_m\rangle^2\;.\label{eq:h2h2}
\end{equation}
Note that the mean is dominated by small marked state energies $\e_m\sim\eta$  while standard deviation is dominated by $\e_m\sim W$.

However for sufficiently large $M$ the mean value of the  sum $\sum_{m=1}^{M}h_m$  is much greater than its standard deviation provided that $\de\ll\eta$
\begin{equation}
\frac{\langle s^h_M\rangle^2-\langle s^h_M\rangle^2}{\langle s^h_M\rangle^2}\simeq \frac{1}{2\pi}\frac{\de}{\eta}\ll 1
\end{equation}
Therefore in the  delocalized phase
\begin{equation}
 \eta \gg \de=\frac{W}{M}\;,\label{eq:Wete}
\end{equation}
the sum $\sum_{m=1}^{M}h_m$ is self-averaging.

It is convenient to  introduce  rescaled variables
\begin{equation}
y_m=\sqrt{h_m\,\eta}\;.\label{eq:y}
\end{equation}
Their  PDF has the form
\begin{equation}
\fp_\eta(y)=\frac{1}{K_\eta y^2\sqrt{1-y^2}}\;.\label{eq:p-y}
\end{equation}
%The subscript indicates that the PDF depends on $\eta$  (\ref{eq:K}). 
Boundaries of the domain of $\pzp_\eta(y)$ are
\begin{equation}
y_{\rm min}=\frac{1}{\sqrt{1+K^2_\eta}}\leq y <y_{\rm max}=1\;.\label{eq:y-domain}
\end{equation}

\section{\label{sec:getax}PDF  of  the  imaginary part of self-energy in self-consistent Born approximation
 }

In this section  we provide details of calculations of self-consistent Born approximation presented in Sec.~\ref{sec:self-consist-B} of the main text.
We study the  PDF of the sum
\begin{equation}
\Si=\cVt^2\sum_{m=1}^{M}\frac{w_m \eta }{(z-\e_m)^2+\eta^2}\;,\label{eq:Ysum}
\end{equation}
 where  $w_m=\cV^2(d_{0m})/\cVt^2$  (see Eqs.~(\ref{eq:q}), (\ref{eq:pzw})) are   random variables sampled from the distribution $g_\infty(w)$ (\ref{eq:g_0(w)}) and marked state energies $\e_m$ obey the dstribution $p_A(\e/W)/W$ (\ref{eq:pA}).
 The sum in (\ref{eq:Ysum}) can be written in the form  
 \begin{equation}
 \Si=\frac{\cVt^2}{\eta}\sum_{m=1}^{M}x_m,\quad x_m=w_m y_m^2\;,   \label{eq:Ywy}
 \end{equation}
 where $y_m$ are   random variables  (\ref{eq:y})  sampled from  the distribution  $\fp_\eta(y)$ (\ref{eq:p-y}).
For $|z|\ll W$ random variables $x_m$ obey  the PDF
 $g_\eta(x)$ such that
\begin{equation}
g_\eta(x)=\int_{y_{\rm min}}^{1}dy \int_{1}^{\infty}dw \,\fp_\eta(y) g_\infty(w)\,\delta(x-w y^2)\;.\label{eq:Px}
\end{equation}
Using (\ref{eq:p-y}) and (\ref{eq:y-domain}) one can show that (cf. also (\ref{eq:pzP}))
\begin{equation}
\lim_{\eta\rightarrow \infty}g_\eta(x)=g_\infty(x)\;.\label{eq:tran}
\end{equation}

In order to calculate the  PDF of the sum $\Si$ (\ref{eq:Ysum})
 in the limit $M\rightarrow \infty$ we use GCLT following the same approach as that in Sec.~\ref{sec:CLT}.
The PDF of the random variable $\Si$ equals
\begin{equation}
\PDF(\Si)\simeq \frac{1}{2\pi}\int_{-\infty}^{\infty}dk\, e^{-i k \Si+M\phi_\eta\(k \cVt^2/\eta\)}\;, \label{eq:PKM1}
\end{equation}
where $ \phi_\eta(u)$ is the characteristic function of the PDF $g_\eta(x)$ (\ref{eq:pzP})   
\begin{equation}
 \phi_\eta(u)=\int_{\frac{1}{1+K^2_\eta}}^{\infty}dx\, g_\eta(x)(e^{i u x}-1)\;. \label{eq:phiK}  
\end{equation}

\subsection{PDF  of  individual terms in the sum}

 \begin{figure}[htb]
  \includegraphics[width= 3.35in]{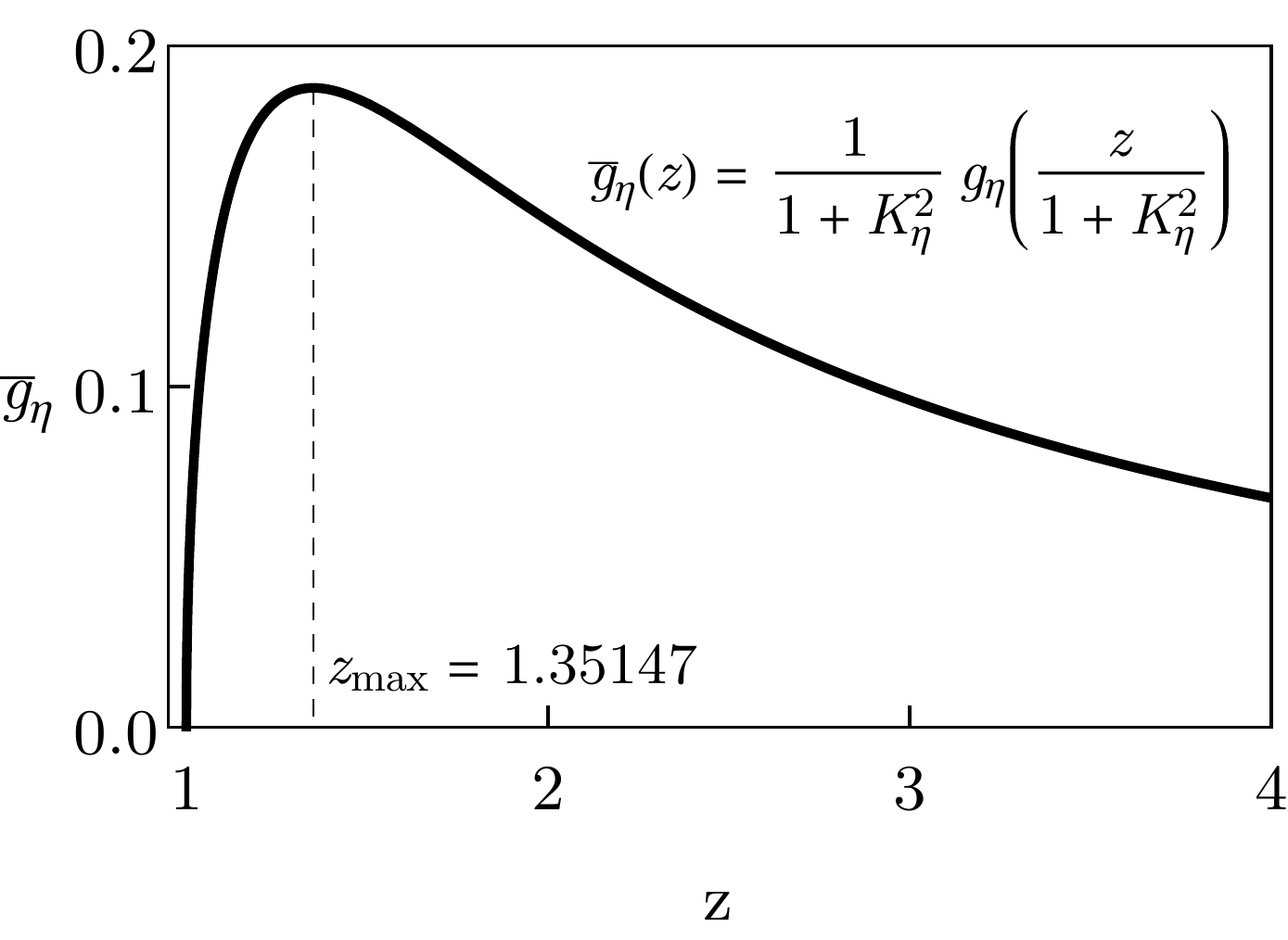}
  \caption{Plot of  $\bar{g}_\eta(z)$  given in (\ref{eq:bargK}) for $K_\eta=\sqrt{30}$.      \label{fig:bargK}}
\end{figure}

After some transformations we get from Eq.~(\ref{eq:Px}) 
\begin{align}
&g_\eta(x)=\frac{1}{x^2 K_\eta\sqrt{2\pi}} \nonumber \\
&\times \int_{\frac{1}{\sqrt{1+K^2_\eta}}}^{\min(1,\sqrt{x})}\frac{dy}{\sqrt{(1-y^2)(\log x^{1/2}-\log y)}}\;.\label{eq:pzP}
\end{align}
The PDF is plotted in Fig.~\ref{fig:pzP}. Its maximum lies very close to the left boundary of its domain
 $x\in[1/(1+K^2_\eta),\infty )$. For $x\ll1$ the PDF $g_\eta(x)$ depends on $x$ in terms of the rescaled parameter $z=x (1+K^2_\eta)$ whose PDF is 
 \begin{equation}
\bar{g}_{\eta}(z) \simeq \frac{{\rm erf}\(\sqrt{\frac{1}{2}\log z}\)  }{z^{3/2}\sqrt{2}}\;.\label{eq:bargK}
 \end{equation}
The plot of  $\bar{g}_{\eta}(z)$ is given  in Fig.~\ref{fig:bargK}), its maximum $z_{\rm max}\simeq 1.35$. Typical values of $x_m\simeq z_{\rm max}/K^2_\eta\ll 1$ correspond to $w_m\sim 1$ and  to a broad  PDF of marked state energies,  $|z-\e_m|\sim W$.

We are interested in the limits (cf. (\ref{eq:MK1}))
\begin{equation}
x\gg1,\quad K_\eta\gg 1\;.\label{eq:xKlim}
\end{equation}
We note that  $\log x\gg |\log y|$ in the denominator of (\ref{eq:pzP}) for all $y$ except for the small interval 
\[\frac{1}{\sqrt{1+K^2_\eta}} \leq  y   \lesssim \frac{1}{x}\;,\]
\noindent
 whose contribution to the integral  neglected.
Expanding the integrand in powers of $(\log x)^{-1/2}$ we get
\begin{equation}
g_\eta(x)\simeq \frac{\pi}{2K_\eta}g_\infty(x)-\frac{\pi\log 2}{2K_\eta\pi^{1/2}x^2\log^{\frac{3}{2}}x}\;,\label{eq:pzPa}
\end{equation}
where function $g_\infty(x)$ is defined in (\ref{eq:g_0(w)}).% Replacing the lower limit of integration by zero for $K\gg1$, we obtain
%\begin{equation}
%g_K(x)\simeq \frac{\pi}{2K}g_0(x)+\cO\left(K^{-1} x^{-2}\log^{-\frac{3}{2}}x\right)\;.
%\label{eq:pzPa}
%\end{equation}
We observe from  (\ref{eq:K}), (\ref{eq:hav}) that 
%\begin{equation}
$\eta \langle  h_m\rangle=  \frac{\pi}{2K_\eta}$. Using the expressions for  $g_\infty$ (\ref{eq:g_0(w)}) and $\langle  h\rangle$ (\ref{eq:hav}) we obtain under the condition (\ref{eq:xKlim}) 
\begin{equation}
g_\eta(x)\simeq  \eta\avg{h} g_\infty \(\frac{x}{\eta\avg{h}} \),\quad x\gg 1\;.\label{eq:pzPaa}
\end{equation}
Given a  large  deviation of   $x_m$  satisfying (\ref{eq:xKlim}),  the conditional PDF of $\eta h_m$  is  narrowly peaked around its  mean value corresponding  to  $|\e_m-z|\sim \eta$.
In contrast, typical values of $x_m$ correspond to  a much broader PDF of   $\e_m\sim W$.  This gives rise to a small factor $\pi/2K_\eta\sim \eta/W$ in the leading order term in (\ref{eq:pzPa}).

% \textcolor{blue}{
%We note that for sufficiently large   $M\gg1$   the terms with  $x_m\gg1$  dominate the sum   in $s_x^M$ (\ref{eq:sxM}) 
%because the probability PDF function $g_K(x)$  has a heavy  tail without a first moment. }

\subsection{Characteristic function of the PDF of the elements in the sum%of   $x_m=\frac{w_m \eta^2}{(z-\e_m)^2+\eta^2}$
}

The relation between the characteristic functions $\phi_\eta(u)$ and $\phi_\infty(u)$ (\ref{eq:phi-inf}) in the limit 
\begin{equation}
|u| \ll 1\;,\label{eq:uKlim}
\end{equation}
 should be the same as the relation (\ref{eq:pzPa}) between the corresponding 
PDFs $g_\eta(x)$ and $g_\infty(x)$ in the limit of large $x$ (\ref{eq:xKlim}).  Here we will establish this directly.
 We break $\phi_\eta(u)$ in two parts
\begin{equation}
\phi_\eta(u)=\phi_\eta^1(u)+\phi_\eta^2(u)\;,\label{eq:phi12}
\end{equation}
where
\begin{equation}
\phi_\eta^1(u)=\int_{\frac{1}{1+K^2_\eta}}^{1}dw\, g_\eta(w)(e^{i u w}-1)\;,\label{eq:phi1}
\end{equation}
\vspace{-0.2in}
\begin{equation}
\phi_\eta^2(u)=\int_{1}^{\infty}dw\, g_\eta(w)(e^{i u w}-1)\;.\label{eq:phi2}
\end{equation}
Expanding  $\phi_\eta^1(u)$ in $u$ we get
\begin{equation}
 \phi_\eta^1(u)\simeq \frac{\pi}{2K_\eta} \, i \zeta_1 u\;, % \frac{\zeta_1}{K\sqrt{\pi}}\,i u
 \end{equation}
 where
 \begin{equation}
 \zeta_1=\frac{2}{\pi^{3/2}} \int_{0}^{1}\frac{dx}{x}\int_{0}^{\sqrt{x}}dy\frac{1}{\sqrt{(1-y^2)\log(x/y^2)}}\;.\label{eq:zeta_1}
 \end{equation}
To calculate $\phi_\eta^2(u)$ in the limit of small $|u|$ we introduce $X\gg1$ such that
\begin{equation}
|u|\ll X |u|\ll 1\;,\label{eq:ineq}
\end{equation}
and write
\begin{equation}
\phi_\eta^2(u) =\phi_\eta^{2,-}(u) +\phi_\eta^{2,+}(u)\;. 
\end{equation}
Here
\begin{align}
\phi_\eta^{2,-}(u) &=\int_{1}^{X}dx \,g_\eta(x)(e^{i u x}-1)\;,\label{eq:phi2m} \\
\phi_\eta^{2,+}(u) &=\int_{X}^{\infty}dx \,g_\eta(x)(e^{i u x}-1)\;.\label{eq:phi2p} 
\end{align}
We use (\ref{eq:ineq}) and expand $\phi_\eta^{2,-}(u) $ in $u$
\begin{align}
\phi_\eta^{2,-}(u) \simeq i u\,\int_{1}^{X}g_\eta(x)dx\;.\label{eq:phi-}
\end{align}
To calculate the term $\phi_K^{2,+}(u)$ we use the approximation (\ref{eq:pzPa})  and write
\begin{align}
\phi_\eta^{2,+}(u) &=\frac{\pi}{2K_\eta} \phi_\infty(u)
-i u \frac{\pi}{2K_\eta}\int_{1}^{X}g_\infty(x)x dx\;.\label{eq:phi+}\\
&-\frac{\pi\log2}{2K_\eta}\int_{X}^{\infty}dx\,\frac{e^{i u x}-1}{\sqrt{\pi} x^2 (\log x)^{3/2}}
\end{align}
where the characteristic function $\phi_\infty$ is defined in   (\ref{eq:phi-inf}).

Combining $\phi_\eta^{2,\pm}(u)$ together and taking the limit $X\rightarrow \infty$ we get after some transdormations  
\begin{align}
\phi_\eta^2(u)\simeq \frac{\pi}{2K_\eta}(\phi_\infty(u)-i \zeta_2 u)
\end{align}
\begin{equation}
\zeta_2=\left(\frac{32}{\pi^3}\right)^{1/2}\int_{0}^{1}dy\left(\frac{\log(1/y)}{1-y^2}\right)^{1/2}\;.\nonumber
\end{equation}
After some transformations one can show that $\zeta_1=\zeta_2$. Therefore terms $\sim u$ in $\phi_\eta^1(u)$ and $\phi_\eta^2(u)$ cancel each other. Combining these two quantities together  in (\ref{eq:phi12}) we finally get 
\begin{equation}
\phi_\eta(u)\simeq \frac{\pi}{2K_\eta}\phi_\infty(u)+{\cal O}\left(\frac{ |u|}{  K_\eta  |\log u |^{3/2}}\right)\;.\label{eq:phi0u}
\end{equation}
As expected, this  relation corresponds to the relation (\ref{eq:pzPa}) between the PDFs $g_k$ and $g_\infty$.

\subsection{GCLT for the sum %$\Si=\cVt^2 \sum_{m=1}^{M}\frac{\eta}{(z-\e_m)^2+\eta^2}$
}

We now  revisit the expression  (\ref{eq:PKM1}) for the PDF of the variable $\Si$ \eqref{eq:Ysum}). In the limit $M\rightarrow\infty$ the
integral over $k$ in the r.h.s of (\ref{eq:PKM1}) is dominated by small values of the argument in $\phi_\eta(k \cVt^2/\eta)$. Then using \eqref{eq:phi0u} and \eqref{eq:K} we get after the change of a variable of integration in (\ref{eq:PKM1})
  \begin{equation}
\PDF(\Si)=\frac{1}{2\pi   \Sti }\int_{-\infty}^{\infty}dq e^{-i q\frac{\Si}{\Sti}+\Omega_\eta\phi_\infty(q/\Omega_\eta)}\;,\label{eq:bPY}
  \end{equation}
where $\Sti$  (\ref{eq:Gdef1}) is  the characteristic  value of imaginary part of self-energy of marked states   obtained in FGR-based calculation in Sec.~\ref{sec:FGR} and quantity $\Omega_\eta$ equals
\begin{equation}
\Omega_\eta=\frac{\pi M}{2 K_\eta}=\frac{\pi\eta}{\de}\;.\label{eq:L-eta}
\end{equation}
 It has a meaning of the typical number of marked states within the non-ergodic miniband of the width $\eta$
 (cf. Eq.~\eqref{eq:Mres} and Fig.~\ref{fig:FGR}).

 We make  a self-consistent assumption (cf. Eq.~(\ref{eq:eta-S}) in the main text) and set 
\begin{equation}
\eta=\Sti\;.\label{eq:eta-S-app}
\end{equation}
Then, one can immediately see that
\begin{equation}
\Omega_\eta=\Omega_{\Sti}=\Omega\;,\label{eq:self-consist-Omega}
\end{equation}
where $\Omega$ is  the typical number of marked states in a mini-band  defined in (\ref{eq:L}).

 Comparing the expression (\ref{eq:bPY}) with (\ref{eq:PM1}) and (\ref{eq:self-consist-Omega}) we represent the random variable $\Si$ in the form
\begin{equation}
\Si  \disteq  \sigma_{\Omega}\Sti \,x+b_{\Omega}\Sti,\quad \PDF(x)=L^{1,1}_{1}(x)\;.
\end{equation}
Here random variable  $x$ obeys a   Levy alpha-stable distribution    (\ref{eq:char}) shown  in Fig.~\ref{fig:st-dist}. The  quantities  $b_\Omega$, $\sigma_\Omega$ are given below
\begin{equation}
\sigma_\Omega=\sqrt{\frac{\pi}{4\log \Omega}}\;,\label{eq:sigmaOmega}
\end{equation}
\vspace{-0.2in}
\begin{equation}
b_\Omega\simeq \sigma_{\Omega}^{-1}-\frac{2}{\pi} \sigma_{\Omega}\log(\sigma_{\Omega}^{-1}) +\frac{2}{\pi} (1-\gamma_{\rm Euler}) \sigma_\Omega\;,\label{eq:bOmega}
\end{equation}
Their dependence on  $\Omega$ is given in the main text, Eqs.~(\ref{eq:bM}),(\ref{eq:sigmaM}), where we should replace  $M$ with $\Omega$.

\begin{figure}[!htb]
\includegraphics[width= 3.35in]{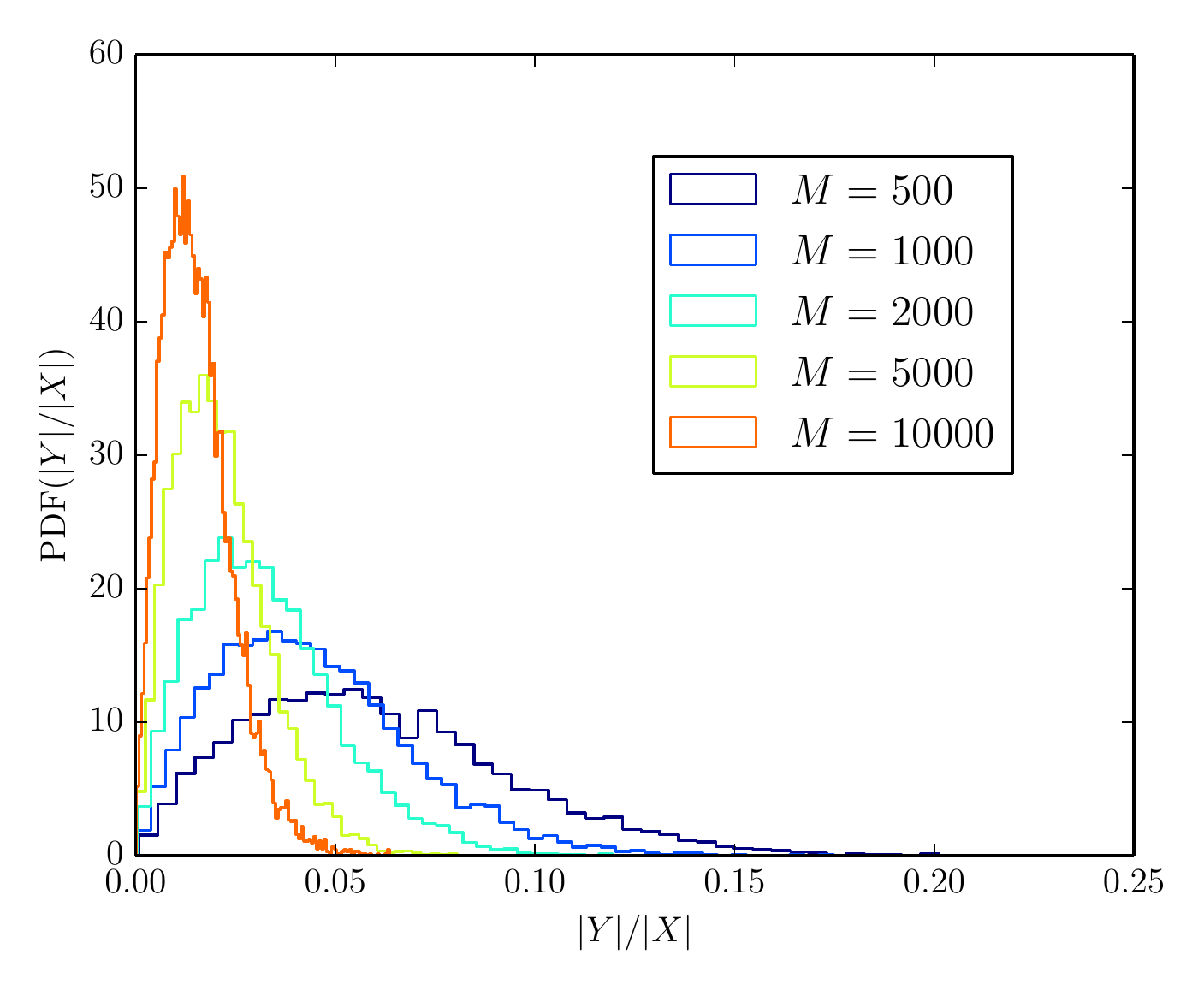}
\caption{Probability distribution of the ratio $\left| Y / X \right|$ defined in Eqs.~(\ref{eq:X-App}),(\ref{eq:Y-App}) for  $\gamma=0.6$. \label{fig:Y/X}} 
\end{figure}
\begin{figure}[htb]
%\includegraphics[width= 3.35in]{lambdas_impurities_0.3}
%\caption{$\gamma=1$}
\includegraphics[width= 3.35in]{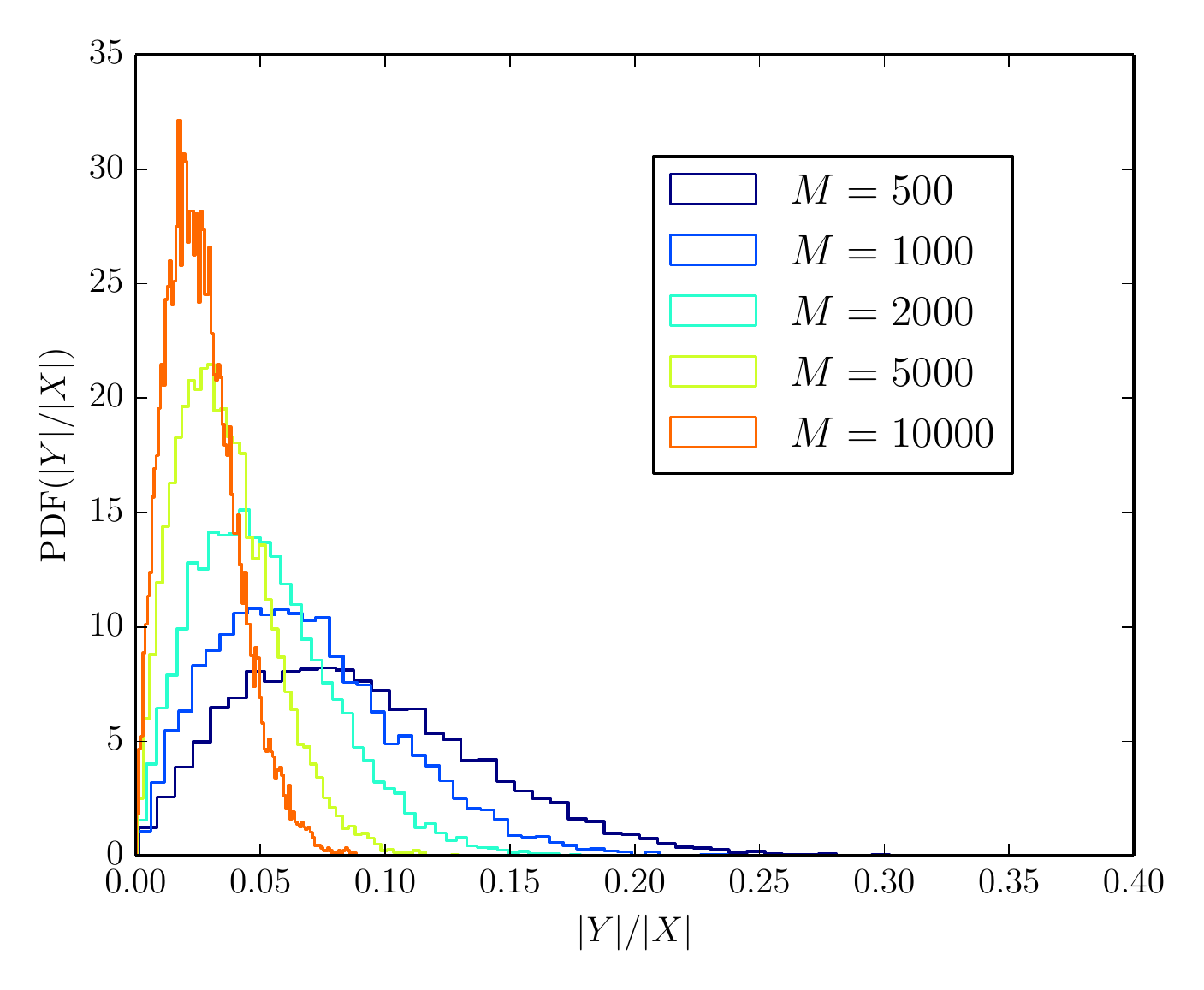}
%\caption{$\gamma=1.4$}
%\includegraphics[width= 3.35in]{lambdas_impurities_0.8}
\caption{The same as in Fig.~\ref{fig:Y/X} but with   $\gamma=1.2$. \label{fig:Y/X-1}} 
\end{figure}

\begin{figure}[htb]
%\includegraphics[width= 3.35in]{lambdas_impurities_0.3}
%\includegraphics[width= 3.35in]{lambdas_impurities_0.3}
%\caption{$\gamma=1$}
%\includegraphics[width= 3.35in]{lambdas_impurities_0.6}
%\caption{$\gamma=1.4$}
\includegraphics[width= 3.35in]{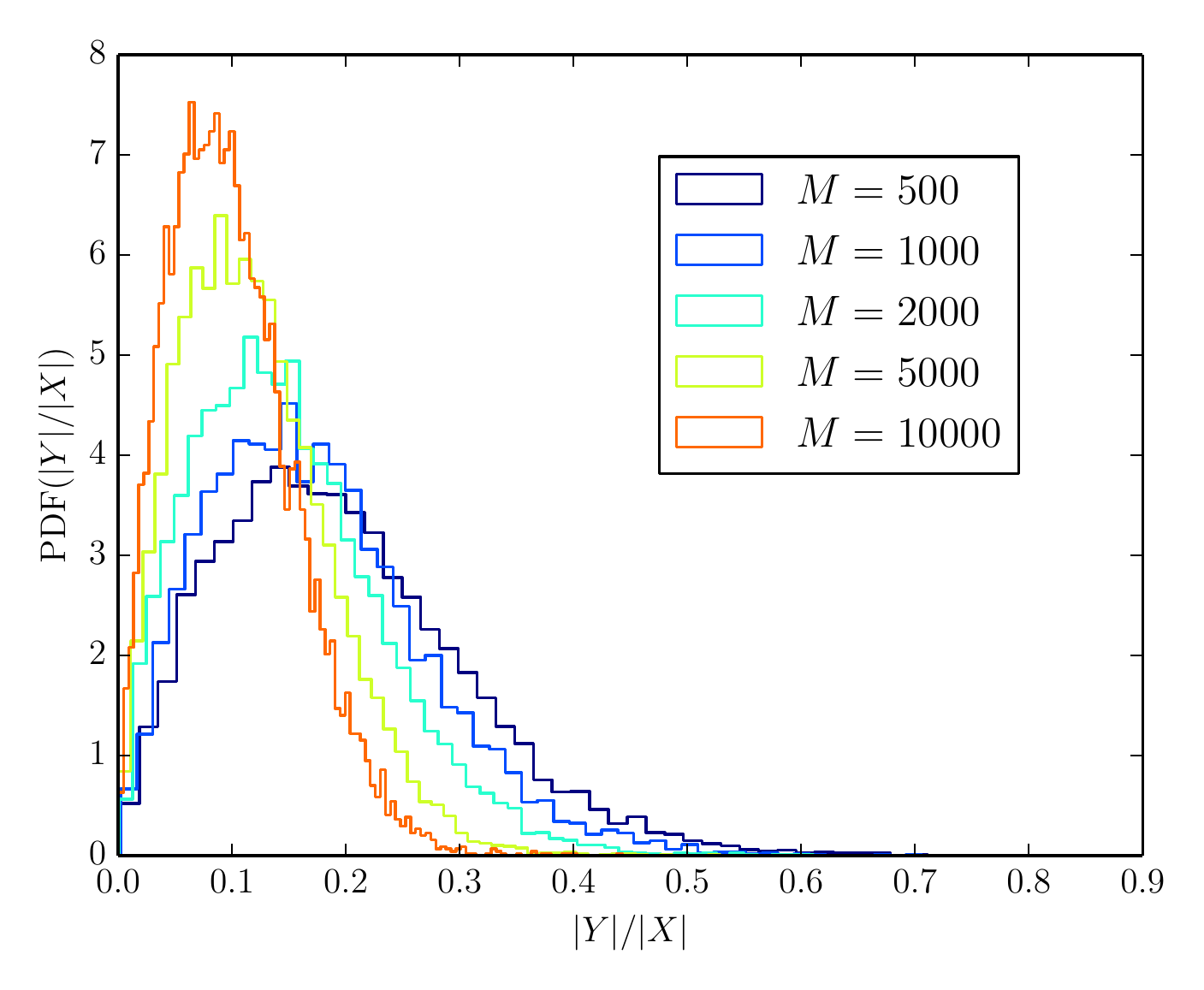}
\caption{The same as in Fig.~\ref{fig:Y/X} but for $\gamma=1.6$. \label{fig:Y/X-2}} 
\end{figure}

\section{\label{sec:Num-Sim-Appendix} Numerical simulations}

In this Section we provide details of the numerical analysis of the ensemble of Hamiltonians introduced in Sec.~\ref{sec:ensembleH} in addition to the results in Sec.~\ref{sec:NumSim}.

\subsection{\label{sec:GreenFunc} Numerical justification of cavity equations}

\begin{figure}[!tb]
%\begin{minipage}[t]{0.48\textwidth}
\includegraphics[width=3.35in]{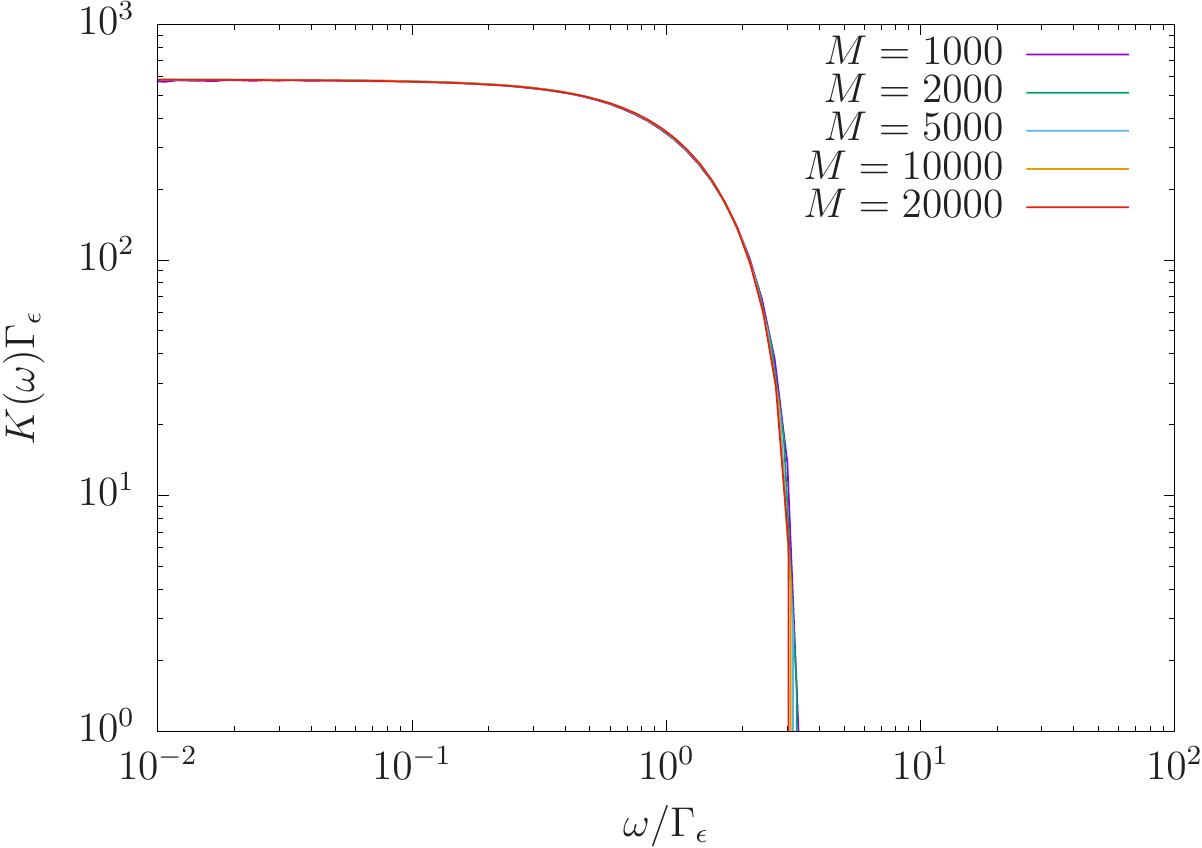}
\caption{$K(\omega)$ rescaled with the characteristic energy $\Gamma_\varepsilon = 2 \Sigma''_{\mathrm{typ}} M^\varepsilon$ where the typical mini-band width is given by Eq.~(\ref{eq:shift}). Here $\gamma=1$  with fitting exponent $\varepsilon =  -0.025$. \label{fig:KOmegaGammas1}}
\end{figure}

%\end{minipage}\hfill%
%\begin{minipage}[t]{0.48\textwidth}

Application of cavity method to the case of the ensemble of dense matrices considered in this paper, see Sec.~\ref{sec:ensembleH}, exploits the similarity between the local structure of the adjacency graph of the Hamiltonian $\scH$ and the Bethe lattice. The derivation of the cavity equations~(\ref{eq:Xc}),(\ref{eq:Yc}) for the case of  $\scH$ outlined in Sec.~\ref{sec:Cavity} neglects off diagonal terms $Y$ in comparison to diagonal $X$, which is justified for graphs with extensive number of neighbors~\cite{cizeau1994theory}, where,
\begin{gather}
X=\frac{1}{M}\sum _j \scH_{1 j}^2   G_{jj}\left(z\right), \label{eq:X-App}\\
Y=\frac{2}{M} \sum _{j\neq k} \scH_{1 j} \scH_{1 k} G_{jk} \left(z \right),\label{eq:Y-App}
\end{gather}
where $G_{ij}$ is the single particle Green function corresponding to the Hamiltonian $\scH$ at energy near the center of the band, introduced in Sec.~\ref{sec:Cavity}.
It has been shown for Levy matrices~\cite{cizeau1994theory} that the ratio $\left|Y/X\right|$ scales to zero with growing matrix size $M$ and therefore can be neglected. This argument could be extended to PBLMs considered in this paper. We confirm the validity of this approximation numerically by analyzing the probability distribution of the ratio $\left|Y/X\right|$ as a function of the matrix size $M$. In Figs.~\ref{fig:Y/X},\ref{fig:Y/X-1}, \ref{fig:Y/X-2} the distribution of $\left|Y/X\right|$ scales towards high weight at vanishing values of $\left|Y/X\right|$ with growing $M$.

\begin{figure}
\includegraphics[width=3.35in]{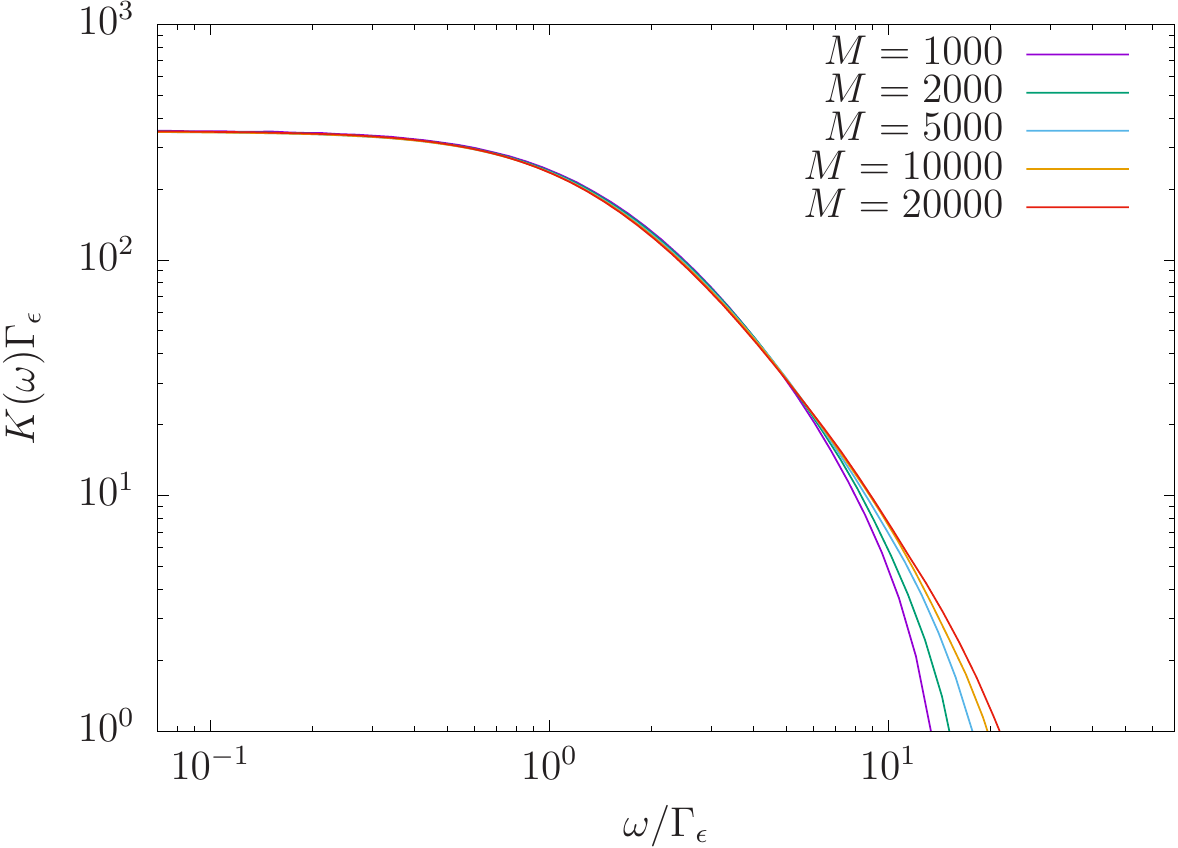}
\caption{The same as in Fig.~\ref{fig:KOmegaGammas1} but with  $\gamma=1.4$ and  fitting exponent $\varepsilon = 0.04$. \label{fig:KOmegaGammas2}}
%\end{minipage}%
%  \vspace{1cm}
%\begin{minipage}[t]{0.48\textwidth}
\end{figure}

%\onecolumngrid

\begin{figure}
\includegraphics[width=3.35in]{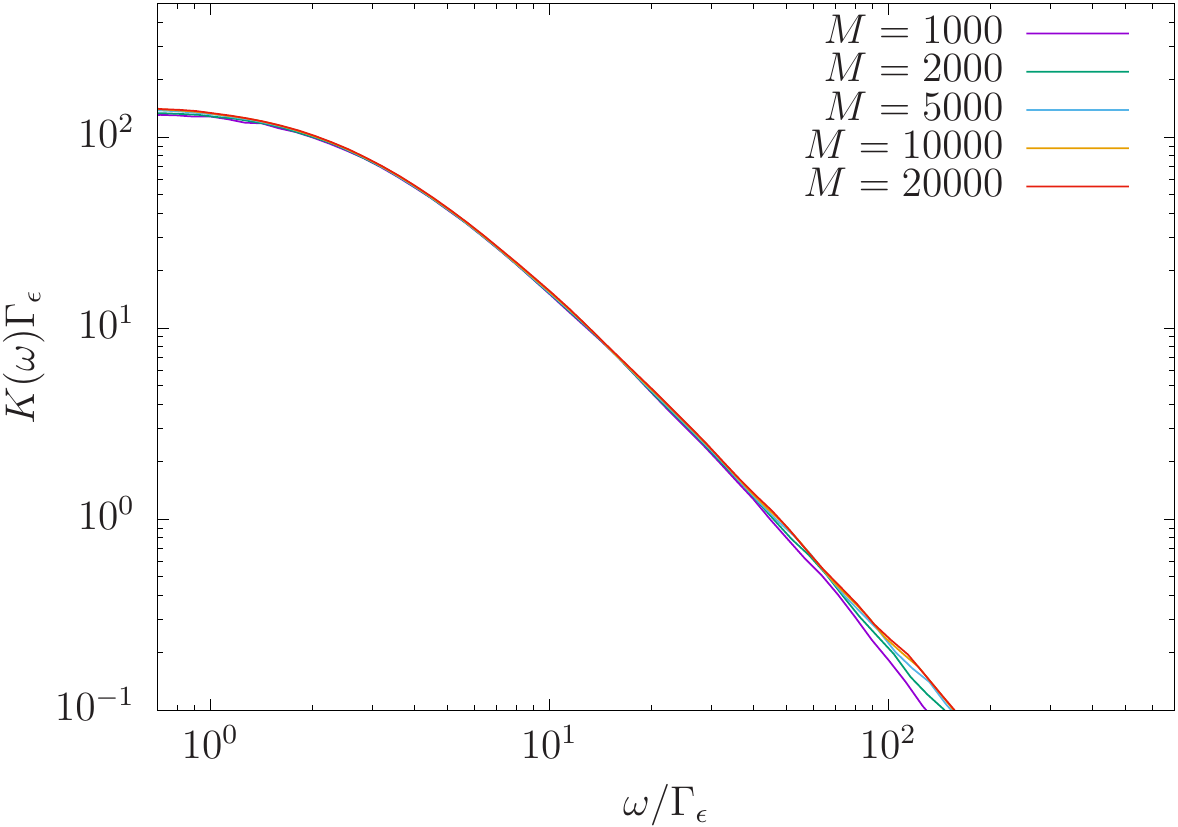}
\caption{The same as in Fig.~\ref{fig:KOmegaGammas1} but with  $\gamma=1.8$ and  fitting exponent $\varepsilon = -0.05$.   \label{fig:KOmegaGammas3}}
%\end{minipage}\hfill%
%\begin{minipage}[t]{0.48\textwidth}
\end{figure}

\begin{figure}
\includegraphics[width=3.35in]{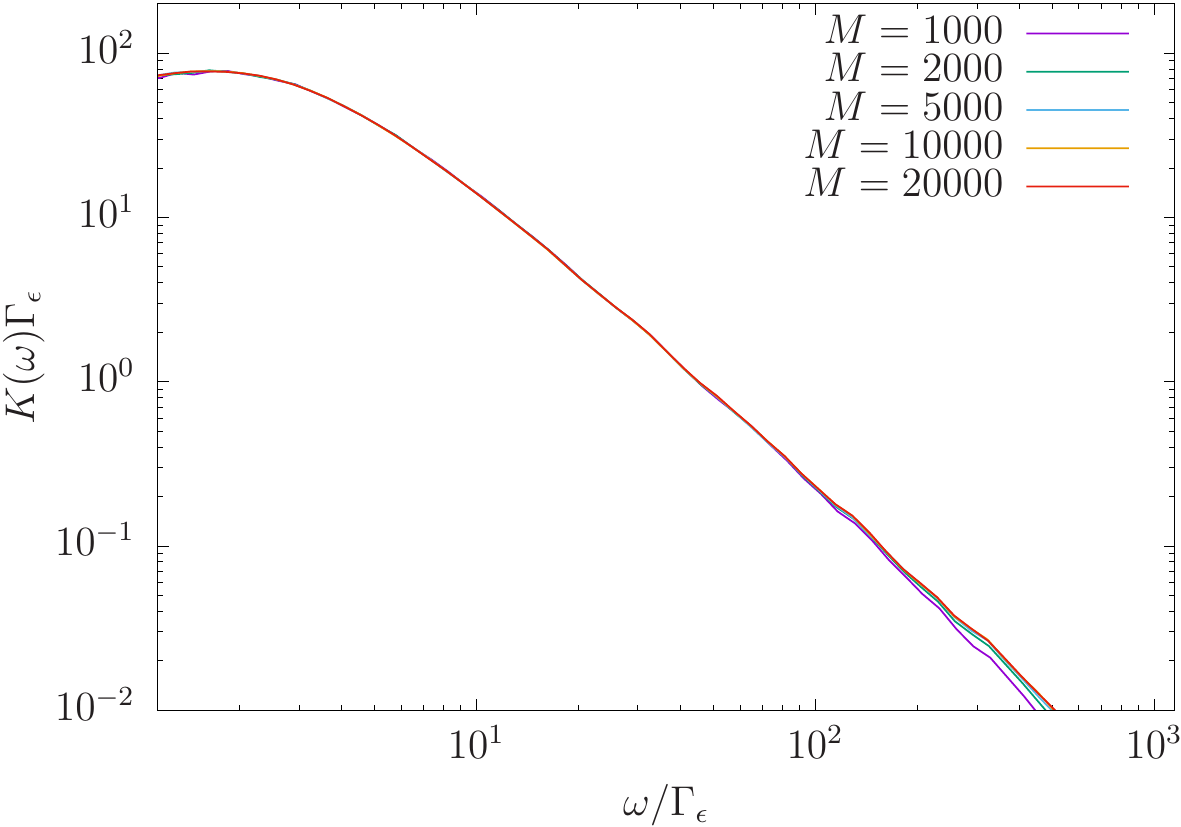}
\caption{The same as in Fig.~\ref{fig:KOmegaGammas1} but with  $\gamma=2$ and  fitting exponent $\varepsilon =-0.055$.  \label{fig:KOmegaGammas4} }
%\end{minipage}%
%\end{widetext}
\end{figure}
%\twocolumngrid

\begin{figure}[h]
\includegraphics[width=3.35in]{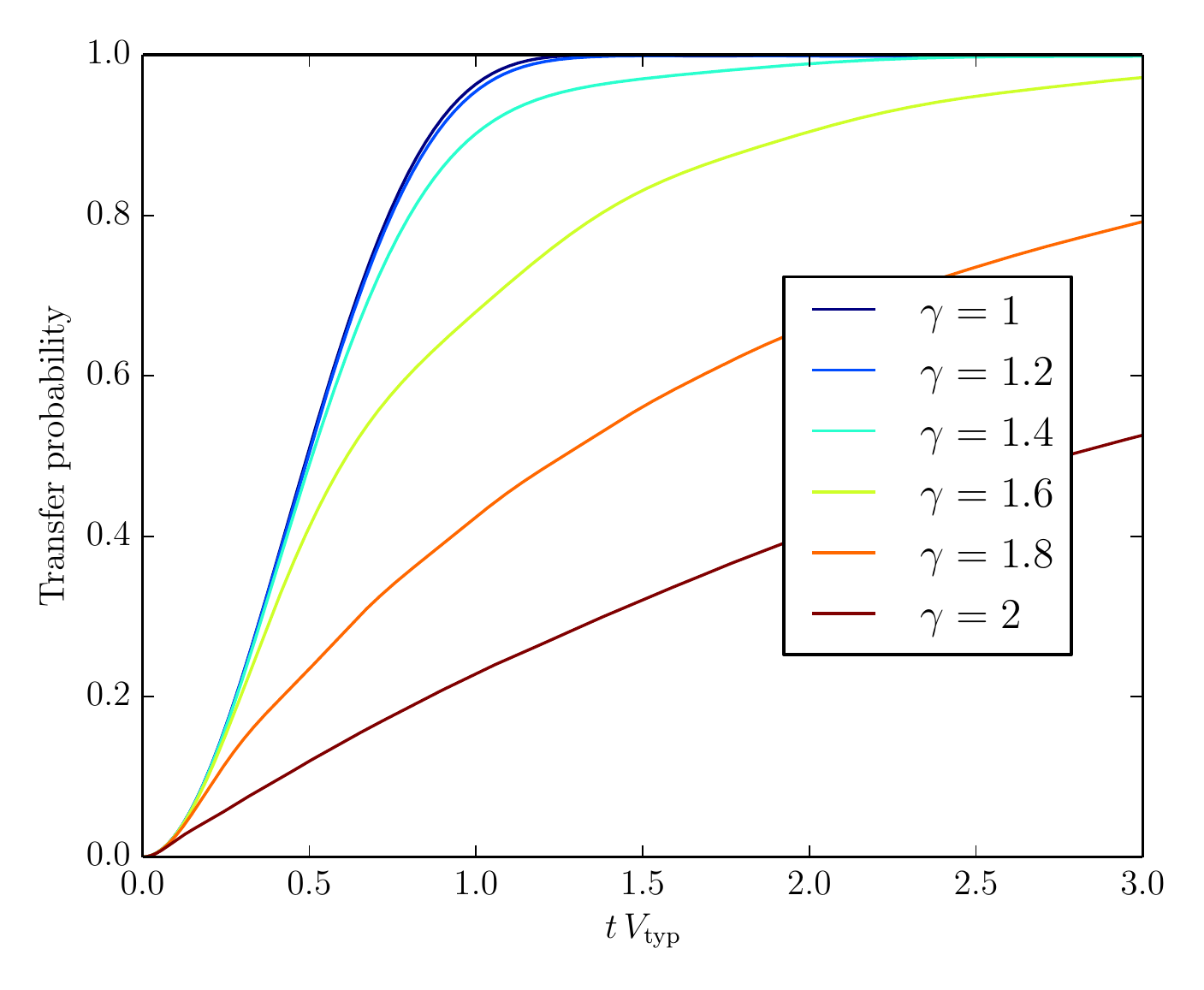}
  \caption{Population transfer probability as a function of time $t$ in units of $1/V_{\mathrm{typ}}$ for various values of parameter $\gamma =2 a$. 
    \label{fig:PTt}}
\end{figure}

\subsection{\label{sec:TransferTime} Numerical analysis of population transfer time}

\subsubsection{\label{sec:PopulationTransfer} Population transfer time from the dynamical correlator }

In addition to Fig.~\ref{fig:k4} in Sec.~\ref{sec:NumSim} of the main text, we perform a similar collapse of the dynamical correlator frequency dependence for different matrix sizes $M$ for a range of different values of $\gamma$. In Figs.~\ref{fig:KOmegaGammas1}-\ref{fig:KOmegaGammas4}  the characteristic energy scale extracted from each set of plots using this procedure $\Gamma_\varepsilon = \Gamma_{\textrm{typ}} M^\varepsilon$ corresponds to the typical mini-band width with the respective value of the parameter $\gamma$. The fitting parameter in the scaling exponent $\varepsilon$ is small for all $\gamma$ we considered and is consistent with finite size effect.

\begin{figure}[!]
\includegraphics[width=3.35in]{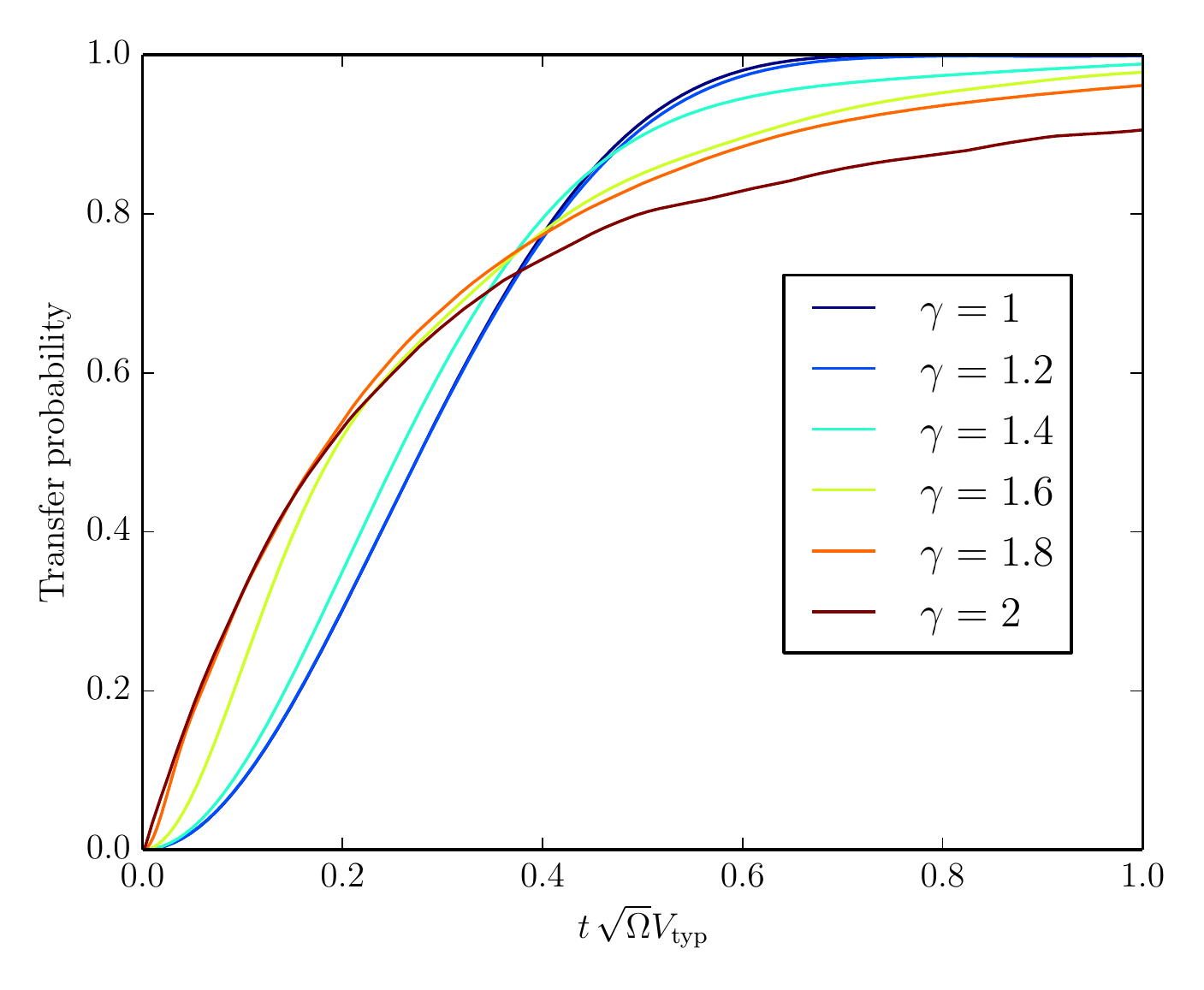}
  \caption{ Population transfer probability as a function of time rescaled with the effective mini-band width $\sqrt{\Omega}$ where the number of states in the mini-band is estimated using Fermi's golden rule $\Omega=M^{2-\gamma}$, see Eq.~(\ref{eq:GM}) of the main text.
    \label{fig:PTtM}}
\end{figure}

\subsubsection{\label{sec:TransferTime1}  Population transfer probability as a function of time }

In the main text we analyzed the complexity of the PT protocol using the solution of the full non-linear cavity equations for the size of the typical mini-band and estimated the number of states in the mini-band using the classical value of the level spacing $W/M$. In this section we analyze the scaling of the population transfer time using exact  numerical time evolution.  We contrast the population transfer time obtained from the characteristic energy scale of the frequency dependence of the dynamical correlator in Figs.~\ref{fig:KOmegaGammas1}-\ref{fig:KOmegaGammas4} with the time dependence of the transfer probability,
\begin{gather}
p(t)=\left| \langle i | \psi(t) \rangle \right|^2,
\end{gather}
where $|i\rangle$ is the initial bitstring and $|\psi(t)\rangle$ is the wave function resulting from the evolution with the impurity band Hamiltonian in transverse field $\scH$, see Sec.~\ref{sec:ensembleH}, for a time $t$, which is the quantity directly observed experimentally.  Note that in Fig.~\ref{fig:PTt} the time scale at which the transfer probability becomes of order one depends strongly on the parameter $\gamma$, reflecting the fact that the characteristic time is determined by the size of the many-body mini-band $\Gamma$ rather than the typical off-diagonal matrix element $V_{\mathrm{typ}}$.  To verify this we rescaled the unit  of time with the square root of the number of states in the mini-band $\sqrt{\Omega}$, a good approximation for the scaling of the mini-band, see Sec.~\ref{sec:FGR} for qualitative discussion and Sec.~\ref{sec:sol} for rigorous results. We observe approximate collapse of the curves for different values of $\gamma$ corroborating the PT runtime scaling presented in the main text as well as the estimate of the number of states in the mini-band.

\end{document}